\pdfoutput=1

\documentclass[british,fleqn, refpage]{article}
\usepackage[T1]{fontenc}
\usepackage[latin9]{inputenc}
\usepackage[a4paper]{geometry}
\usepackage{babel}
\usepackage{array}
\usepackage{verbatim}
\usepackage{prettyref}
\usepackage{booktabs}
\usepackage{units}
\usepackage{url}
\usepackage{multirow}
\usepackage{amsmath}
\usepackage{wasysym}
\usepackage{graphicx}
\usepackage{setspace}
\usepackage{esint}
\usepackage[authoryear]{natbib}
\usepackage{xargs}[2008/03/08]
\usepackage{nomencl}
\makenomenclature
\usepackage{subscript}
\makeatletter

\providecommand{\tabularnewline}{\\}

\addto\captionsenglish {}
\usepackage{url}
\date{} 

\usepackage{enumitem}
\setlist{itemsep=0em, topsep=0em}

\setdescription{labelwidth=-2em, leftmargin=2em}

\usepackage[compact, small]{titlesec}
\usepackage{multicol} 
\usepackage{slashbox} 

\usepackage{tocloft}

\usepackage{appendix}

\renewcommand*\nompreamble{Page where symbol is defined is listed.\begin{multicols}{2}}
\renewcommand*\nompostamble{\end{multicols}}

\usepackage[gen]{eurosym}

\usepackage[margin=10pt,font=small,labelfont=bf, format=hang]{caption}
\usepackage{threeparttable} 

\usepackage{color}

\usepackage{prettyref}
\newrefformat{cha}{Chapter~\ref{#1}} 
\newrefformat{sec}{Section~\ref{#1}} 
\newrefformat{sub}{Section~\ref{#1}} 
\newrefformat{tab}{Table~\ref{#1}} 
\newrefformat{fig}{Figure~\ref{#1}} 
\newrefformat{eq}{Equation~\ref{#1}} 
\newrefformat{app}{Appendix~\ref{#1}} 
\newrefformat{ass}{(p~\pageref{#1})}  

\usepackage{fixltx2e} 

\usepackage{units}

\AtBeginDocument{\setlength\abovedisplayskip{8pt}}
\AtBeginDocument{\setlength\belowdisplayskip{8pt}}
\AtBeginDocument{\setlength\abovedisplayshortskip{8pt}}
\AtBeginDocument{\setlength\belowdisplayshortskip{8pt}}

\usepackage[unicode=true, pdfusetitle, bookmarks=true, bookmarksnumbered=true, bookmarksopen=true, bookmarksopenlevel=2,
 breaklinks=true,pdfborder={0 0 0},backref=false,colorlinks=false]{hyperref}

\def\lofarCost{reference class LOFAR}\def\LofarCost{Reference class LOFAR}
\def\lofarDig{reference class LOFAR, all-digital beamforming}\def\LofarDig{Reference class LOFAR, all-digital beamforming}
\def\lofarRF{reference class LOFAR, RF tile beamforming}
\def\skadsCost{bottom-up SKADS}\def\SkadsCost{Bottom-up SKADS}
\def\skadsDig{bottom-up SKADS, all-digital beamforming}
\def\skadsRF{bottom-up SKADS, RF tile beamforming}
\def\refClass{reference class}
\def\botUp{bottom-up}
\def\uvc{(\emph{u},\,\emph{v}) coverage}
\def\uv{(\emph{u},\,\emph{v})}

\def\ICRAR{International Centre for Radio Astronomy Research}
\def\SPDO{SKA Program Development Office}

\def\DRM{Design Reference Mission}
\def\DRMi{DRM\textsubscript{1}} 

\def\SKAi{SKA\textsubscript{1}}
\def\SKAii{SKA\textsubscript{2}}
\def\SKAiLow{SKA\textsubscript{1}-low}
\def\SKAlow{SKA-low}

\def\CostStrategy{\textit{Draft SKA costing strategy}}

\def\hlsd{high level system description}
\def\cpf{central processing facility}
\def\stan{signal transport and networks}
\def\nip{non-imaging processor}\def\Nip{Non-imaging processor}
\def\niping{non-imaging processing}
\def\CoDR{Concept Design Review}
\def\nre{non-recurring engineering}
\def\pdf{probability distribution function}
\def\fov{field of view}  
\def\rof{radio over fibre}
\def\paf{phased array feed} 
\def\spfeed{single-pixel feed}
\def\aa{aperture array}
\def\laa{low-frequency \aa{}}
\def\fpga{field-programmable gate array}
\def\MWA{Murchison Widefield Array}
\def\LWA{Long Wavelength Array}
\def\SKADS{SKA Design Studies} 

\makeatother

\begin{document}
\global\long\def\e#1{\times10{}^{#1}}
\global\long\def\tp{(\theta,\phi)}
\def\THPBW{Half-power (3dB) beamwidth}\global\long\def\HPBW{\theta_{{\rm HP}}}
\global\long\def\Aeff{A_{{\rm e}}}
\global\long\def\Ag{A_{{\rm g}}}
\global\long\def\ltrans{\lambda_{{\rm trans}}}
\def\Tlmin{Minimum wavelength}\global\long\def\lmin{\lambda_{{\rm min}}}
\def\Tlmax{Maximum wavelength}\global\long\def\lmax{\lambda_{{\rm max}}}
\global\long\def\Nin{N_{{\rm inputs}}}
\def\TNin{Number of processing unit inputs}\global\long\def\Ninfn#1{N_{{\rm inputs\operatorname-#1}}}
\def\TNbitDig{Number of bits out of digitiser}\global\long\def\NbitDig{N_{{\rm bit\operatorname-dig}}}
\def\TNbitCFB{Number of bits out of coarse filterbank}\global\long\def\NbitCFB{N_{{\rm bit\operatorname-CFB}}}
\def\TchBW{Spectral resolution (channel width)}\global\long\def\chBW{\Delta\nu_{{\rm ch}}}
\def\TchBWcoarse{Coarse filterbank spectral resolution (channel width)}\global\long\def\chBWcoarse{\Delta\nu_{{\rm ch\operatorname-CFB}}}
\def\TchBWfine{Fine filterbank spectral resolution (channel width)}\global\long\def\chBWfine{\Delta\nu_{{\rm ch\operatorname-FFB}}}
\def\TNch{Number of channels}\global\long\def\Nch{N_{{\rm ch}}}
\def\TNchCoarse{Number of channels per coarse filterbank}\global\long\def\NchCoarse{N_{{\rm ch/CFB}}}
\def\TDt{Temporal resolution (integration time)}\global\long\def\Dt{\Delta t}
\def\TOst{Single station beam FoV}\global\long\def\Ost{\Omega_{{\rm st}}}
\def\TOtile{Single tile beam FoV}\global\long\def\Otile{\Omega_{{\rm tile}}}
\def\TOstproc{Station processed FoV}\global\long\def\Ostproc{\Omega_{{\rm proc}}}
\def\TOreq{Required processed field of view}\global\long\def\Oreq{\Omega_{{\rm req}}}
\def\TBW{Processed bandwidth}\global\long\def\BW{\Delta\nu}
\def\TFFst{Station filling factor}\global\long\def\FFst{FF_{{\rm st}}}
\def\Ttmax{Maximum scan angle}\global\long\def\tmax{\theta_{{\rm max}}}
\def\Tfmax{Maximum frequency}\global\long\def\fmax{\nu_{{\rm max}}}
\def\Tfmin{Minimum frequency}\global\long\def\fmin{\nu_{{\rm min}}}
\def\Tfsplit{Dual-band frequency split}\global\long\def\fsplit{\nu_{{\rm split}}}
\global\long\def\ftarget{\nu_{0}}
\def\Tftrans{Dense--sparse transition frequency}\global\long\def\ftrans{\nu_{{\rm transition}}}
\def\TNpol{Number of polarisations}\global\long\def\Npol{N_{{\rm pol}}}
\def\TAonT{Sensitivity metric}\global\long\def\AonT{A/T}
\global\long\def\AonTunit{{\rm m^{2}K^{-1}}}
\def\TTsys{System temperature}\global\long\def\Tsys{T_{{\rm sys}}}
\def\TTrec{Receiver temperature}\global\long\def\Trec{T_{{\rm rec}}}
\def\TTsky{Sky temperature}\global\long\def\Tsky{T_{{\rm sky}}}
\def\Tdee{Inter-element spacing}\global\long\def\dee{d}
\def\Tdeemin{Minumum inter-element spacing}\global\long\def\deemin{d_{{\rm min}}}
\def\Tdeemax{Maximum inter-element spacing}\global\long\def\deemax{d_{{\rm max}}}
\def\Tdeeavg{Average inter-element spacing}\global\long\def\deeavg{d_{{\rm avg}}}
\global\long\def\deeavgLow{d_{{\rm avg[L]}}}
\global\long\def\deeavgHigh{d_{{\rm avg[H]}}}
\global\long\def\deeavgSingle{d_{{\rm avg[S]}}}
\def\TAee{Antenna element effective area}\global\long\def\Aee{A_{{\rm e\operatorname-e}}}
\global\long\def\Age{A_{{\rm g\_e}}}
\global\long\def\Aeei{A_{{\rm e\_e},\, i}}
\def\TDDe{Antenna element directivity}\global\long\def\DDe{\mathcal{D}_{{\rm e}}}
\def\TGGe{Antenna element gain}\global\long\def\GGe{G_{{\rm e}}}
\global\long\def\Ne{N_{{\rm e}}}
\global\long\def\Nt{N_{{\rm tile}}}
\def\TNet{Number of elements per tile}\global\long\def\Net{N_{{\rm e/tile}}}
\def\TNbt{Number of dual polarisation tile beams}\global\long\def\Nbt{N_{{\rm b\operatorname-tile}}}
\def\TAgst{Station geometric area}\global\long\def\Agst{A_{{\rm g\operatorname-st}}}
\def\TAest{Station effective area}
\global\long\def\Aest{A_{{\rm e\operatorname-st}}}
\def\TDst{Station diameter}\global\long\def\Dst{D_{{\rm st}}}
\def\TDstLow{Low-band station diameter}\global\long\def\DstLow{D_{{\rm st[L]}}}
\def\TDstHigh{High-band station diameter}\global\long\def\DstHigh{D_{{\rm st[H]}}}
\def\TDstSingle{Single-band station diameter}\global\long\def\DstSingle{D_{{\rm st[S]}}}
\global\long\def\DDst{\mathcal{D}_{{\rm st}}}
\def\TLestavg{Average element--station link length}\global\long\def\Lestavg{\overline{L_{{\rm e\operatorname-st}}}}
\def\TLtilestavg{Average tile--station link length}\global\long\def\Ltilestavg{\overline{L_{{\rm tile\operatorname-st}}}}
\def\TNst{Number of stations in the array}
\global\long\def\Nst{N_{{\rm st}}}
\global\long\def\NstLow{N_{{\rm st[L]}}}
\global\long\def\NstHigh{N_{{\rm st[H]}}}
\global\long\def\NstSingle{N_{{\rm st[S]}}}
\def\TNtst{Number of tiles per station}\global\long\def\Ntst{N_{{\rm tile/st}}}
\def\TNest{Number of elements per station}\global\long\def\Nest{N_{{\rm e/st}}}
\global\long\def\NestLow{N_{{\rm e/st[L]}}}
\global\long\def\NestHigh{N_{{\rm e/st[H]}}}
\global\long\def\NestSingle{N_{{\rm e/st[S]}}}
\def\TNbst{Number of dual polarisation beams per station}\global\long\def\Nbst{N_{{\rm b\operatorname-st}}}
\def\TNbstLow{Number of dual polarisation beams per low-band station}\global\long\def\NbstLow{N_{{\rm b\operatorname-st[L]}}}
\def\TNbstHigh{Number of dual polarisation beams per high-band station}\global\long\def\NbstHigh{N_{{\rm b\operatorname-st[H]}}}
\def\TNbstSingle{Number of dual polarisation beams per single-band station}\global\long\def\NbstSingle{N_{{\rm b\operatorname-st[S]}}}
\def\TNbstavg{Average $\Nbst$ over the band}\global\long\def\Nbstavg{\overline{N_{{\rm b\operatorname-st}}}}
\def\TbeamBW{Beam--bandwidth product}\global\long\def\beamBW{\Nbstavg\Delta\nu}
\def\TNbstavgSingle{Average $\Nbst$ over the single-band}\global\long\def\NbstavgSingle{\overline{N_{{\rm b\operatorname-st[S]}}}}
\def\TNbstavgDual{Average $\Nbst$ over the dual-band}\global\long\def\NbstavgDual{\overline{N_{{\rm b\operatorname-st[D]}}}}
\def\TNbstavgLow{Average $\Nbst$ over the low-band}\global\long\def\NbstavgLow{\overline{N_{{\rm b\operatorname-st[L]}}}}
\def\TNbstavgHigh{Average $\Nbst$ over the high-band}\global\long\def\NbstavgHigh{\overline{N_{{\rm b\operatorname-st[H]}}}}
\def\TNbstavgRatio{Ratio of$\Nbstavg$ between the dual and single-band implementations}\global\long\def\NbstavgRatio{\Nbstavg{\rm (dual:single)}}

\def\TKst{Station beam taper}\global\long\def\Kst{\mathcal{K}_{{\rm st}}}
\def\TDarr{Array diameter}\global\long\def\Darr{D_{{\rm arr}}}
\def\TNbarr{Number of (phased or tied) array beams}\global\long\def\Nbarr{N_{{\rm b\operatorname-arr}}}
\def\TAearr{Array effective area}\global\long\def\Aearr{A_{{\rm e\operatorname-arr}}}
\def\TRsample{Digitiser sampling rate}\global\long\def\Rsample{R_{{\rm sample}}}
\def\TRdig{Data rate from a digitiser}\global\long\def\Rdig{R_{{\rm dig}}}
\def\TRst{Total data rate from a station}\global\long\def\Rst{R_{{\rm st}}}
\def\TRcorrin{Data rate into the correlator}\global\long\def\Rcorrin{R_{{\rm corr\operatorname- in}}}
\def\TRcorrout{Data rate out of the correlator}\global\long\def\Rcorrout{R_{{\rm corr\operatorname- out}}}
\def\TRcorroutRatio{Ratio of$\Rcorrout$ between the dual and single-band implementations}\global\long\def\RcorroutRatio{\Rcorrout{\rm (dual:single)}}
\def\TPcorr{Correlator processing load}\global\long\def\Pcorr{P_{{\rm corr}}}
\def\TPopen{Imaging processing load for an `open loop' algorithm}\global\long\def\Popen{P_{{\rm open}}}
\def\TNbstavgRatio{Ratio of$\Popen$ between the dual and single-band implementations}\global\long\def\PopenRatio{\Popen{\rm (dual:single)}}
\def\TNbstavgRatio{Ratio of$\Pclosed$ between the dual and single-band implementations}\global\long\def\PclosedRatio{\Pclosed{\rm (dual:single)}}
\def\TCfix{Fixed cost}\global\long\def\Cfix#1{C_{{\rm fix#1}}}
\def\TCvar{Variable cost}\global\long\def\Cvar#1{C_{{\rm var#1}}}
\def\TCtot{Total cost}\global\long\def\Ctot{C_{{\rm total}}}
\def\TCblock{Block cost}\global\long\def\Cblock{C_{{\rm block}}}

\noindent \begin{center}
\textbf{Cost-effective aperture arrays for SKA Phase 1: single or
dual-band}?\textbf{ }\\
Tim Colegate%
\footnote{\ICRAR{}, Curtin University, Perth, Australia (tim.colegate@icrar.org)%
}, Peter Hall\textsuperscript{1}, Andre Gunst%
\footnote{ASTRON, Dwingeloo, the Netherlands %
}\textsuperscript{,}%
\footnote{\SPDO{}%
}\\
 27/02/12
\par\end{center}
\begin{abstract}
An important design decision for the first phase of the Square Kilometre
Array is whether the low frequency component (\SKAiLow{}) should
be implemented as a single or dual-band aperture array; that is, using
one or two antenna element designs to observe the 70--450\,MHz frequency
band. This memo uses an elementary parametric analysis to make a quantitative,
first-order cost comparison of representative implementations of a
single and dual-band system, chosen for comparable performance characteristics.
A direct comparison of the \SKAiLow{} station costs reveals that
those costs are similar, although the uncertainties are high. The
cost impact on the broader telescope system varies: the deployment
and site preparation costs are higher for the dual-band array, but
the digital signal processing costs are higher for the single-band
array. This parametric analysis also shows that a first stage of analogue
tile beamforming, as opposed to only station-level, all-digital beamforming,
has the potential to significantly reduce the cost of the \SKAiLow{}
stations. However, tile beamforming can limit flexibility and performance,
principally in terms of reducing accessible \fov{}. We examine the
cost impacts in the context of scientific performance, for which the
spacing and intra-station layout of the antenna elements are important
derived parameters. We discuss the implications of the many possible
intra-station signal transport and processing architectures and consider
areas where future work could improve the accuracy of \SKAiLow{}
costing. 
\end{abstract}
\begin{spacing}{1}
\setlength{\cftbeforesecskip}{0.1em} 
\tableofcontents
\end{spacing}  

\pagebreak{}

\section{Introduction\label{sec:introduction}}

\emph{}The release of a high-level concept design \citep{GarCor10}
for the first phase of the Square Kilometre Array (\SKAi{}) represented
an important milestone for the telescope. It describes a baseline
telescope implementation with aperture array (AA) and dish receptors,
to observe the 70--450\,MHz and 0.45--3\,GHz frequency ranges respectively.
Further design details have been added through the subsequent development
of a preliminary system description \citep{Dewbij10} and a high level
system description (\citealp{DewHal11-HLSD}, hereafter HLSD). However,
an outstanding question in the SKA community is whether the 70--450\,MHz
frequency range (\SKAiLow{}) should be observed with an array composed
of a single wideband antenna element design (single-band implementation),
or with two arrays, each observing approximately half the fractional
bandwidth (dual-band implementation). 

\emph{}The choice of a single or dual-band implementation is a key
design decision for \SKAiLow{}. The system descriptions present an
overview of the telescope as a complex set of inter-connected parts
(sub-systems). Although the design of the system and its sub-systems
have been refined through \CoDR{}s (CoDRs), most of the recent system-level
studies (including the CoDRs) for the SKA have assumed a single-band
implementation for \SKAiLow{} (e.g. \citealp{AAV11-Concept} and
references therein).~Unfortunately, this means that the effects of
a dual-band implementation on the system design are not currently
well documented. However, the dual-band approach used in the LOFAR
telescope gives many insights into a putative \SKAiLow{} instrument.

Furthermore, the \emph{SKA AA CoDR panel report} \citep{DewLon11-AAreport}
recommends that the impact of the dual-band option on system design
should be considered. Indeed, not meeting the \SKAiLow{} requirements
with a single-band implementation is identified as a risk in \citet{EsArd11-Risk}.
While the\emph{ AA Concept Descriptions} document \citep{AAV11-Concept}
makes a quantitative technical analysis of the single and dual-band
approaches, to date there has been no comparison of the cost-effectiveness
of each approach. Such a comparison is important, because the chosen
approach has consequences for antenna element design, manufacture
and deployment costs, and influences the downstream signal processing
costs.

\emph{}This memo uses an elementary parametric analysis to determine
if a dual-band implementation, with twice as many antenna elements,
is significantly more expensive than the single-band implementation
described in the HLSD. We use simple algebraic equations to model
the cost and performance of each implementation, and consider the
cost drivers of the single and dual-band \SKAiLow{} at two levels.
The first level is simply the cost of the hardware required specifically
for the \laa{} sub-systems (`stations'). But because these stations
are inter-linked with other sub-systems to realise \SKAiLow{} as
a telescope, the effect of design choices within the stations is considered
throughout the system. Thus our second, higher level, analysis incorporates
costs which differ between the two implementations, such as those
of the correlator, imaging processor and \nip{} sub-systems, as well
as site-related costs specific to \SKAiLow{}. 

The representative single and dual-band implementations are chosen
for comparable sensitivity, field of view (FoV) and survey speed performance.
The single-band implementation is that which is described in the \SKAi{}
HLSD, but there is no similar guidance to the design of the dual-band
implementation. The illustrative comparison in this memo uses a canonical
form of the dual-band implementation, composed of two single-band
arrays, each of which simultaneously observes approximately equal
bandwidth ratios of~2.5:1. This produces a low-band array (70--180\,MHz)
with the same physical layout as the HLSD, and an additional high-band
array (180--450\,MHz) with a 0.75\,m spacing between antenna elements.
Alternative implementations could, for example, have overlapping bands
or a 50\,MHz minimum frequency. However, consideration of the system
implications of such comparisons is beyond the scope of the present
document.

Our parametric modelling shows that the smaller inter-element spacing
for the high-band array is a key driver in reducing the cost of the
dual-band implementation, via reduced digital data transport and processing
loads throughout the system. Although 0.75\,m inter-element spacing
is arbitrarily chosen as being half that used for the low-band and
single-band arrays, such a spacing maintains similar sensitivity performance
to the single-band array, at least for the antenna elements described
in the HLSD. 

The goal of the trade-off and decision making processes for the SKA
is to refine the design options by linking performance, cost and risk
to science returns \citep{Ste11-SEMP,Dew10-Trade-off}. This work
is not an analysis of expected telescope performance and total cost,
nor are the examined systems optimised for performance and cost. However,
by drawing upon the existing documented studies of \aa{}s and the
SKA, our analysis is intended to assist these trade-off and decision
making processes.

\section{Document structure}

\prettyref{sec:parametric-cost-modelling} outlines the parametric
cost modelling approach, along with the models and cost data sources
used. \prettyref{sec:representative-systems} details representative
single and dual-band \SKAiLow{} implementations and compares the
station sub-system costs. Other selected \SKAiLow{} sub-system costs,
which vary between implementations, are considered in \prettyref{sec:variable-cost-system-implications}.
\prettyref{sec:discussion} discusses the performance and cost trends,
uncertainties and the relevance to \SKAii{}. \prettyref{sec:other-considerations}
investigates some topical additional trade-offs: smaller station diameter,
reduced beam-bandwidth product and changed intra-station architecture.
Recommendations for further work are made in \prettyref{sec:further-work}
and conclusions set out in \prettyref{sec:conclusions}. A list of
symbols is given in Appendix~A and a summary of major assumptions
listed in \prettyref{app:assumptions}.

Although the comparisons and trade-offs are progressively developed
in each section, there may be aspects of the system which are of interest
to particular readers. These are cross-referenced as follows:
\begin{itemize}
\item station hardware sub-systems

\begin{itemize}
\item cost data sources: \prettyref{sub:cost-data-sources}
\item derived unit costs and models: \prettyref{app:models-SKAlow}
\item station design details: \prettyref{sub:station-design-details}
\item representative implementation costs: \prettyref{sub:station-sub-systems}
\item RF tile beamforming vs. all-digital beamforming: \prettyref{sub:RF-beamforming-cost-reduction}
\item uncertainty in representative implementation costs: \prettyref{sub:risk-uncertainty}
\item alternative intra-station architectures: \prettyref{sub:alternative-architectures-examples}
\item station diameter variation: \prettyref{app:Nst-Nest-trade-station} 
\item reduced processed FoV through a fixed beam--bandwidth product: \prettyref{app:reduced-FoV-station-hardware-costs}
\end{itemize}
\item variable system costs

\begin{itemize}
\item site-related costs and models: \prettyref{app:models-SKA-site-related} 
\item \cpf{} sub-system costs and models: \prettyref{app:central-processing} 
\item representative implementation costs: \prettyref{sec:variable-cost-system-implications}
\item station diameter variation: \prettyref{app:Nst-Nest-trade-system}
\item reduced processed FoV through a fixed beam--bandwidth product: \prettyref{app:reduced-FoV-system-implications}
\end{itemize}
\item station power demand

\begin{itemize}
\item power demand  models: \prettyref{app:power-demand}
\item representative single and dual-band implementations: \prettyref{sub:power}
\item alternative intra-station architectures: \prettyref{sub:alternative-architectures-examples}
\item station diameter variation: \prettyref{app:Nst-Nest-trade-system}
\end{itemize}
\item general cost trends

\begin{itemize}
\item single vs. dual-band comparison: \prettyref{sub:cost-trends}
\item station diameter variation: \prettyref{sub:Nst-Nest-trade}
\item reduced processed FoV through a fixed beam--bandwidth product: \prettyref{sub:reducing-FoV}
\end{itemize}
\item station performance

\begin{itemize}
\item single vs. dual-band comparison: \prettyref{sub:performance-trends}
and \prettyref{app:station-performance}
\item hierarchical beamforming \prettyref{sub:hierarchical-beamforming-performance}
\end{itemize}
\end{itemize}

\section{Parametric cost modelling\label{sec:parametric-cost-modelling}}

The key to successful parametric analysis is to have a scalable model
which is sufficiently general, but still describes the system with
enough completeness and accuracy. To describe the actual cost of the
system, the model should provide a cost estimate and associated uncertainty.
Early in the project, the uncertainties are large, and as the project
progresses and the cost models are refined, the uncertainties reduce.
This is known as the `cone of uncertainty', wherein the cost estimate
eventually converges on the actual cost \citep{GAO09}. 

Defining a scalable model for the SKA is challenging because the project
is currently in the design definition phase, where many design options
and architectures are available, and a complete set of requirements
is still being developed. For these reasons, the model for the parametric
analysis is necessarily simple; in this case it is captured in a dozen
scalable blocks. Although such a model has high uncertainties, it
provides useful insight at this point and also indicates prime areas
for further study.

\subsection{Parametric analysis\label{sub:parametric-analysis}}

\begin{table}[!t]
\caption{Recommended use of cost estimation methodologies at various stages
in a project. SKA is currently in the design definition stage. Adapted
from \citet{NAS08}. \label{tab:cost-methods}}
\begin{tabular}{>{\raggedright}p{0.14\textwidth}>{\centering}p{0.14\textwidth}>{\centering}p{0.14\textwidth}>{\centering}p{0.14\textwidth}>{\centering}p{0.14\textwidth}>{\centering}p{0.14\textwidth}}
\hline 
 & Early concept definition & Design definition & Detailed design & Construction and deployment & Operations, support and disposal\tabularnewline
\hline 
Parametric & {\large $\CIRCLE$} & {\large $\CIRCLE$} & {\large $\RIGHTcircle$} & {\large $\RIGHTcircle$} & {\large $\Circle$}\tabularnewline
Reference class (analogous) & {\large $\CIRCLE$} & {\large $\RIGHTcircle$} & {\large $\RIGHTcircle$} & {\large $\RIGHTcircle$} & {\large $\Circle$}\tabularnewline
Bottom-up (engineering) & {\large $\RIGHTcircle$} & {\large $\RIGHTcircle$} & {\large $\CIRCLE$} & {\large $\CIRCLE$} & {\large $\CIRCLE$}\tabularnewline
\multicolumn{6}{l}{$\CIRCLE$ Primary\qquad{} $\RIGHTcircle$ Applicable\qquad{} $\Circle$
Not applicable}\tabularnewline
\hline 
\end{tabular}
\end{table}

Parametric analysis is a useful systems engineering tool, enabling
understanding and exploration of performance and cost trends, and
of key trade-offs. It involves defining the system with a set of variable
parameters, and then modifying one or more of these parameters to
model the effect on the performance and cost of the system. In a complex
system such as the SKA, many parameters are closely inter-linked. 

In terms of parametric cost estimation, the \emph{2008 NASA cost estimating
handbook} \citep{NAS08} lists parametric cost models as one of three
cost estimation methodologies. The second methodology is analogous
(or reference class) costing, where cost data from similar sub-systems
(or projects) is used. The cost data is adjusted, depending on the
relative complexity of the projects, technological improvements, inflation
and other factors. The third is an engineering (bottom-up) cost estimate,
which builds up a cost estimate from all the individual cost elements
in the system. The \botUp{} approach requires a good understanding
of all the costs involved (`cost coverage'), as outlined in the
\CostStrategy{} \citep{McC10-cost-strategy}. \prettyref{tab:cost-methods}
compares the project stages where these methodologies are most useful;
the SKA is currently in the design definition stage. Parametric costing
is the primary method in the earlier stages of the project, facilitating
high-level trade-offs when there is insufficient data for a detailed
approach. A \botUp{} costing is more useful later in the project,
as greater design detail and actual cost data are accumulated \citep{NAS07}.

A complete parametric analysis would allow for the design to be optimised
for cost while still meeting the system requirements (scientific,
environmental and operational requirements). These requirements are
being developed in the current design definition phase and will be
documented in the requirements specifications \citep{Ste11-SEMP}.
Importantly, the design will be optimised for the life-cycle cost,
which is the total cost of ownership over the lifetime of the system
\citep{NAS07}. However, all costs in the life-cycle need to be considered
to make the like-for-like trade-off. For the SKA, costs such as construction,
site operations, power infrastructure and software development must
be considered in life-cycle costing \citep{McC10-cost-strategy}.
These are additional to the hardware and operations costs of the sub-systems,
on which most of the SKA costing focus has been placed thus far. There
are also some SKA project overhead costs, such as contingency, taxes and system integration costs,
which do not contribute to current cost estimates \citep{McC10-cost-strategy}.
Not all these life-cycle costs are currently available or sufficiently
understood, nor are the requirements specifications complete. For
these reasons, this work does not give a final cost, but instead makes
a comparative cost analysis of the single and dual-band \SKAiLow{}
implementations.

\subsection{Scalable parametric cost models\label{sub:parametric-models}}

The parametric cost model encodes the telescope system design using
a small number of blocks, where each block describes the quantity
and cost scaling relationships of one or more sub-systems. Relatively
simple algebraic equations are used to describe the cost of these
blocks as a function of one of more variables (parameters). The scalable
models enable performance and cost exploration and trade-offs; these
are much more difficult to make with, for example, a \botUp{} cost
estimate that assumes a specific design. However, the parametric equations
themselves are derived from one or more cost estimates, and \refClass{}
and \botUp{} cost estimates can be suitable sources of data.

The cost data sources usually have itemised costs; each cost item
is assigned to a particular block, producing an aggregate cost for
each block. Because this aggregate cost is for a given set of parameters
(as per the HLSD in this case), solving the parametric equation determines
the value of the cost coefficients. The cost of the block can then
be expressed as a function of its variable parameters, and the total
cost of the model calculated from the summation of the quantity and
cost product of each block.

For example, the parametric equation of a particular block is given
by
\begin{equation}
\Cblock=\Cfix{}+\Cvar{}\, x,
\end{equation}
where $\Cblock$\nomenclature[sCblock]{$\Cblock$}{\TCblock{}} is
the cost of a single `instance' (occurrence) of the block, $\Cfix{}$
and $\Cvar{}$ are the cost coefficients (unit costs) and $x$ is
a scaling parameter. Say an aggregate cost estimate of the block gives
$C_{{\rm block}}=\euro200$, for $x=8$. Using a reasonable estimate
of what proportion of cost is fixed (say 20\%), the unit costs for
the block can be determined: $\Cfix{}=\euro40$ and $\Cvar{}=\euro20$.
The cost of the block can then be determined for other values of $x$
(within reasonable design limitations). The total cost $\Ctot$\nomenclature[sCtot]{$\Ctot$}{\TCtot{}}
is simply the summation of the product of the quantity ($N$) and
cost ($C$) of each block:
\begin{equation}
\Ctot=N_{{\rm blockA}}C_{{\rm blockA}}+N_{{\rm blockB}}C_{{\rm blockB}}+\ldots\label{eq:cost-total}
\end{equation}

Although these algebraic equations do not capture all the nuances
of a design, they do provide a scalable, first-order cost estimate.
Also, as is done in the present analysis, a direct comparison of different
cost data sources can be made by running the model with the different
unit costs derivations, but using the same parametrisation. This highlights
areas of the design where costs differ, thus are most uncertain and
require further investigation.

The \SKAi{} high level system description \citep[HLSD,][]{DewHal11-HLSD}
is a useful starting point to create a parametric cost model for \SKAiLow{},
because it defines a system hierarchy which describes how the sub-systems
relate to each other. For example, the \laa{}s, \stan{}, signal
processing, computing and software, and infrastructure are all immediate
sub-systems of the telescope system. Although the HLSD will evolve
as the system requirements are refined, it forms a `representative
system' as a common basis for the analysis of sub-system performance
and cost. 

To model the differences between the single and dual-band \SKAiLow{},
we decompose the \laa{} sub-system into another level of sub-systems.
As mentioned in \prettyref{sec:introduction}, the term `\SKAiLow{}'
encompasses hardware specifically related to the \laa{} sub-system
as well as other SKA sub-systems. Although not explicitly defined
in the HLSD, the \laa{} sub-system approximately describes the hardware
for the \SKAiLow{} stations. 

\begin{figure}[!t]
\noindent \begin{centering}
\includegraphics[width=0.3\textwidth]{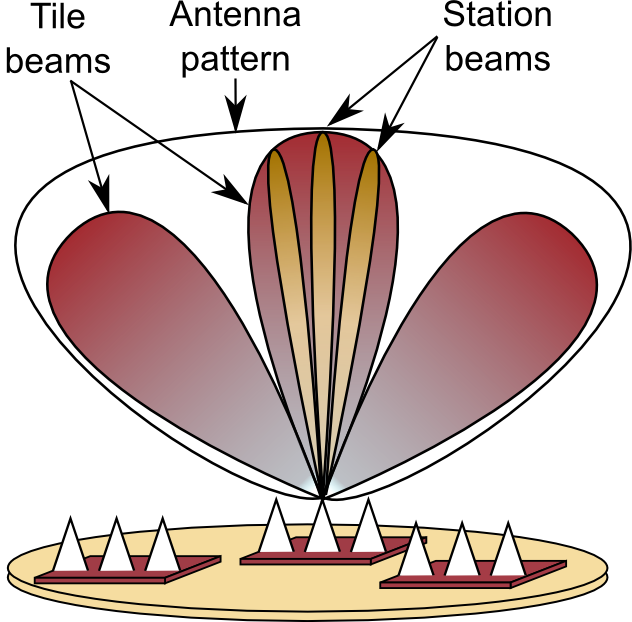}
\par\end{centering}

\noindent \centering{}\caption{Aperture array schematic, showing the antenna elements (triangles),
tiles (squares) and the station (disc), and their corresponding beam
pattern.\label{fig:beam-schematic}}
\end{figure}

The \SKAiLow{} antenna elements are grouped into `tiles' and `stations',
as shown in \prettyref{fig:beam-schematic}, to reduce the data transport
and processing loads. This grouping of antenna elements may affect
the physical layout or may only change the signal processing architecture;
the partitioning of the processing is termed hierarchical beamforming.
The `processed FoV' is synthesised from the formation of multiple
station beams, which are cross-correlated in the telescope\textquoteright{}s
signal processing system. 

The HLSD also describes the geographical layout (configuration) of
the stations. Approximately half the AA stations are located in a
closely packed `core' region, and the others placed with exponentially
increasing density away from the core. A similar layout applies to
the dishes. Because the AA and dish cores are densely packed, these
are located nearby to each other but separate\label{ass:low-high-cores}.
At larger radii from the core, the stations and dishes are co-located,
so that the data transport, timing signals and power distribution
infrastructure can be shared.

Although the \SKAiLow{} sub-systems are captured in a small number
of blocks to maintain clarity, it is important that the quantity of
each of these blocks still scales correctly when the parameters are
varied. This approach is similar to some of the previous SKA costing
efforts \citep{ChiCol07,BolAle09}, except the blocks in this analysis
describe the system at a higher level. Because only a small number
of blocks are used, most of the results in this analysis have been
calculated using a spreadsheet. However, the parametric models have
been developed with a view to transferring them to SKACost, an SKA
performance and cost modelling tool. (See \citealp{ForBol10} for
a description of the tool.) SKACost will allow further exploration
of trade-offs in the \SKAiLow{} design space, and enable statistical
treatment of uncertainties. SKACost is used to make the preliminary
uncertainty analysis in \prettyref{sub:statistical-uncertainty-analysis}.

\begin{figure}[!t]
\centering{}\includegraphics[width=1\textwidth]{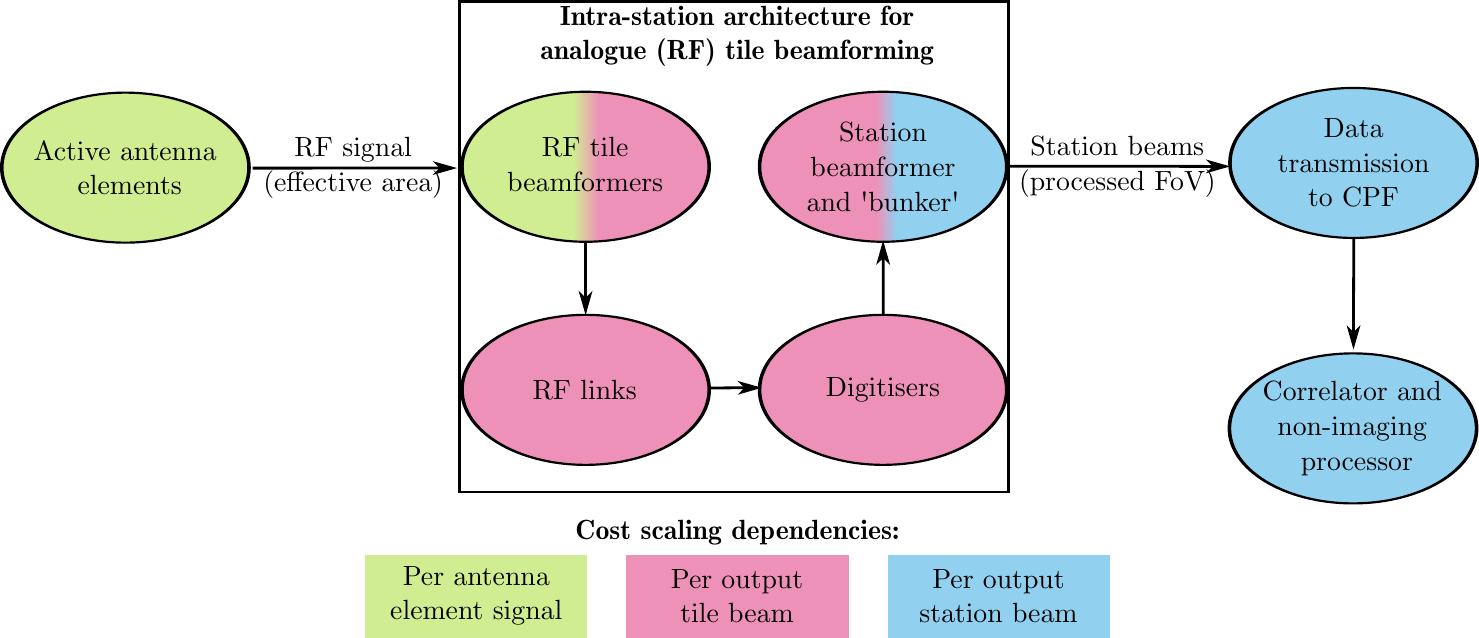}\caption{Conceptual diagram showing the general cost scaling dependencies of
key blocks (ovals), for the RF tile beamforming intra-station architecture.
The gradient indicates that both scaling dependencies apply.\label{fig:RF-tile-architectures-conceptual}}
 
\end{figure}

The blocks (sub-systems) follow the elemental signal path, in an approach
similar to \citet{Hal04a} and \citet{HorChi04}. The blocks used
in this analysis to describe the elemental signal path are:
\begin{description}
\item [{active~antenna~element:}] reception and amplification at the
antenna element
\item [{RF~tile~beamformer:}] analogue beamforming of the elements in
a tile. Digital tile beamforming is also possible, see \prettyref{sub:digital-hierarchical-beamforming}
\item [{RF~link:}] analogue signal transport from the active antenna element
or the tile beamformer
\item [{digitiser:}] digitisation
\item [{station~beamformer:}] coarse channelisation (filterbank) and digital
beamforming of elements or tiles in the station
\item [{station~`bunker':}] controlled environment and infrastructure
at the station processing node to house the beamformer hardware
\item [{digitiser--bunker~link:}] digital signal transport from the digitiser
to the bunker 
\item [{central~processing~facility~(CPF):}] central signal processing
and science computing sub-systems (specifically, the correlator, imaging
and non-imaging processing). The CPF serves both the AA and dish arrays.
Some parts of the CPF may be located off-site
\item [{data~transmission~to~CPF~(station--CPF~link):}] digital signal
transmission from the station processing node to the central processing
facility (excludes network infrastructure).
\end{description}
These blocks assume a time to frequency domain transformation and
cross-correlation `FX' correlator architecture, being the most cost-effective
architecture for the SKA (see \prettyref{app:FX-architecture}).\label{ass:FX-correlator} 

The blocks are combined to form a specific signal transport and processing
architecture within a station, shown conceptually in \prettyref{fig:RF-tile-architectures-conceptual}.
A station will have some number of antenna elements (to realise effective
area, hence sensitivity) and produce some number of station beams
(to form the processed FoV). The rectangular box in \prettyref{fig:RF-tile-architectures-conceptual}
encompasses the intra-station signal transport and processing architecture.
The exact path of the signal from one block to the next depends on
the architecture chosen; this is discussed in \prettyref{sub:signal-architectures}.

\prettyref{fig:RF-tile-architectures-conceptual} also shows the general
cost scaling dependencies of each block, as relevant to the single
versus dual-band comparison. The dependencies are linear parametric
equations, with the key variable parameters being the number of antenna
element signals, output beams formed by the tile beamformer and output
beams formed by the station beamformer. For the digital signal transport
links, the number of signals or beams transmitted is a proxy for the
data rate transmitted. The scaling relationships and unit costs of
each block are detailed in \prettyref{app:models-SKAlow}. The two
intra-station signal transport and processing architectures analysed
in this work are discussed in the next section.

\subsection{Cost data sources \label{sub:cost-data-sources}}

\begin{table}[!t]
\centering{}\caption{Sub-system cost data sources\label{tab:cost-sources}}
\begin{threeparttable}%
\begin{tabular}{>{\raggedright}p{0.28\textwidth}cc>{\centering}p{0.15\textwidth}>{\centering}p{0.1\textwidth}}
\toprule 
 & \multicolumn{4}{c}{Cost data source}\tabularnewline
\cmidrule{2-5} 
Sub-system & SKADS\textsuperscript{a}  & LOFAR\textsuperscript{a} & Signal Processing CoDR\textsuperscript{b} & Professional opinion \tabularnewline
\midrule
Active antenna element & {\large $\times$} & {\large $\times$} &  & \tabularnewline
RF tile beamformer &  & {\large $\times$} &  & \tabularnewline
RF links & {\large $\times$} & {\large $\times$} &  & \tabularnewline
Digitiser & {\large $\times$} & {\large $\times$} &  & \tabularnewline
Digitiser--bunker links & {\large $\times$} &  &  & \tabularnewline
Station beamformer & {\large $\times$} & {\large $\times$} &  & \tabularnewline
Station bunker & {\large $\times$} & {\large $\times$} &  & \tabularnewline
Data transmission to CPF & {\large $\times$}\textsuperscript{c} &  &  & \tabularnewline
Correlator &  &  & {\large $\times$} & \tabularnewline
\Nip{} &  &  & {\large $\times$} & \tabularnewline
Correlator--computing data transport &  &  &  & {\large $\times$}\tabularnewline
Computing (imaging processor) &  &  &  & {\large $\times$}\tabularnewline
Deployment  &  &  &  & {\large $\times$}\tabularnewline
Site preparation &  &  &  & {\large $\times$}\tabularnewline
\bottomrule
\end{tabular}\begin{tablenotes}
\small
\item[a] \citet{FauVaa11-Deployment}
\item[b] \citet{Tur11-Cost}
\item[c] \citet{BolAle09}
\end{tablenotes}
\end{threeparttable}
\end{table}

\begin{table}[!t]
\caption{Costing methodologies and data sources \label{tab:methodologies-sources-matrix} }

\centering{}%
\begin{tabular}{llc>{\centering}p{0.15\textwidth}}
\toprule 
 &  & \multicolumn{2}{c}{Cost data source}\tabularnewline
\cmidrule{3-4} 
 &  & SKADS & LOFAR \tabularnewline
\midrule
\multirow{2}{*}{Costing methodology} & Bottom-up & {\large $\times$} & \tabularnewline
 & Reference class &  & {\large $\times$}\tabularnewline
\midrule
 & Tile beamforming & digital & RF\tabularnewline
Intra-station architecture & Tile--station signal transport  & digital & RF\tabularnewline
 & Station beamforming & digital & digital\tabularnewline
\bottomrule
\end{tabular}
\end{table}

The cost data sources for this analysis are shown in \prettyref{tab:cost-sources}.
Two cost estimates were developed in the AA CoDR \citep{FauVaa11-Deployment}
for the \SKAiLow{} stations; one based on the \SKADS{} (SKADS) work
\citep[e.g.][]{FauAle10}, the other extrapolated from the existing
Low Frequency Array (LOFAR) telescope%
\footnote{\url{www.lofar.org}%
}. Although these estimates describe stations which achieve similar
sensitivity and FoV performance, they present two alternative intra-station
signal transport and processing architectures, and also use different
cost estimation methodologies and assumptions, as summarised in \prettyref{tab:methodologies-sources-matrix}. 

The principal architectural differences between the two estimates
are how and where the digitisation and hierarchical beamforming is
performed. The SKADS architecture uses all-digital beamforming, where
both the tile and station beamforming are done digitally, while the
LOFAR architecture uses analogue tile beamforming. These are the architectures
considered in our study; the cost and performance implications of
some other intra-station architectures on the \SKAiLow{} sub-system
costs are discussed in \prettyref{sub:signal-architectures}.

\begin{figure}[!t]
\includegraphics[width=1\textwidth]{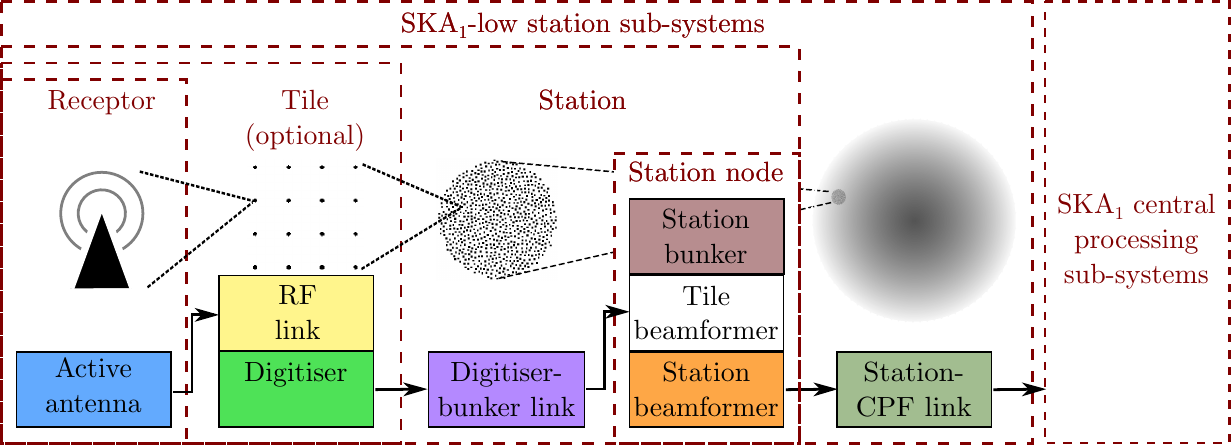}

\caption{Schematic of the all-digital beamforming architecture, showing the
signal path through the \SKAiLow{} sub-systems to the central processing
facility. Two or more stages of hierarchical beamforming are assumed
to occur within the station beamformer block.\label{fig:schematic-all-dig}}
\end{figure}

\begin{figure}[!t]
\includegraphics[width=1\textwidth]{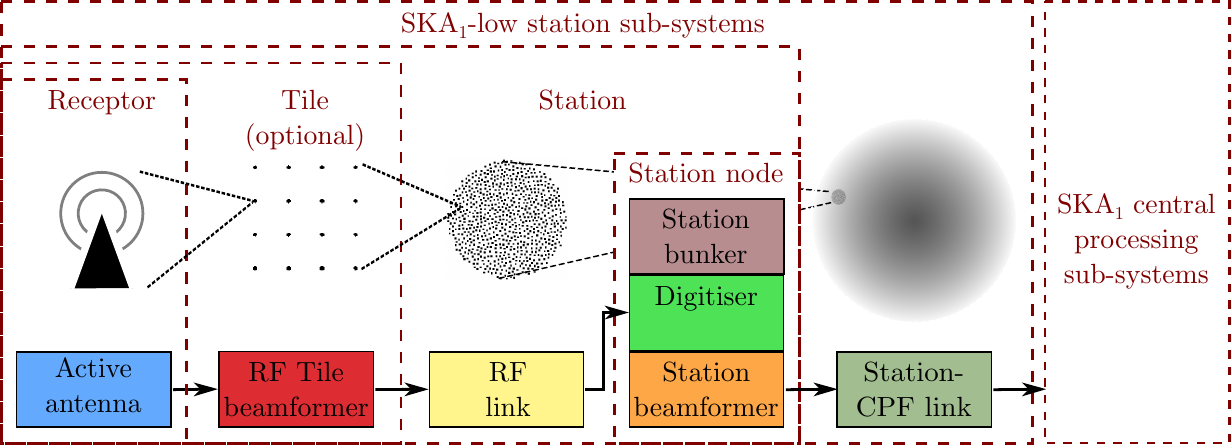}

\caption{Schematic of the RF tile beamforming architecture, showing the signal
path through the \SKAiLow{} sub-systems to the central processing
facility.\label{fig:schematic-RF-tile}}
\end{figure}

The parametric models for the stations are based on these two cost
estimates and their architectures. The all-digital beamforming architecture,
as broken down into the sub-system blocks for this analysis, is shown
in \prettyref{fig:schematic-all-dig}. In this architecture, the signals
are digitised close to the antenna elements, but no beamforming occurs
at the tiles. Instead, two or more stages of hierarchical beamforming
are assumed to occur within the station beamformer block. This differs
slightly from \citet{FauVaa11-Deployment}, where a first stage of
beamforming is done at the tile. \prettyref{fig:schematic-RF-tile}
shows the sub-system blocks for the second architecture, which uses
a first stage of analogue (RF) beamforming at the tile. In this architecture,
the analogue signals are transported to the station processing node
and digitisation occurs at that node. 

Alongside the all-digital and RF tile beamforming architectures, we
consider the different cost estimation methodologies and assumptions
used for the two estimates in \citet{FauVaa11-Deployment}. The cost
estimate for the all-digital beamforming architecture primarily uses
the \botUp{} cost method described earlier; we term this the `\skadsCost{}'
estimate. The cost estimate for the RF tile beamforming architecture
uses LOFAR station costs to make an analogous cost estimate of the
\SKAiLow{} station, hence we term it the `\lofarCost{}' estimate.
Although it is an example of a \refClass{} cost estimate, it is implemented
at a more detailed sub-system level, instead of at the system level
where the \refClass{} estimation methodology is often applied. 

Understanding the epoch of the source cost estimate is also important,
especially in regards to the digital hardware. Digital technology
advancements can often be generalised by exponential laws; the most
common being Moore's law, where the cost of an equivalent digital
product halves every 1.5--2 years. Thus the year for which the cost
data applies, and the technical capability of the hardware to which
it applies, are relevant factors. The \skadsCost{} estimate makes
the assumption that technology advances reduce costs and that \SKAiLow{}
construction commences in 2016. In contrast, the \lofarCost{} estimate
is based on 2007 technology and is only for an output bandwidth of
approximately 50\,MHz. We make the assumption that by 2016, newer
technology will allow for the processing of the full 380\,MHz bandwidth
for the same cost.\label{ass:LOFAR-50MHz-BW}

The unit costs for the station sub-systems are derived from the cost
data in \citet{FauVaa11-Deployment}, as described in \prettyref{app:models-SKAlow}.
The derivation method is that described earlier, where the cost data
is aggregated for each block, and the unit costs are derived from
these block aggregates and the parameters in the HLSD (see \prettyref{tab:single-system-details}).
The costs, especially some of the digital costs, are necessarily simplistic
in order to transcend the design details and multitude of options
for each sub-system; however, the simplifications do introduce another
level of uncertainty into the analysis. The costing of the dual-band
array uses a similar parametrisation process, but applies multipliers
(discounts) to the unit costs to account for the design differences
between sub-systems of the single-band and low and high-band arrays,
as discussed further in \prettyref{sub:station-design-details}.

The \skadsCost{} unit costs are consistently lower than the \lofarCost{}
unit costs. From \prettyref{tab:blocks-cost} of \prettyref{app:models-SKAlow},
the relative difference in cost is a factor of 2--3 for the active
antenna element and fixed cost portion of the station bunker, 4.5
for the digitiser, 26 for the station beamformer and 43 for the variable
cost portion of the station bunker. While useful for this first-order
analysis, these cost differences show that more work is required to
confirm the accuracy and precision of the unit cost derivation and
the cost estimates themselves, as discussed in Sections \ref{sub:risk-uncertainty}
and \ref{sec:further-work}. Note that the \botUp{} methodology does
not necessarily produce a lower cost estimate than the \refClass{}
methodology. Adjustments such as technology improvements and learning
curves for mass production can be used to change the cost estimate
\citep{NAS08}. Thus the relative costs also depend on what cost
adjustments are applied to each cost estimate, as well as the method
of arriving at the cost data.

Many of the station sub-systems have a cost data source from both
the \skadsCost{} and \lofarCost{} estimates (\prettyref{tab:cost-sources}).
Not all aspects of each cost estimate are comparable because of the
different architectures used. However, the active antenna element,
digitiser, station beamformer and station infrastructure sub-systems
are comparable and these costs comprise most of the \SKAiLow{} station
costs. These comparable sub-systems mean that the unit costs derived
from the \skadsCost{} and \lofarCost{} estimates can be applied
to either the all-digital or RF tile beamforming architectures, allowing
four different cost scenarios to be modelled:
\begin{itemize}
\item \lofarRF{}
\item \skadsRF{}
\item \skadsDig{}
\item \lofarDig{}
\end{itemize}
Because the \skadsRF{} and \lofarDig{} architectures are extrapolated,
they are less optimised for technical performance and cost than the
other two scenarios; this is an extra source of uncertainty for those
scenarios. Uncertainties are analysed in \prettyref{sub:risk-uncertainty}.

The costs discussed thus far are for the station sub-system hardware.
\prettyref{sec:variable-cost-system-implications} considers hardware
cost for the \cpf{} sub-systems. But even with these costs included,
this does not represent the total telescope cost.\label{ass:costs-not-price-to-build}
The \CostStrategy{} \citep{McC10-cost-strategy} discusses some of
the other costs to be considered for the SKA. The sub-system hardware
cost is included in the present analysis, although with some caveats
discussed in \prettyref{app:models-SKAlow}. The costs excluded in
the analysis are: 
\begin{itemize}
\item sub-system hardware operations$^{\dagger}$
\item temporary construction and integration facilities
\item site operations infrastructure
\item construction$^{\dagger}$ (including network trenching)
\item annual fibre costs
\item antenna siting costs (inclusive of foundations)
\item land acquisition
\item power infrastructure$^{\dagger}$
\item software development

$^{\dagger}$these costs can depend on the intra-station architecture,
see \prettyref{sub:signal-architectures}. 

\end{itemize}
Also listed in the \citet{McC10-cost-strategy} are project overheads,
which are outside the scope of this analysis.

However, many of the excluded costs remain approximately constant
between the single and dual-band implementations. Hence \prettyref{sec:variable-cost-system-implications}
makes some zeroth-order estimates of those excluded costs that will
vary between implementations, namely site preparation and antenna
element deployment costs. This allows for a comparison to be made
in the absence of all the cost information.

\section{Single and dual-band representative implementations\label{sec:representative-systems}}

This analysis is based on a single and dual-band representative system,
rather than an optimised system. Most of the recent \SKAi{} design
work, as presented in the sub-system concept design reviews, has been
developed with the \SKAi{} high level system description in mind.
For this reason, the single-band \SKAiLow{} in the HLSD is used as
the starting point for a comparison of the single and dual-band implementations.
No dual-band implementation is described in detail in the\emph{ AA
Concept Descriptions} document; in our study, the canonical (not optimised)
dual-band design has been chosen so that its scientific performance
will be similar to the single-band system. 

\begin{figure}[!t]
\noindent \centering{}\includegraphics[width=0.55\textwidth]{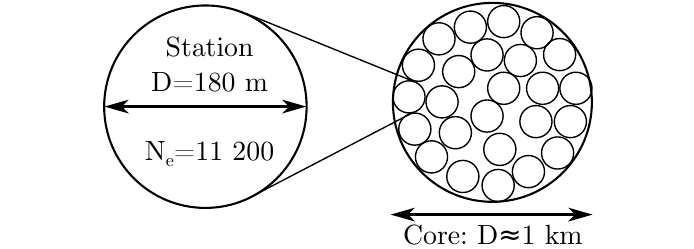}\caption{Representation of a single-band station within the densely packed
core region.\label{fig:single-station}}
\end{figure}
\begin{table}[!t]
\begin{centering}
\caption{Single-band \SKAiLow{} system details, as per the HLSD except where
noted.\label{tab:single-system-details} }
\begin{threeparttable}%
\begin{tabular}{>{\raggedright}p{0.4\textwidth}>{\centering}p{0.25\textwidth}}
\toprule 
Frequency range & 70--450\,MHz\tabularnewline
Number of stations $\Nst$\nomenclature[sNst]{$\Nst$}{\TNst} & 50\tabularnewline
Average spacing between elements $\deeavg$\nomenclature[sd_eeavg]{$\deeavg$}{\Tdeeavg{}} & 1.5\,m ($\lambda/2$ at 100\,MHz) \tabularnewline
\TNet{}\textsuperscript{a} $\Net$\nomenclature[sNet]{$\Net$}{\TNet{}} & 16\tabularnewline
Station beam taper\textsuperscript{b} $\Kst$\nomenclature[sK~st]{$\Kst$}{\TKst{}} & 1.02\textsuperscript{c}\tabularnewline
Number dual-polarisation beams per station (averaged over the band)
$\Nbstavg$ & 210\textsuperscript{d}\tabularnewline
Dense-sparse transition\textsuperscript{e} & 115\,MHz\tabularnewline
Gain of an isolated element  & 6.2\,dBi\tabularnewline
\bottomrule
\end{tabular}\begin{tablenotes}
\small
\item[a] For the RF first-stage beamforming architecture.
\item[b] Where station beam FoV $\Ost=\pi/4(\Kst\lambda/\Dst)^{2}$.\nomenclature[s~Ost]{$\Ost$}{\TOst{}} 
\item[c] A uniform aperture distribution is assumed \citep{RohWil04}, whereas the HLSD specifies $\Kst=1.3$.
\item[d] This differs from the 160 beams specified in the HLSD.
\item[e] See \prettyref{app:sensitivity-requirements}.
\end{tablenotes}
\end{threeparttable}
\par\end{centering}

\noindent \centering{}
\end{table}

\subsection{ \texorpdfstring{\SKAiLow{}}{SKA1-low} station design details \label{sub:station-design-details}}

Using a dual-band implementation with similar performance characteristics
to the single-band implementation ensures the like-for-like comparison.
In particular, the low-band array (70--180\,MHz) has the same physical
layout as the single-band array to achieve the same sensitivity at
the lower frequencies. The high-band array (180--450\,MHz) has the
same number of antenna elements as the low and single-band arrays,
so despite the smaller inter-element spacing, sensitivity is maintained
as equivalent to the single-band at most frequencies (see \prettyref{app:sensitivity-requirements}),
while reducing the geometrical area occupied by the station. The required
processed field of view (FoV) is a minimum of $\Oreq=\unit[20]{deg^{2}}$\nomenclature[s~Oreq]{$\Oreq$}{\TOreq{}},
observed concurrently across the frequency band\label{ass:concurrent-FoV}.
Dual-polarisation ($\Npol=2$), or full-Stokes signals, are assumed
throughout this analysis.\nomenclature[sNpol]{$\Npol$}{\TNpol{}}\label{ass:dual-pol}

The representative single-band implementation is that which is described
in the HLSD. The pertinent features of the system are shown in \prettyref{fig:single-station}
(station diameter $\Dst$\nomenclature[sDst]{$\Dst$}{\TDst{}} and
number of elements per station $\Nest$\nomenclature[sNest]{$\Nest$}{\TNest{}})
and \prettyref{tab:single-system-details}. As shown in \prettyref{app:constant-FoV},
the required FoV $\Oreq$ is synthesised from an average number of
station beams over the band $\Nbstavg$\nomenclature[sNbstavg]{$\Nbstavg$}{\TNbstavg};
enough processing is costed to form these beams. An irregular intra-station
element layout of approximately uniform element distribution is assumed.\label{ass:intra-station-layout}

\begin{figure}[!t]
\noindent \centering{}\includegraphics[width=0.55\textwidth]{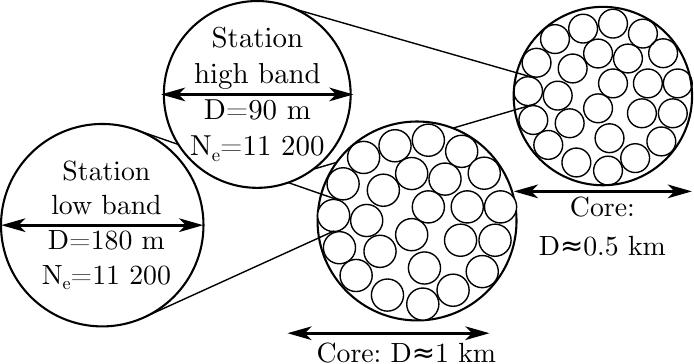}\caption{Representation of a low and high-band station within two densely packed
core regions.\label{fig:dual-station}}
\end{figure}

\begin{table}[!t]
\centering{}\caption{Dual-band \SKAiLow{} system details.\label{tab:dual-system-details} }
\begin{threeparttable}%
\begin{tabular}{>{\raggedright}p{0.45\textwidth}>{\centering}p{0.17\textwidth}>{\centering}p{0.17\textwidth}}
\toprule 
 & Low band & High band\tabularnewline
\midrule 
Frequency range & 70--180\,MHz & 180--450\,MHz \tabularnewline
Number of stations $\Nst$ & 50  & 50\tabularnewline
Average spacing between elements $\deeavg$ & 1.5\,m ($\lambda/2$ at 100\,MHz) & 0.75\,m ($\lambda/2$ at 200\,MHz) \tabularnewline
\TNet{}\textsuperscript{} $\Net$ & 16 & 16\tabularnewline
Station beam taper\textsuperscript{} $\Kst$ & 1.02  & 1.02 \tabularnewline
Number dual-polarisation beams per station (averaged over the respective
band)  & 44 & 70\tabularnewline
Dense-sparse transition & 115\,MHz & 230\,MHz\tabularnewline
Gain of an isolated element\textsuperscript{a} & 6.2\,dBi & 6.2\,dBi\tabularnewline
Observing mode & \multicolumn{2}{c}{Simultaneous (i.e. 70--450\,MHz)}\tabularnewline
\bottomrule
\end{tabular}\begin{tablenotes}
\small
\item[a] Gain is assumed to be the same for each band, to ensure that the first-order station $\AonT$ estimates are comparable. Actual gain values will depend on the antenna element designs.
\end{tablenotes}
\end{threeparttable}\label{ass:antenna-element-gain}
\end{table}

The key differences in the dual-band implementation are the separate
low and high-band stations, and the average inter-element spacing
of 0.75\,m in the high band. \prettyref{fig:dual-station} and \prettyref{tab:dual-system-details}
show the details of this system. The stations and the two cores are
assumed to be separate, as portrayed in \prettyref{fig:dual-station}.
Rather than the two cores shown in the HLSD (the second core being
composed of dishes), an \SKAi{} with a dual-band \SKAiLow{} implementation
would have three cores. The separated cores means each core can be
densely packed, resulting in a higher filling factor. Such densely
packed cores allow for more efficient searches of pulsars and other
high time resolution events (see \prettyref{app:NIP}). However, other
science applications for the high-band (180--450\,MHz) would require
evaluation to ensure that the array configuration composed of the
smaller, 0.5\,km diameter high-band core remains suitable. Consideration
of other configurations, such as interspersed or interleaved stations,
is beyond the scope of the present analysis.

The cost of the dual-band implementation is estimated by costing the
low and high-band stations separately. This means determining $\Ctot$
in \prettyref{eq:cost-total} for each band, and then summing the
costs. To determine $\Ctot$ for each band, the unit cost of every
block is given as some fraction of the single-band cost, as detailed
in \prettyref{app:dual-band-costs}. These costs are considered to
be reasonable estimates but are not based on detailed investigation.

For this analysis, infrastructure such as housing for the station
processing node is not shared.\label{ass:no-shared-station-infrastructure}
However, it is assumed that stations are co-located beyond the core,
hence the trenching and cables for the data transmission and power
to these stations can be shared.\label{ass:trenching-cabling-co-location}
Costing the systems separately ensures clarity for comparison purposes;
an actual implementation could share some infrastructure and possibly
signal processing units, while still being capable of observing the
full 70--450\,MHz frequency range. Some of the implications of station
co-location are discussed in \citet{AAV11-Concept} and \citet{FauAle10}.

\subsection{Comparison of the station sub-systems\label{sub:station-sub-systems}}

\begin{figure}[!t]
\begin{centering}
\includegraphics[height=0.68\textwidth]{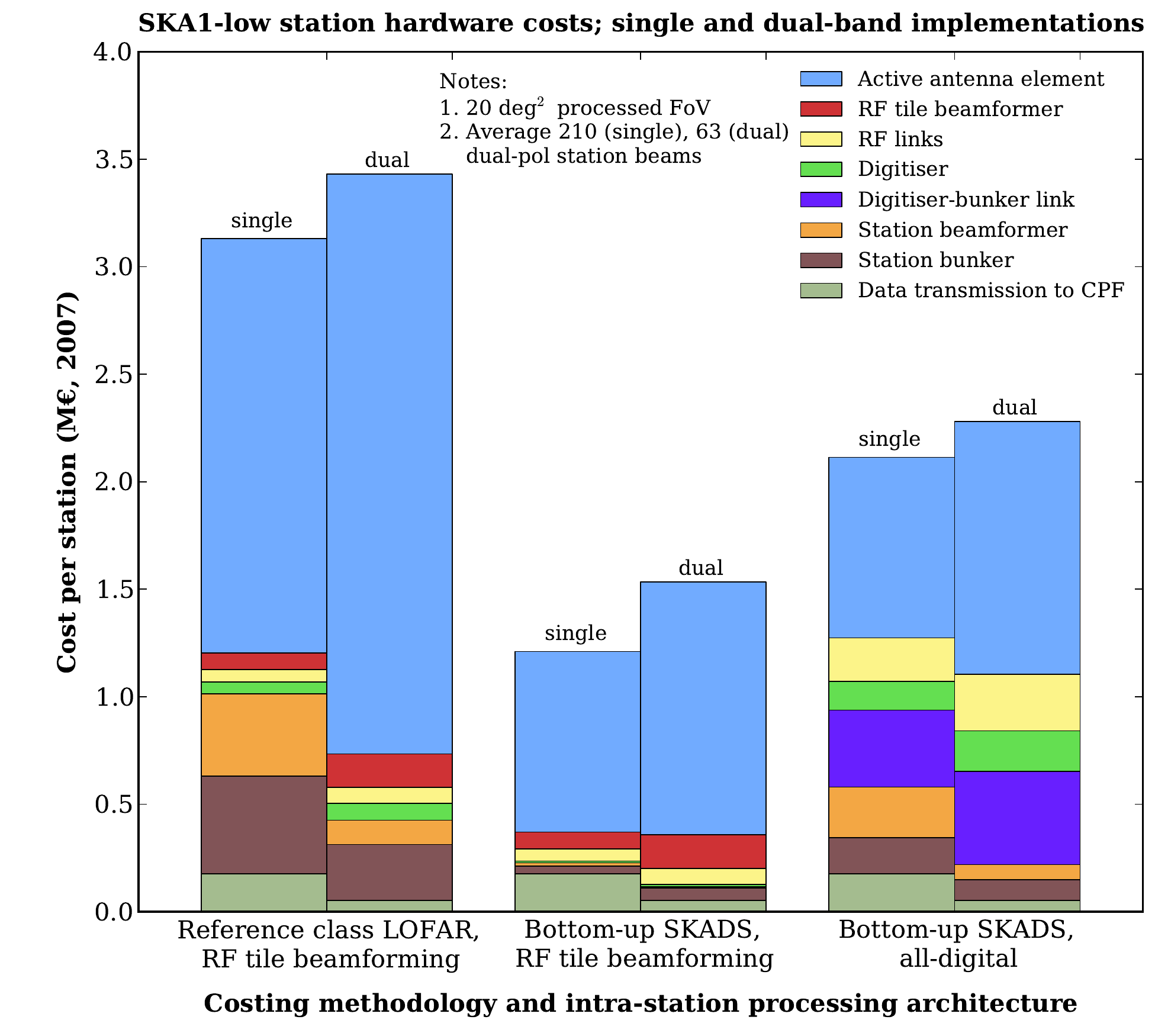}\includegraphics[height=0.68\textwidth]{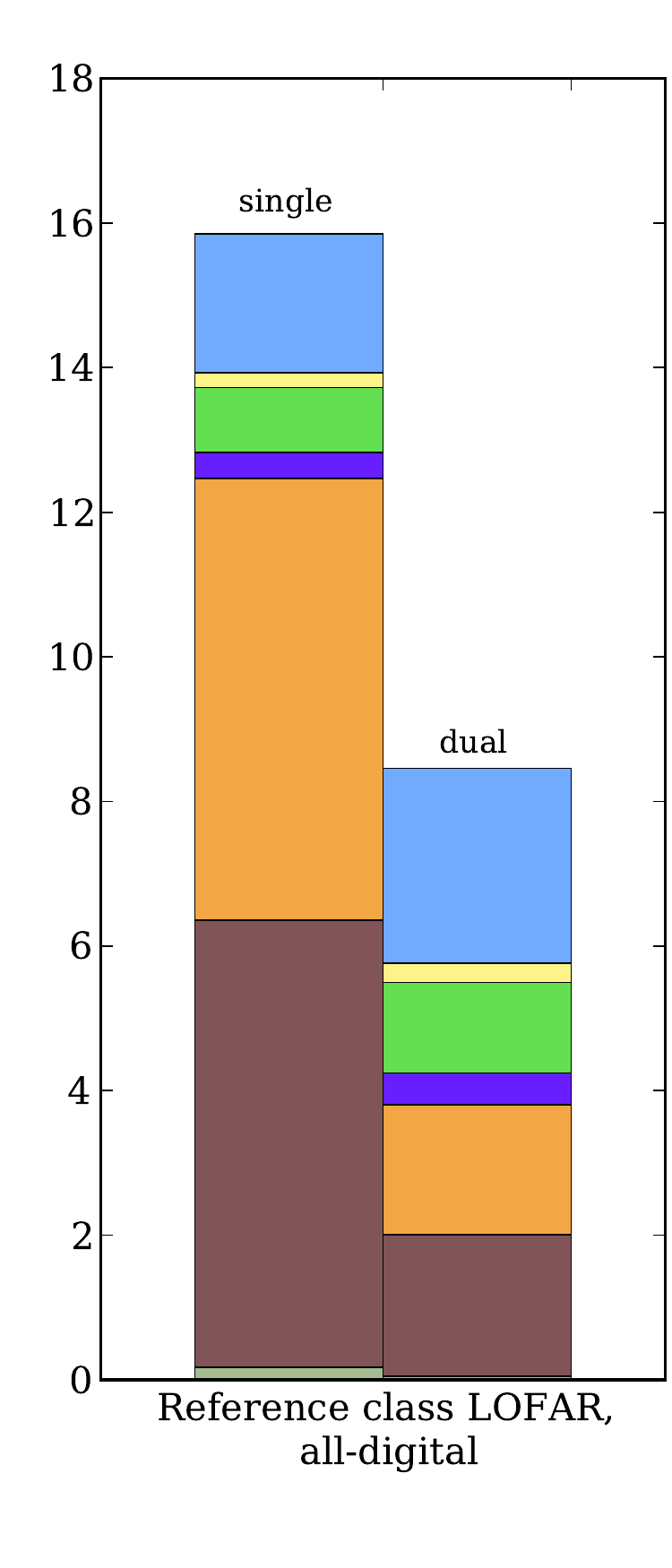}
\par\end{centering}

\caption{\SKAiLow{} station hardware cost, for permutations of single or dual-band
implementations, costing methodologies and intra-station processing
architectures. (\LofarDig{}, right, has a different y-axis scale.)
The \lofarRF{} and \skadsDig{} scenarios are derived from the cost
estimates and architectures in \citet{FauVaa11-Deployment}; the other
two scenarios are the extrapolations from those derivations. Station
bunker refers to the station beamformer housing and infrastructure
(racks, power supply etc.). Sub-systems are colour-coded to match
Figures \ref{fig:schematic-all-dig} and \ref{fig:schematic-RF-tile}.\label{fig:station-costs}}
\end{figure}

\prettyref{fig:station-costs} plots the \SKAiLow{} station hardware
costs for the four different cost scenarios (\prettyref{sub:cost-data-sources}),
each scenario being a combination of:
\begin{itemize}
\item \lofarCost{} or \skadsCost{} cost estimate, and
\item RF tile beamforming with subsequent digital station beamforming or
all-digital beamforming.
\end{itemize}
These cost scenarios are plotted for both the single and dual-band
implementations. The costs shown here are for the \SKAiLow{} station
sub-systems, from the active antenna element up to and including the
data transmission to the \cpf{} (see Figures \ref{fig:schematic-all-dig}
and \ref{fig:schematic-RF-tile}). 

The sub-system costs which dominate the station hardware cost in the
single-band implementation differ between the RF tile beamforming
and all-digital architectures. For the all-digital architecture, the
active antenna element cost is less than half the total station cost.
With RF tile beamforming, the majority of the cost is active antenna
element costs. All the other blocks have a decreased cost, due to
the factor of 16 decrease in the number of signal chains after the
RF tile beamformer (with the exception of the station--CPF link, which
remains constant). This implies that understanding the active antenna
element costs is more important for the RF tile beamforming architecture.
The dual-band array displays the similar broad trend as the single-band
array, where the active antenna element cost is more dominant in the
RF tile beamforming architecture than the all-digital architecture. 

Comparing the single and dual-band implementations shows the driving
costs for each. Because the dual-band implementation has twice the
number of stations, hence twice as many antenna elements and therefore
signal chains, the active antenna element costs are significantly
higher. However, the increase is less than double, because of the
cost discounts applied in \prettyref{tab:blocks-cost-dual} (\prettyref{app:dual-band-costs}).
These discounts reflect the less onerous requirements on the components
in the dual-band implementation, due to the smaller fractional bandwidth
and the 180\,MHz frequency split between the low and high bands.

The opposing trend is that the smaller average number of beams across
the band puts downward pressure on the dual-band cost. This is reflected
in the lower cost of the station beamformer, station bunker and station--CPF
data transmission sub-systems. The cost reduction is significant for
the all-digital architecture, but is less pronounced for the RF tile
beamforming architecture, where the reduced number of digital signal
paths already decreases the digital processing and data transport
costs somewhat.

A conspicuous feature of \prettyref{fig:station-costs} is that the
station beamforming and related bunker cost in the \lofarCost{} estimate
is significantly larger than the \skadsCost{} estimate. This is due
to the large difference in unit cost estimates (see \prettyref{app:beamforming-computational-cost}).
The most likely reasons for this cost discrepancy are related to the
type of beamforming processing technology and architecture, and the
technology advancements which have been assumed. For example:
\begin{itemize}
\item The LOFAR beamformer is not optimised for processing the larger number
of inputs, hence the unit cost derivation may over-estimate the processing
cost.
\item The unit cost derived from the \skadsCost{} estimate already includes
a cost discount for hierarchical beamforming (see \prettyref{app:hierarchical-beamforming}). 
\item The LOFAR station beamformer uses \fpga{} (FPGA) processors, while
the SKADS design \citep{FauAle10} uses more customised processing
chips. 
\item The technology advancements may be more optimistic for the SKADS estimate
than those assumed for the LOFAR estimate \prettyref{ass:LOFAR-50MHz-BW},
and production would be for larger quantities than for LOFAR. 
\end{itemize}
The station beamformer cost is an example of the potential for further
investigation to determine: the accuracy and uncertainties of each
of the cost data sources and their subsequent derivation into unit
costs, the accuracy of the first-order station beamformer model, the
requirements that contribute to the beamformer cost and design solutions
or trade-offs to reduce this cost.

\subsection{Cost reduction from analogue (RF) tile beamforming\label{sub:RF-beamforming-cost-reduction}}

While not the main focus of this work, the parametric analysis allows
a preliminary cost comparison to be made between the two intra-station
architectures: RF tile beamforming and all-digital beamforming. The
results show that for tiles composed of 16 elements, a first stage
of analogue tile beamforming significantly reduces the station hardware
cost. The cost reduction is irrespective of whether a single or dual-band
system is implemented, or the \skadsCost{} or \lofarCost{} cost
estimate is used. 

The station costs for RF tile and all-digital beamforming architecture
can be compared in \prettyref{fig:station-costs}, for the single
and dual-band implementations and \lofarCost{} and \skadsCost{}
cost estimates. \prettyref{tab:RF-vs-all-digital-BF} makes a direct
comparison for each implementation and cost estimate combination.
The cost of the \SKAiLow{} sub-systems with analogue tile beamforming
is 20--67\,\% of the all-digital beamforming; a factor of approximately
1.5--5 reduction in cost. The cost reduction, both in relative and
absolute (euro) terms, is most significant when the cost of the digital
sub-systems is high (e.g. the single-band implementation of the \lofarCost{}
cost estimate). However, no cost reduction applies to the sub-systems
in the central processing facility; because those sub-systems act
on station beams, their costs are independent of the intra-station
architecture.

\begin{table}[!t]
\centering{}\caption{\SKAiLow{} station cost for RF tile beamforming, as a percentage
of all-digital beamforming cost.\textsuperscript{a}\label{tab:RF-vs-all-digital-BF}}
\begin{threeparttable}%
\begin{tabular}{lc>{\centering}p{0.15\textwidth}}
\toprule 
\multirow{2}{*}{Cost estimate} & \multicolumn{2}{c}{Implementation}\tabularnewline
\cmidrule{2-3} 
 & Single-band & Dual-band \tabularnewline
\midrule
\LofarCost{} & 20\,\% & 41\,\%\tabularnewline
\SkadsCost{} & 57\,\% & 67\,\%\tabularnewline
\bottomrule
\end{tabular}\begin{tablenotes}
\small
\item[a] Percentage shown applies to that cost estimate and implementation combination.
\end{tablenotes}
\end{threeparttable}
\end{table}

The cost reduction is due to the FoV accessible at the station beamformer
(and inherent observational flexibility) being restricted early on
in the signal path; the number of digital signal chains is reduced
by a factor equal to the number of elements per tile, in this case
16. The cost of the digitiser and station beamformer blocks is thus
reduced, as can be seen in \prettyref{fig:station-costs}. This is
because fewer digitiser and digitiser--bunker links are required,
and fewer inputs into the station beamformer reduces the processing
load. The performance reduction due to tile beamforming is further
discussed in \prettyref{sub:hierarchical-beamforming-performance}.

\citet{FauAle10} present a qualitative comparison of the all-digital
and RF tile beamforming approaches. The all-digital beamforming is
more flexible, in terms of generating multiple beams, RFI excision
and calibration of the antenna elements, if needed. With upgraded
digital signal transport and processing, the correlation of every
antenna element is also be possible; whereas only the RF beamformed
tiles can be correlated, not the individual elements. The main disadvantage
for the all-digital beamforming approach is cost, as outlined, as
well as the increased power demand of the digital components, and
the distribution (or local generation) of this extra power throughout
the array. This is further discussed in \prettyref{sub:signal-architectures},
in the context of the intra-station architecture.

\section{System implications of variable costs\label{sec:variable-cost-system-implications}}

 The results thus far present the costs of the \SKAiLow{} station
hardware for each scenario and representative implementation. However,
there are other cost implications on the telescope system, some of
which are considered in this section. To put the costs analysed here
in the context of the whole SKA budget, the simplest comparison between
the single and dual-band implementations is to say that some costs
remain approximately constant, such as project overheads, and some
costs vary between the two implementations, such as the \SKAiLow{}
stations costed in \prettyref{sec:representative-systems}. 

The variable costs considered in this analysis to most significantly
impact on the total system cost are:
\begin{itemize}
\item station sub-systems
\item antenna element deployment
\item site preparation
\item central processing sub-systems
\item power provision and distribution.
\end{itemize}
There is no published work on the relative cost between these, hence
the full effect of each implementation on the total cost is difficult
to determine. However, the different attributes of the two implementations,
as shown in \prettyref{tab:attribute-comparison}, gives some insight
into the cost trends. \prettyref{app:models-SKA-other} makes a more
detailed analysis of these variable costs and estimates some zeroth-order
costs.

\begin{table}[!t]
\centering{}\caption{Attributes of the dual-band implementation compared to the single-band.\label{tab:attribute-comparison}}
\begin{threeparttable}%
\begin{tabular}{l>{\centering}p{0.15\textwidth}}
\toprule 
Attribute & Percent of single-band \tabularnewline
\midrule
Number of antenna elements & 200\,\%\textsuperscript{a}\tabularnewline
Physical area & 125\,\%\tabularnewline
Average number of station beams\textsuperscript{b} & 30\,\%\tabularnewline
\bottomrule
\end{tabular}\begin{tablenotes}
\small
\item[a] Half these elements are physically smaller than the single-band elements.
\item[b] Formed over the full 70--450\,MHz band to achieve $\unit[20]{deg^{2}}$.
\end{tablenotes}
\end{threeparttable}
\end{table}

\subsection{Antenna element deployment and site preparation costs}

The cost of deploying the antenna elements (i.e. building the array
on-site) will be higher for the representative dual-band array, because
twice the number of antenna elements need to be deployed. Because
the element size is defined by being sufficiently electrically large
at the lowest observing frequency, it is reasonable to expect that
the size of the low-band element will be similar to the single-band
element, such that comparable antenna gain is obtained. However, the
high-band element will be significantly smaller, and cheaper manufacturing
and deployment options may be available. This means that the increased
deployment cost for the dual-band array would be less than 200\,\%.
Deployment is further discussed in \citet{FauVaa11-Deployment}. 

On a related topic, it is reasonable to expect that some fraction
of the site preparation cost will increase with the physical area
occupied by \SKAiLow{}. However, whether this cost is significant
relative to the total site preparation costs is not known. Site-related
data is being collected as a part of the site selection process \citep{SchAle11-PEP}
and is not currently available. We have used the initial deployment
and site preparations cost estimates outlined in \prettyref{sub:variable-cost-system-comparison}
to illustrate their potential significance.

\subsection{Central processing facility sub-systems\label{sub:central-processing-costs}}

The central processing facility is a broad term to encompass the signal
processing and science computing sub-systems in the HLSD. The processing
is centralised, because it acts on signals from all the antennas (AAs
and dishes) in the array. However, it does not necessarily imply that
all these sub-systems will be on-site; the on-site processing is required
to sufficiently reduce the rate of data sent to the off-site processing.
The principal sub-systems are the correlator and imaging processor,
and the non-imaging processor. The correlator and \nip{} costs are
derived the Signal Processing \CoDR{}, as collated in \citet{Tur11-Cost}.
 These costs focus on the processing units required (the sub-system
hardware), rather than total cost of the sub-systems. The parametric
cost equations of the major sub-systems within the central processing
facility are derived in \prettyref{app:central-processing}; the assumption
being that these sub-systems operate on a `per beam' basis.\label{ass:central-processing} 

\prettyref{tab:central-processing-comparison} summarises the relative
costs between the representative single and dual-band implementations.
The difference in cost between the two implementations is due to the
larger beam size of the smaller (90\,m) diameter high-band station.
This reduces the number of station beams required to fill $\unit[20]{deg^{2}}$
FoV, resulting in lower \cpf{} costs. To ensure comparable performance
between the single and dual-band implementations, $\unit[20]{deg^{2}}$
FoV over the full 380\,MHz bandwidth is correlated and imaged. Following
\citet{AleBre09}, the cost of the imaging processor is assumed to
be dominated by the cost of the data buffer rather than the imaging
operations; these imaging processing costs are discussed further in
\prettyref{app:imaging}. The non-imaging processor acts on phased
or `tied' array beams formed from the densely packed core stations,
and is further described in \prettyref{app:NIP}. Only array beams
formed with the high-band core of the dual-band implementation are
considered, because the current required frequency range for pulsar
surveys with the \nip{} is 0.3--3\,GHz~\citep{SKA11-DRM20}. 

A key aspect of the correlator--imaging data transport and imaging
processor cost is that for \SKAiLow{}, the correlator frequency resolution
requirement is derived from the more stringent science requirements,
rather than from the requirement to keep radial \uv{} smearing below
an acceptable threshold (see \prettyref{app:correlator-frequency-resolution-integration}).
If instead the latter dominates, then the correlator output data rate
becomes independent of station diameter, for a fixed processed FoV.
In that case, the imaging processor is therefore the same for the
single and dual-band implementations.

\prettyref{tab:central-processing-comparison} only applies to items
such as processing components (and associated hardware and cooling),
the cost of which, as a first-order approximation, scales linearly
with the processing load. Correlator and non-imaging processing costs
are summarised in \citet{Tur11-Cost}, representing a range of processor
technologies and architectures. Those costs are highly dependent on
the processing technologies and the trade-off between efficient processing
devices with larger development (\nre{}) costs, and less efficient
but more flexible processing devices.

\begin{table}[!t]
\centering{}\caption{Dual-band implementation central processing sub-system costs, compared
to the single-band.\label{tab:central-processing-comparison}}
\begin{threeparttable}%
\begin{tabular}{l>{\centering}p{0.15\textwidth}}
\toprule 
Central processing sub-system & Percent of single-band \tabularnewline
\midrule
Correlator & 30\,\%\tabularnewline
Correlator--imaging data transport\textsuperscript{a}  & 53\,\%\tabularnewline
Imaging processor\textsuperscript{a, b} & 53\,\%\textsuperscript{}\tabularnewline
Non-imaging processor (NIP)\textsuperscript{c} & 25\,\%\tabularnewline
\bottomrule
\end{tabular}\begin{tablenotes}
\small
\item[a] Required correlator frequency resolution is derived from the science requirements (\prettyref{app:correlator-frequency-resolution-integration}).
\item[b] Cost dominated by  data buffer (\prettyref{app:imaging}).
\item[c] Only the high-band core is used in the NIP, and the processing for the AAs, not the dishes, dominates the cost (\prettyref{app:NIP}).
\end{tablenotes}
\end{threeparttable}
\end{table}

\subsection{Overall  \texorpdfstring{\SKAiLow{}}{SKA1-low}  costs\label{sub:variable-cost-system-comparison}}

\begin{figure}[!tp]
\begin{centering}
\includegraphics[width=0.95\textwidth]{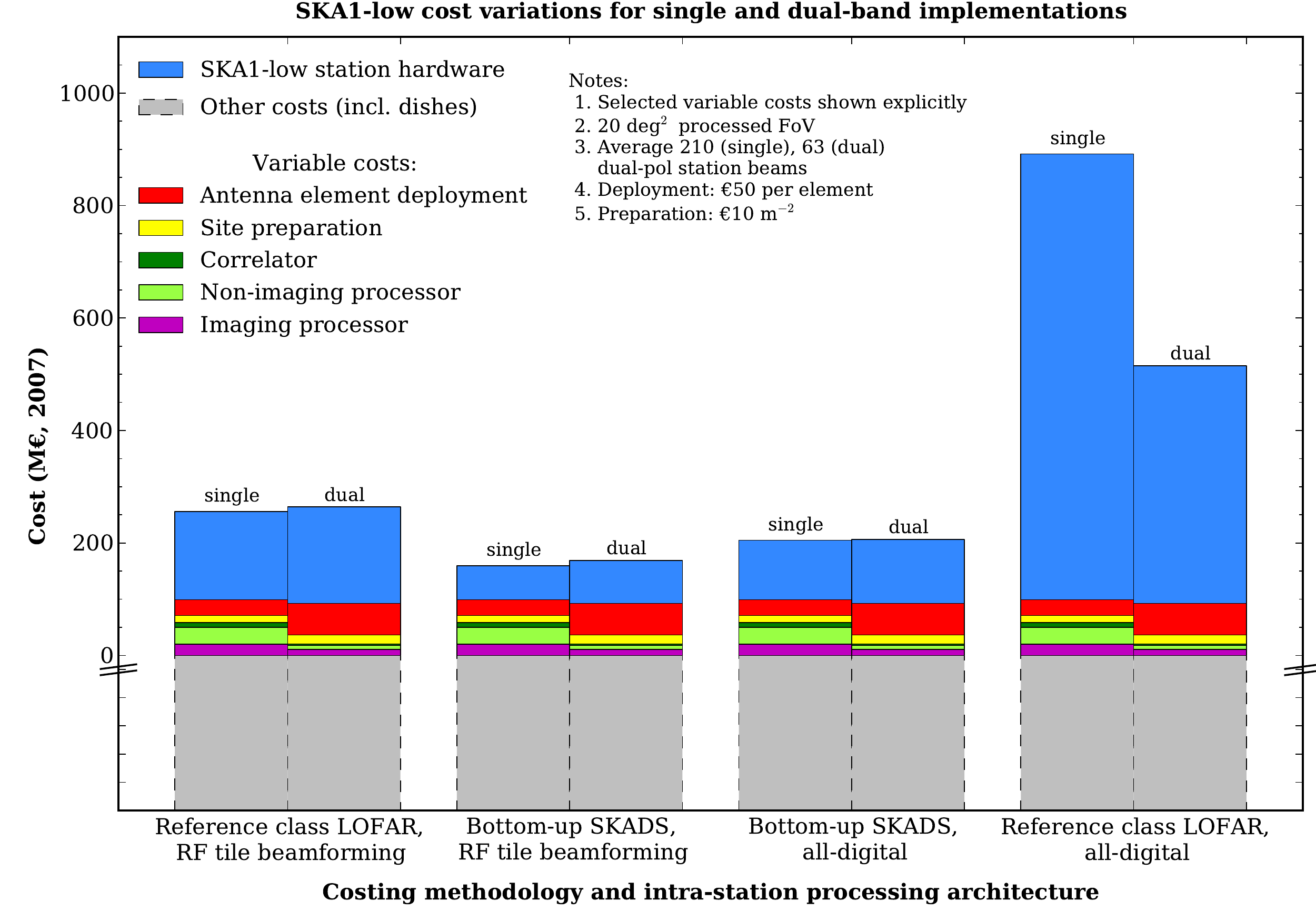}
\par\end{centering}

\centering{}\caption{Significant variable costs (excluding power) for the representative
single and dual-band implementations of \SKAiLow{}, for the different
cost estimates and intra-station architectures. A deployment cost
of \euro50 per element and site preparation cost of \euro$\unit[10]{m^{-2}}$
is assumed. `Other costs' is a placeholder for the costs which do
not differ between implementations.\label{fig:total-cost-50dep-10site}}
\end{figure}

\begin{figure}[!tp]
\begin{centering}
\includegraphics[width=0.95\textwidth]{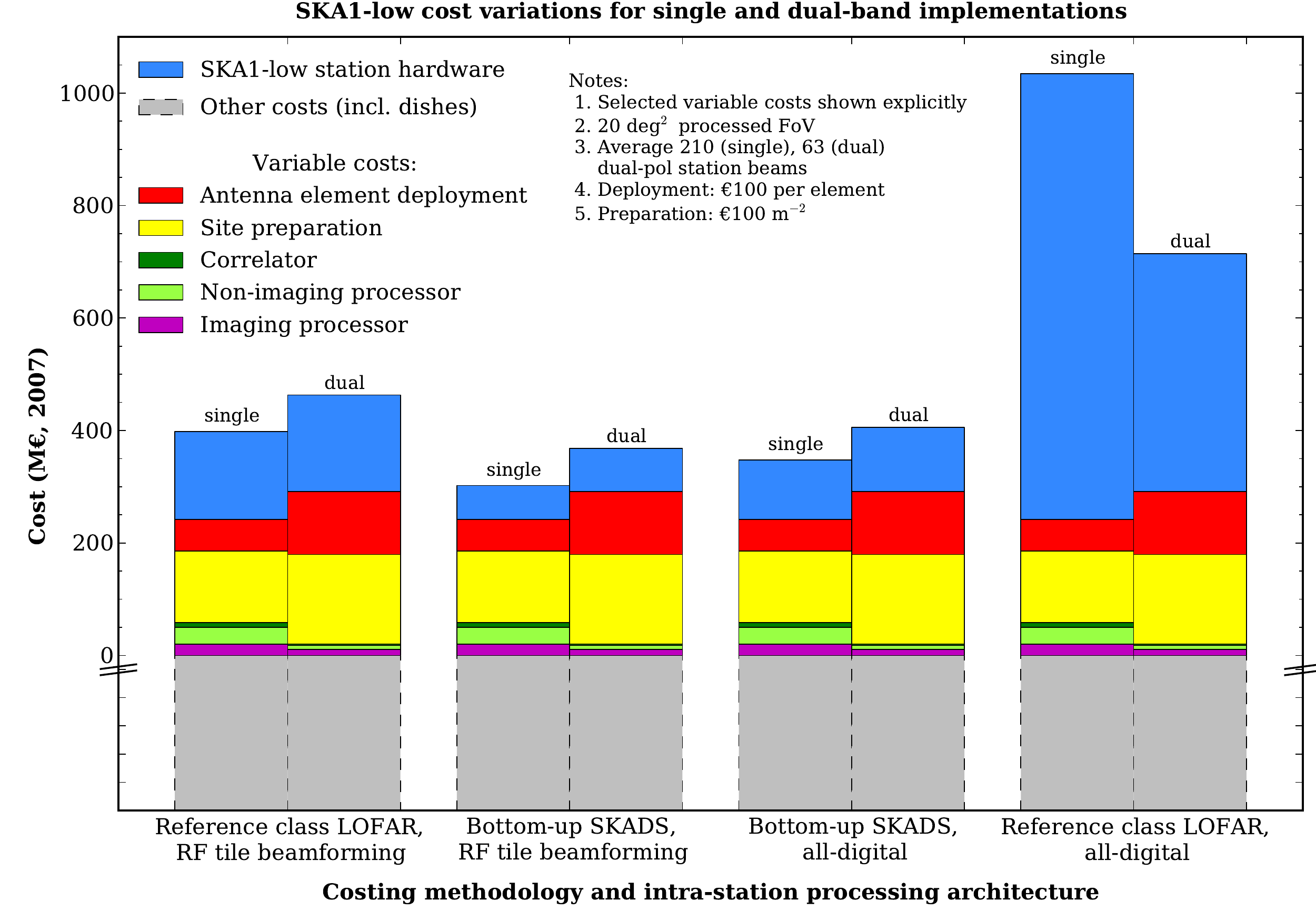}
\par\end{centering}

\centering{}\caption{As per \prettyref{fig:total-cost-50dep-10site}, except a deployment
cost of \euro100 per element and site preparation cost of \euro$\unit[100]{m^{-2}}$
is assumed.\label{fig:total-cost-100dep-100site}}
\end{figure}

Although obtaining accurate total costs of the single and dual-band
implementations is not yet possible, some zeroth-order estimates can
be used to illustrate the system-level costs outlined in this section.
\prettyref{fig:total-cost-50dep-10site} plots these significant variable
system costs (excluding power) for single and dual-band implementations.
The plot includes the four different station cost scenarios, reflecting
the different cost estimates, and intra-station signal transport and
processing architectures. The variable and `other' costs remain
unchanged  for each scenario; they are independent of the intra-station
architecture and station cost estimates. The correlator--imaging data
transport cost is not significant (<1\,\%) and is not plotted. To
indicate the sensitivity of the comparison to changes in the deployment
and site preparation costs, \prettyref{fig:total-cost-100dep-100site}
is a similar plot, but the deployment cost is doubled to \euro100
per antenna element, and the site preparation cost increased by an
order of magnitude to \euro$\unit[100]{m^{-2}}$.

\subsection{Power costs\label{sub:power}}

\begin{figure}[!t]
\begin{centering}
\includegraphics[width=0.75\textwidth]{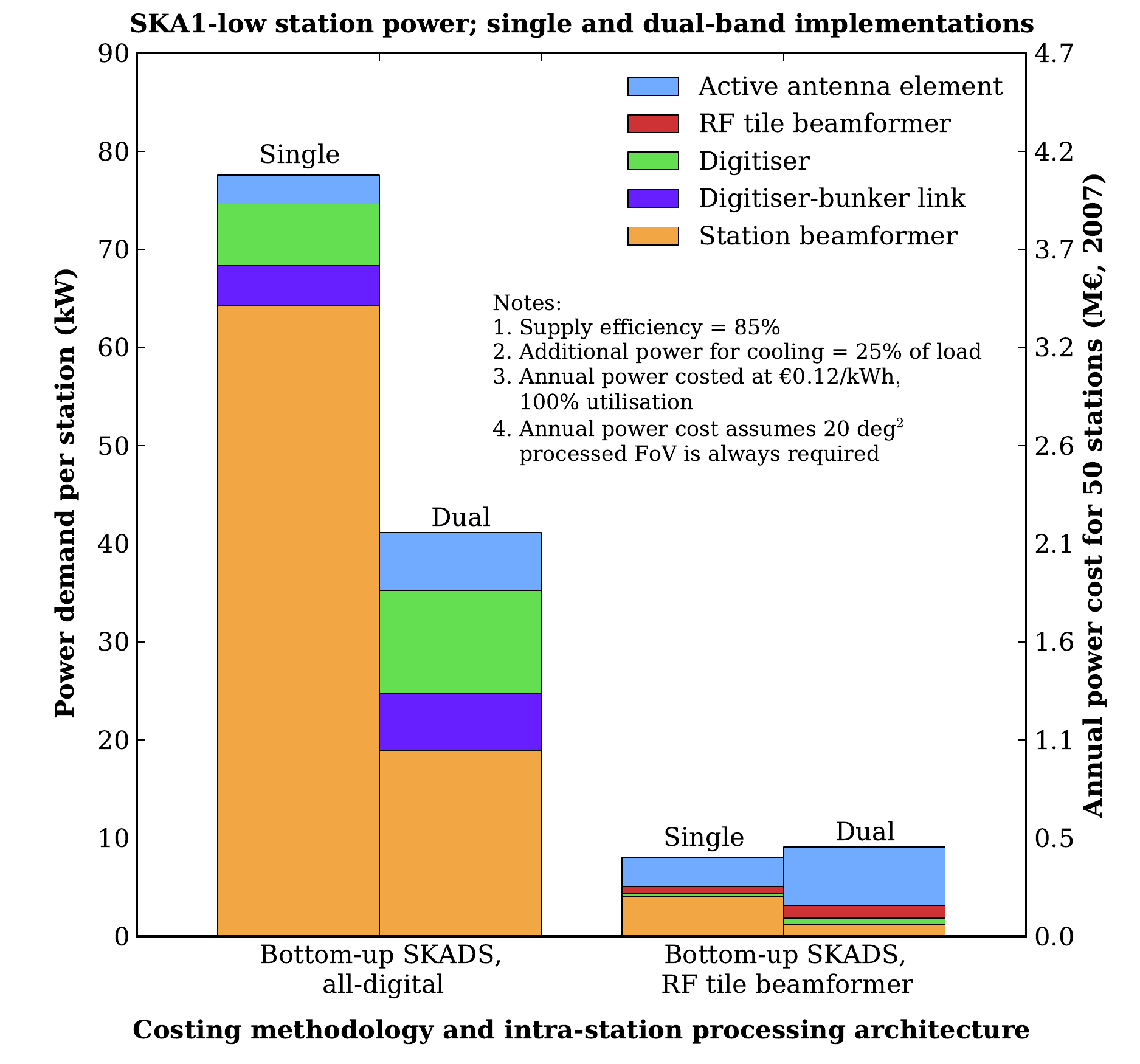}
\par\end{centering}

\caption{Single \SKAiLow{} station power demand estimate and annual power
cost for single and dual-band implementations, and RF tile beamforming
and all-digital intra-station processing. As per \citet{FauVaa11-Deployment}:
i) the power supply efficiency is assumed to be 85\,\% (a 1/0.85
increase in power demand); and ii) an additional 25\,\% power demand
is applied to account for cooling the hardware. Power consumption
is costed at \euro0.12 per kWh \citep{Hal11} and assumes 100\,\%
(continuous) utilisation of the hardware. \label{fig:power}}
\end{figure}

Investigations for the provision and distribution of power for the
SKA are on-going \citep[e.g.][]{Hal11}, alongside analyses of the
power demand of the telescope sub-systems. Although the details of
supplying and distributing power are beyond the scope of this document,
simplified power demand estimates are possible with the parametric
models. To make a zeroth-order estimate of the power costs for the
\SKAiLow{} stations, it is reasonable to expect that the power demand
of each sub-system is linearly proportional to one or more of the
following:
\begin{itemize}
\item number of antenna elements
\item digital processing load
\item number of station beams formed, hence data rate transmitted from the
stations to the CPF.
\end{itemize}
\prettyref{fig:power} shows power demand, estimated from the \skadsCost{}
power budget in \citet{FauVaa11-Deployment}. The `unit power demand'
is parametrised, as was done for the cost data, and listed in \prettyref{app:power-demand}.
The power budget is for the all-digital station architecture, but
is extrapolated to the RF tile beamforming architecture by including
an estimate for the RF beamformer power demand.

The station power demand depends on both the implementation and intra-station
architecture. The dual-band implementation reduces the station beamformer
power demand, but the demand of the other station sub-systems is increased.
For the all-digital architecture, this results in a significantly
lower power demand for the dual-band implementation. But for the RF
tile beamformer architecture, there is little difference between
implementations. Rather, the demand reduction from RF tile beamforming
is the dominant effect.

The method, hence cost, for supplying the power demand of sub-systems
within the station will depend on the intra-station architecture,
as discussed in \prettyref{sub:alternative-architectures-examples}.
The intra-station power distribution cost and power supply inefficiencies
depend on load and distance, making the cost specific to each intra-station
architecture. For example, in the RF tile beamforming architecture,
power at the antenna elements is supplied via the RF link to the station.
In contrast, the all-digital architecture transmits data via optic
fibre, hence requires additional power distribution cabling (which
is not costed here). Some architectures do not even have intra-station
distribution costs, because the active antenna elements are self-powered.
Bearing these caveats in mind, \prettyref{fig:power} assumes an 85\,\%
supply efficiency, and includes a power consumption cost of \euro0.12
per kWh for illustrative purposes. 

For the \cpf{}, the absolute power demand depends on the technologies
used; in general, there is an inverse relationship between the flexibility
and power efficiency of the processing~\citep{Hal11}. But given
that both cost and power demand increase with the amount of processing
required, the values in \prettyref{tab:central-processing-comparison}
can be taken as an indication of the relative power demand between
the single and dual-band implementation: the \cpf{} power demand
is lower for the dual-band implementation.

\section{Discussion of principal analysis\label{sec:discussion}}

The merits of single and dual-band implementations of \SKAiLow{}
can be considered in terms of performance, cost and design flexibility.
This section summarises the cost and performance trends, and the cost
estimation uncertainties. We also discuss the challenge of designing
a single-band implementation which meets the sensitivity requirements
at the lower frequencies and achieves a sufficiently high filling
factor (a measure of array sparsity), which may assist in calibration
at the higher frequencies.

\subsection{Cost trends\label{sub:cost-trends}}

The cost difference between the single and dual-band implementations
is fundamentally a comparison between costs that scale with the number
of signal paths prior to station beamforming and costs that scale
with the number of station beams. The single-band station is very
sparse at high frequencies, requiring many more beams than the dual-band
station to meet a given FoV requirement. This not only increases the
station beamformer and bunker cost, but also the station--\cpf{}
(CPF) data transmission and \cpf{} sub-system costs. The dual-band
implementation has twice as many signal paths prior to station beamforming,
which increases cost. However, this cost is less than double that
of the single-band, primarily because the smaller fractional bandwidth
entails less demanding design specifications for the active antenna
elements. 

The dual-band array decreases the total number of station beams to
be formed, thereby reducing the corresponding signal processing costs.
The decrease in the number of beams comes about from the smaller high-band
station diameter. However, this is predicated on the system having
sufficient processing capacity to concurrently form and process the
number of beams corresponding to $\unit[20]{deg^{2}}$ FoV across
whole 70--450\,MHz band. If less processing capacity is required
for the high-band array, the cost advantage when implementing the
dual-band array is lessened. This is further investigated in \prettyref{sub:reducing-FoV}.

A related effect is the inter-element spacing chosen for the high-band
array. A smaller inter-element spacing further reduces diameter, and
therefore cost. However, this ultimately begins to reduce the high-band
array sensitivity, as discussed in \prettyref{app:sensitivity-requirements}.
The frequency split of the dual-band array is another factor. The
key metric of the average number of beams per station ($\Nbstavg$)
not only depends on inter-element spacing, but also the frequency
split between the low and high-band arrays, as shown in \prettyref{app:constant-FoV-dual-band}.
The frequency split also affects the cost of the active antenna elements
and digitiser (\prettyref{app:dual-band-costs}), as they depend on
minimum and maximum frequencies of each band, and the fractional bandwidth.

For the representative single and dual-band implementations, there
are not significant differences in station hardware costs, except
for the \lofarDig{} scenario. But the variable costs which impact
the system (\prettyref{sec:variable-cost-system-implications}) may
be a discriminating factor between implementations. As discussed,
the dual-band implementation lowers the cost of the CPF sub-systems.
However, the increased number of active antenna elements and the extra
area required for these elements increases the deployment and site
preparation costs. 

Power costs also bear careful consideration; capital cost may be significant
and power is a major operations cost. Our study shows the station
power demand of the dual-band implementation equals that of the single-band
when RF tile beamforming is used, and is less than the single-band
with the use of all-digital beamforming. The RF tile beamforming itself
significantly reduces power demand for both implementations. For the
CPF, the dual-band implementation reduces the power demand.

There are also some cost trends which depend on the intra-station
architecture, but which do not affect the other variable costs which
impact the system. The noteworthy trend is the significantly lower
station cost for RF tile beamforming compared to all-digital beamforming.
In fact, the differences in station cost between these architectures
is more significant than the differences between the single and dual-band
implementations. Also, \citet{AAV11-Concept} discusses a dual-band
implementation which shares a common processing system to reduce processing
requirements, although exactly how this sharing occurs is not specified.
While not specifically costed in this analysis, alternative intra-station
architectures (including shared processing) are discussed in \prettyref{sub:signal-architectures}.

\subsection{Performance trends\label{sub:performance-trends}}

The performance characteristics of the canonical dual-band implementation
used in this analysis is comparable to the single-band implementation
described in the \SKAi{} high level system description. This is achieved
by using separate low and high-band arrays, observing simultaneously,
to create a $\unit[20]{deg^{2}}$ processed FoV over the 70--450\,MHz
band, while maintaining a sensitivity curve across the band which
is similar to the single-band implementation. However, these representative
implementations are not optimised for performance.

The performance and design flexibility of the implementations is compared
in \citet{AAV11-Concept} and potential performance issues for the
single-band system are identified. In particular, an antenna element
design suitable for the wide fractional bandwidth of about 6.5\,:\,1
is required. The antenna must then be matched to the low noise amplifiers
(LNAs) across the frequency band, which is more difficult for the
wide fractional bandwidth. These steps are less challenging for the
dual-band implementation, with the 2.5\,:\,1 fractional bandwidth
in each band. 

Another key performance issue is the frequency-dependent sensitivity
($\AonT$) curve; the frequency range for which the sensitivity is
optimised is a high-level design issue \citep{AleHal10}. Figure 23
of the HLSD makes a simple estimation of sensitivity for \SKAiLow{};
this curve is replicated in \prettyref{app:sensitivity-requirements}
(Figures \ref{fig:sensitivity-simple} and \ref{fig:sensitivity-simple-highband}).
A key feature of the curve is that the sensitivity is lowest at the
minimum (70\,MHz) and maximum (450\,MHz) frequencies of the band,
with a peak between 100 and 200\,MHz; there is also a factor of 3
difference between the highest and lowest sensitivity values. 

\prettyref{app:sensitivity-requirements} shows the effect of inter-element
spacing on sensitivity. In the single-band implementation, the inter-element
spacing must be chosen to best satisfy requirements across a larger
band, resulting in sub-optimal spacing at the lowest and highest frequencies
in the band. At 70\,MHz, the sensitivity is limited by the 1.5\,m
inter-element spacing. At 450\,MHz, the 1.5\,m spacing means that
more beams are required to form the $\unit[20]{deg^{2}}$ processed
FoV, increasing the processing costs, as discussed in \prettyref{sub:cost-trends}. 

The dual-band implementation uses a smaller inter-element spacing
in the high band, thus it can achieve beamformer cost savings in the
high-band, while maintaining similar sensitivity to the single-band
implementation. The only exception is the reduced sensitivity between
180 and 230\,MHz. Another aspect of the dual-band flexibility is
that a different number of antenna elements could potentially be used
in the low and high band arrays, to better suit sensitivity requirements.\textbf{}

A single-band implementation may also present a greater calibration
challenge than a dual-band array. As detailed in \prettyref{app:filling-factor-calibration},
station calibration requires a sufficiently high filling factor so
that enough calibration sources are detectable within a station beam
FoV. At 400\,MHz, the filling factor of the single-band station is
too low to detect the required number of calibration sources. The
representative dual-band implementation solves this problem because
the filling factor for the high-band station is greater.

\subsection{Risk and uncertainty\label{sub:risk-uncertainty}}

The modelling of uncertainties is an often overlooked aspect of the
performance and cost analysis. However, these uncertainties can manifest
themselves as project risks. As mentioned in \prettyref{sec:parametric-cost-modelling},
the parametric cost estimate should provide an associated uncertainty
estimate, which is large in the early stages of the project. As the
project progresses, the cone of uncertainty reduces and the cost estimate
eventually converges on the actual cost. This section discusses the
station hardware cost uncertainties and makes an initial uncertainty
estimate for the \lofarRF{} scenario.

\subsubsection{\SKAiLow{} station uncertainties}

The range of cost estimates for the \SKAiLow{} station hardware give
some indication of how sensitive the single and dual-band implementations
are to different cost projections. \prettyref{tab:total-cost-comparison}
provides a summary of the single and dual-band station costs plotted
in \prettyref{fig:station-costs}. All costs are normalised to the
cost of the \lofarRF{} scenario of the single-band array. For each
cost scenario, the table shows the normalised total for each implementation
and the ratio of the dual-band station cost to the single-band. For
three of the scenarios, the dual-band station is more expensive than
the single band, while for the fourth, the opposite is true.

\begin{table}[!t]
\centering{}\caption{Comparison of \SKAiLow{} station cost for the single and dual-band
implementations. \label{tab:total-cost-comparison} }
\begin{threeparttable}%
\begin{tabular}{@{}l>{\raggedright}m{0.13\textwidth}>{\raggedright}p{0.24\textwidth}>{\centering}p{0.12\textwidth}>{\centering}p{0.12\textwidth}>{\raggedright}m{0.12\textwidth}}
\cmidrule{2-6} 
 & \multirow{2}{0.13\textwidth}{Cost scenario } & \multirow{2}{0.24\textwidth}{} & \multicolumn{2}{c}{Implementation\textsuperscript{a}} & \multirow{2}{0.12\textwidth}{\centering{}Dual to single-band ratio }\tabularnewline
\cmidrule{4-5} 
 &  &  & Single-band  & Dual-band & \tabularnewline
\cmidrule{3-6} 
\cmidrule{2-2} & \multirow{2}{0.13\textwidth}{RF tile beamforming} & \LofarCost{} & 100\,\% & 110\,\% & \centering{}108\,\%\tabularnewline
 &  & \SkadsCost{} & 39\,\% & 49\,\% & \centering{}127\,\%\tabularnewline
\cmidrule{3-6} 
\cmidrule{2-2} & \multirow{2}{0.13\textwidth}{All-digital beamforming} & \SkadsCost{} & 68\,\% & 73\,\% & \centering{}110\,\%\tabularnewline
 &  & \LofarCost{} & 506\,\% & 270\,\% & \centering{}53\,\%\tabularnewline
\cmidrule{2-6} 
\end{tabular}\begin{tablenotes}
\small
\item[a] Percentage of the single-band, \lofarRF{} scenario.
\end{tablenotes}
\end{threeparttable}
\end{table}

In \prettyref{tab:total-cost-comparison}, there is a factor of~7.5
difference between the \lofarCost{} and \skadsCost{} cost estimates
of the all-digital, single-band implementation, and a factor of~3.7
for the dual-band. For the scenarios with RF tile beamforming, the
difference is a factor of~2.6 and 2.2 for the single and dual-band
respectively. Some of the unit costs only have a single data source
(estimate) and are used for both cost estimates; those costs do not
contribute to the cost differences. Some potential reasons for the
cost differences were discussed in \prettyref{sub:cost-data-sources},
but this is an area for further investigation.

Importantly for the single and dual-band comparison, the range in
the relative station hardware costs between the two representative
implementations (as indicated by the final column of \prettyref{tab:total-cost-comparison})
is much lower than the range of absolute station hardware costs.
This implies that the uncertainty of the single and dual-band station
cost comparison is less than that of the individual station cost estimates.

\subsubsection{Sensitivity analysis: cost drivers}

The parametric model allows for sensitivity analysis of the inputs,
to determine which blocks in the system significantly affect the total
cost. By identifying these cost drivers, extra attention can be paid
to them during the design phase, thereby reducing risk \citep{NAS08}.
Sensitivity analyses can also involve assessing the impact of changed
requirements.

Within the context of the \SKAiLow{} stations, an inspection of \prettyref{fig:station-costs}
gives an indication of the cost drivers. The dominant blocks are the
active antenna element, digitiser--bunker links and the station beamformer
and bunker. For both the single and dual-band implementations, the
active antenna element is the largest cost block, except in the \lofarDig{}
scenario where the station beamformer and bunker costs dominate. For
the \skadsDig{} scenario, the digitiser--bunker links are the second
most costly block. However, the large variation of the cost of many
of the blocks between the scenarios indicates that the cost estimates
require further refinement before a conclusive set of cost drivers
can be determined. The drivers also depend on the intra-station
architecture, as discussed in \prettyref{sub:signal-architectures}.

Additionally, the other variable costs listed in \prettyref{sec:variable-cost-system-implications}
have the potential to be cost drivers in the comparison of single
and dual-band implementations, and could exceed the most costly individual
station sub-system. For example, Figures \ref{fig:total-cost-50dep-10site}
and \ref{fig:total-cost-100dep-100site} in effect form a rudimentary
analysis of the sensitivity of the \SKAiLow{} cost to changes in
variable antenna element deployment and site preparation costs.

\subsubsection{Statistical uncertainty analysis\label{sub:statistical-uncertainty-analysis}}

\begin{figure}[!t]
\begin{centering}
\includegraphics[width=0.9\textwidth]{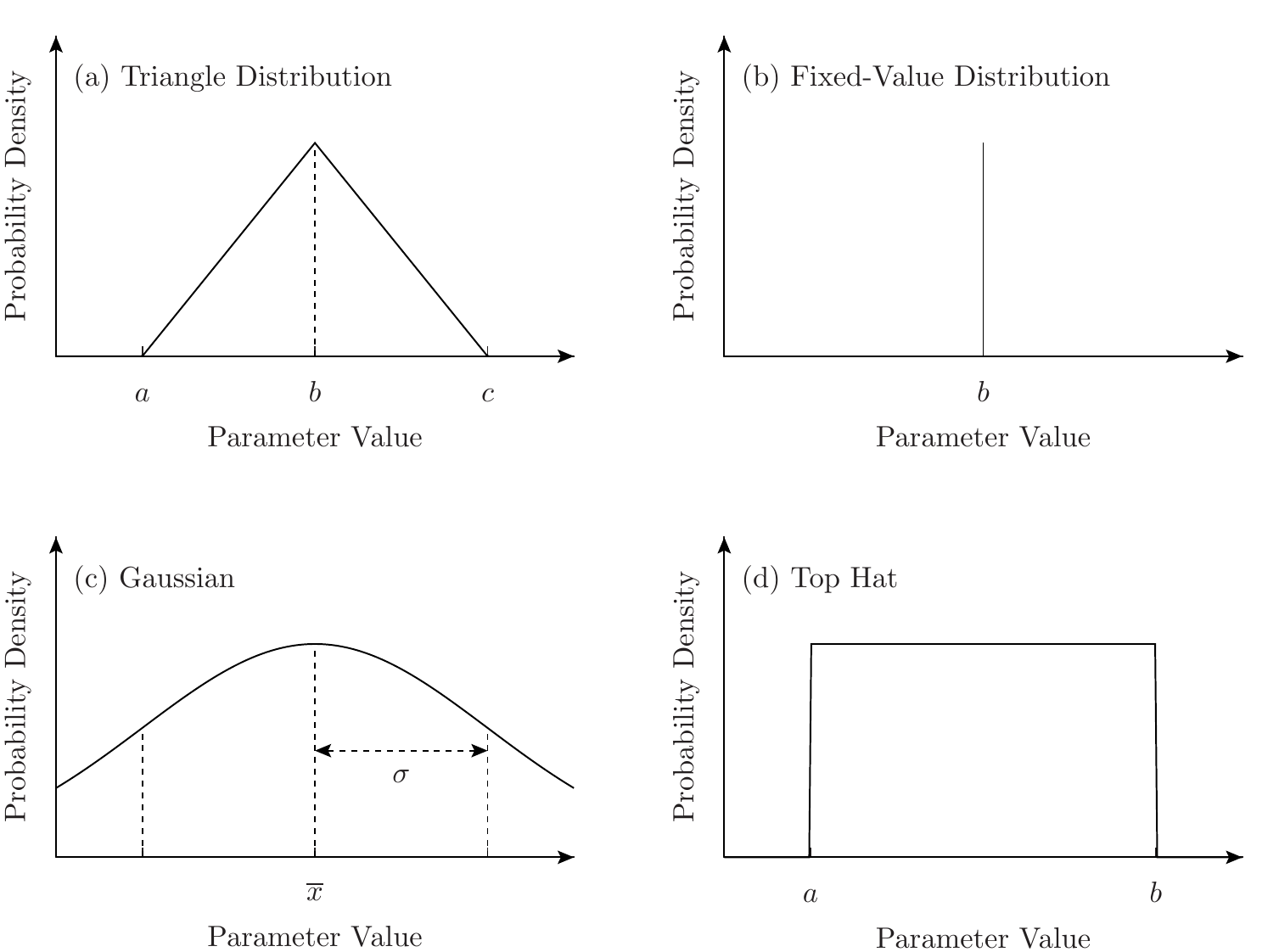}
\par\end{centering}

\caption{Input \pdf{}s (PDFs) available in SKACost. The PDF of a value may
be described by (a) minimum $a$ and maximum $c$ possible values,
with a peak-likelihood value of $b$; (b) single-value at $b$; (c)
value with mean $\overline{x}$ and and standard deviation $\sigma$;
or (d) value ranging between minimum $a$ and maximum $b$. Source:
\citet{ForBol10}\label{fig:PDFs}}
\end{figure}

Monte Carlo simulation provides a statistical approach for uncertainty
analysis, using \pdf{}s (PDFs) for each `uncertain' input in the
parametric model. An input, for example, may be a variable parameter
(such as number of elements per station) or a unit cost. The simulation
then randomly picks values from these input PDFs and calculates a
point cost estimate. Over thousands of iterations, this results in
a new PDF which is the cost estimate with a statistical distribution
\citep{NAS08}. Monte Carlo uncertainty analysis has been implemented
in SKACost \citep{ChiCol07,ForBol10}; \prettyref{fig:PDFs} shows
the input PDFs available in SKACost. 

Although the input unit cost distributions require further expert
attention, we make an initial estimate of the uncertainty of the
\lofarRF{} scenario. For this estimate, we only consider the uncertainties
of the two largest costs; the active antenna elements and the station
beamformer processing. For the active antenna element (which includes
the ground plane), a triangular probability distribution function
is applied, where the minimum unit cost is that derived from the \skadsCost{}
estimate (\euro75 per element), the most likely unit cost is that
derived from the \lofarCost{} estimate (\euro172 per element) and
the maximum cost is twice that (\euro344 per element). For the station
beamformer processing and the portion of the bunker which scales with
the amount of processing, a top-hat cost distribution is applied,
because the most likely unit cost is unknown at present. The top-hat
probability distribution function is used when only the minimum and
maximum values are known, and the true cost could lie anywhere, with
equal probability, between these limits. We use the unit processing
cost derived from the \skadsCost{} estimate (\euro0.1 per beamformer
input per output beam) as the minimum value and the \lofarCost{}
estimate (\euro2.6 per beamformer input per output beam) as the maximum
value. 

\begin{figure}[!t]
\centering{}\includegraphics[width=0.65\textwidth]{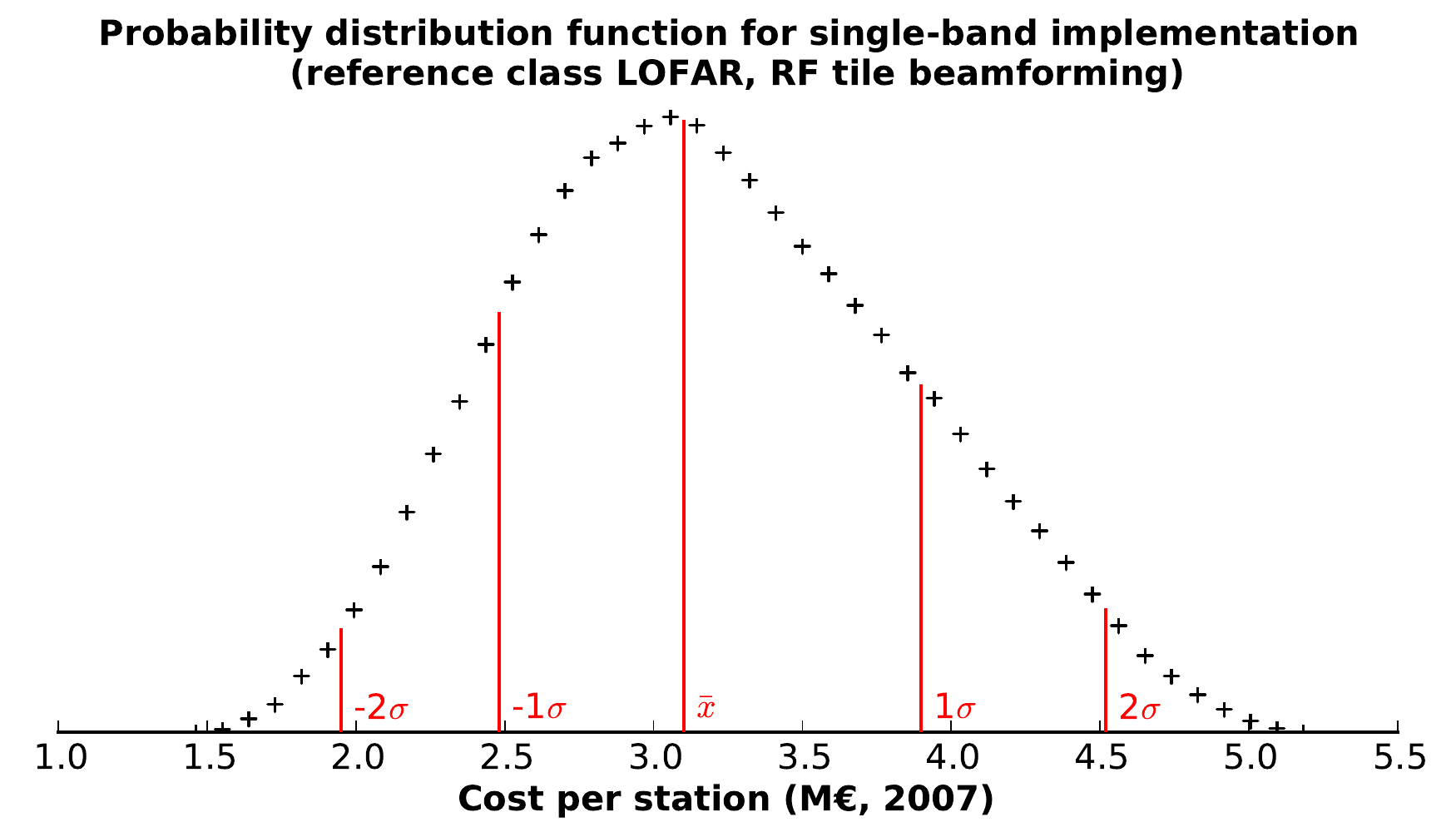}\caption{Probability distribution function for the single-band implementation
of the \lofarRF{} scenario, as calculated using SKACost. The crosses
show the probability density for each sample bin and the red lines
show the percentiles equivalent to the mean and 1 and 2 standard deviations
of a Gaussian distribution. Refer to the text for the input \pdf{}s
used. \label{fig:lofar-PDF}}
\end{figure}

\begin{figure}[!t]
\begin{centering}
\includegraphics[height=0.68\textwidth]{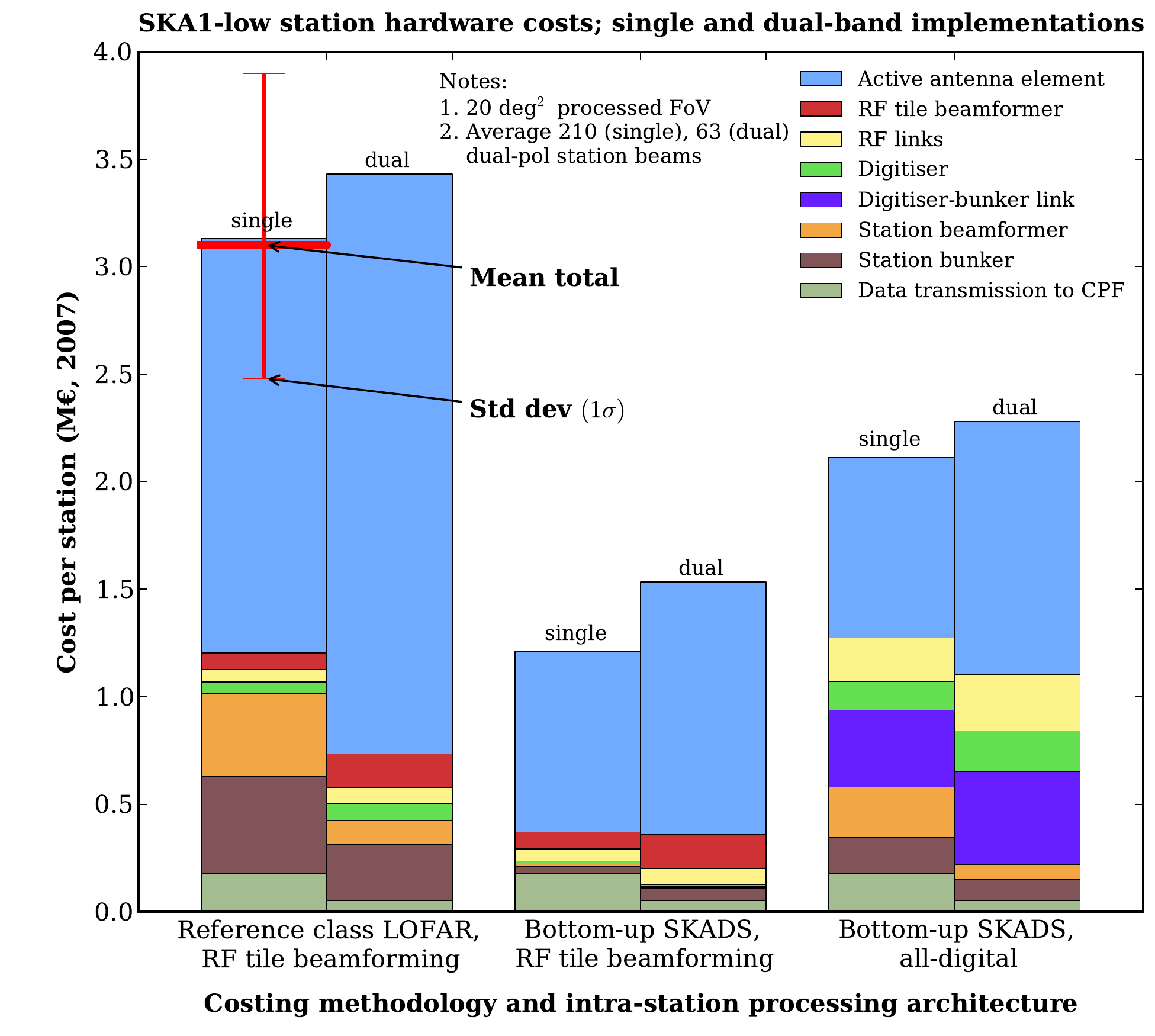}\includegraphics[height=0.68\textwidth]{AA/SKAlowSubsystems-1}
\par\end{centering}

\caption{\SKAiLow{} station hardware cost as per \prettyref{fig:station-costs},
but with the mean and $1\sigma$ error bars from \prettyref{fig:lofar-PDF}
shown for the \lofarRF{} scenario.\label{fig:Cost-SKAlow-sub-systems-err}}
\end{figure}

\prettyref{fig:lofar-PDF} shows the resulting PDF of the Monte Carlo
analysis of the \lofarRF{} scenario with the above-mentioned input
PDFs. The station cost is \euro3.1\,$(+0.8,-0.6)$\,M (these percentiles
are equivalent to the mean and 1 standard deviation of a Gaussian
curve, 50\,\% and 15.9\,\%, 84.1\,\% respectively). \prettyref{fig:Cost-SKAlow-sub-systems-err}
plots this mean and error onto the station hardware cost break-down.
For comparison, the single-value station cost plotted in \prettyref{fig:station-costs}
is \euro3.1\,M. With due consideration of the PDFs of the major
costs, similar uncertainty analyses can be made of the other station
scenarios, and also the other variable costs (\prettyref{sec:variable-cost-system-implications}).

\subsection{Relevance to SKA Phase 2}

This parametric modelling analysis is applicable to SKA Phase 2 (\SKAii{}),
although the unit costs and scientific requirements will differ. \SKAii{}
will most likely use the low-frequency receptors defined for \SKAi{}.
If the intra-station signal transport and processing architecture
is similar, then the parametric equations for the station hardware
costs (\prettyref{app:models-SKAlow}) will still apply. The main
difference will be the cost for each block. Although the cost of the
digital blocks will reduce, the FoV requirements for \SKAii{} will
be much higher. Thus, it is possible that any digital cost reductions
will be offset by the extra digital processing needed to form many
more station beams to achieve the required FoV. For the other attributes
shown in \prettyref{tab:attribute-comparison} (number of antenna
elements, physical area and average number of station beams), the
ratio between the two implementations still holds, regardless of required
FoV. The same applies to the \cpf{} costs (\prettyref{tab:central-processing-comparison}). 

The \SKAii{} science requirements must also be kept in mind when
comparing the single and dual-band implementations. These requirements
may place a difference emphasis on the frequency-dependent sensitivity
and FoV performance, compared to \SKAi{}; this may affect the optimal
inter-element spacing. While shorter term gains might be made by optimising
the design for \SKAi{}, there may be implications for \SKAii{} cost
and performance which necessitate further investigation.

\section{Supplementary analyses \label{sec:other-considerations}}

In this section we extend our analysis to include some topical factors
important to the design of the \SKAiLow{} system.

\subsection{Varying station diameter\label{sub:station-diameter}\label{sub:Nst-Nest-trade}}

The HLSD discusses a potential change to the \SKAiLow{} system,
where diameters smaller than 180\,m are used, but the number of stations
$\Nst$ is increased to maintain the \SKAiLow{} array effective area
$\Aearr$\nomenclature[sAearr]{$\Aearr$}{\TAearr{}}, thus sensitivity.
The problem is posed as a trade-off between \aa{} station diameter
and the number of stations (where $\Nst\propto1/\Dst^{2})$. But as
\prettyref{app:trade-Dst-Nst} shows, station diameter is not an independent
parameter, because $\Dst\propto\sqrt{\Nest}\deeavg$, where $\Nest$
is the number of elements per station and $\deeavg$ is the average
inter-element spacing. Assuming $\deeavg$ remains constant, a more
exact description is that the number of elements per station is traded
with the number of stations. 

To illustrate the cost sensitivities of this $\Nst$ vs. $\Nest$
trade-off, \prettyref{app:station-diameter-example} considers a simple
comparative example, where the diameter of every single-band, low-band
or high-band station is halved (reducing $\Nest$ by a factor of 4),
which results in $\Nst$ increasing by a factor of 4 to maintain $\Aearr$.
The inter-element spacing for each band remains constant.

The hardware cost of the \SKAiLow{} stations shows no dominant trend
between the smaller stations and their full-sized counterparts (\prettyref{fig:cost-SKAlow-stations-smaller},
p \pageref{fig:cost-SKAlow-stations-smaller}). For both the single
and dual-band implementations, reducing the station diameter:
\begin{itemize}
\item decreases the total station beamformer cost and variable bunker cost
\item decreases the total cost of links from the antenna element or tile
to the processing bunker
\item increases the total fixed bunker cost.
\end{itemize}
The effect of these trends on the total cost of the stations depends
on the implementation, intra-station architecture and cost data source.
However, given the cost uncertainty discussed in \prettyref{sub:risk-uncertainty},
there is no dominant trend for the station hardware cost of the smaller
stations relative to their full-size counterparts; only the single-band
implementation of the \lofarDig{} scenario shows a significant change
in total station hardware cost. (In this case a reduction in cost.)

Of the variable costs outlined in \prettyref{sec:variable-cost-system-implications},
only the correlator and imaging processing costs change with the $\Nst$
vs. $\Nest$ trade-off; they increase by a factor of 4 and 8 respectively
(\prettyref{fig:total-cost-50dep-10site-smaller-stations}, p \pageref{fig:total-cost-50dep-10site-smaller-stations}).
The deployment and site preparation costs are unchanged, because the
total number of antenna elements in the array remains constant. Although
the correlator and the imaging processor become significant costs
for the half-diameter station example, these are zeroth-order estimates.
If smaller stations are being considered, more detailed investigation
is required as to the accuracy of using the correlator output data
rate as a proxy for the imaging processor cost scaling, and of the
correlator and processor cost estimates themselves.

Cost is not the only factor to consider; there is also the effectiveness
in meeting science requirements. For example, meeting \uvc{} requirements
 can depend on increasing $\Nst$ \citep[e.g.][]{BolSca09,LalLob09}.
But distributing the extra stations to improve \uvc{} will likely
have additional infrastructure costs \citep{BolMil11-Config}. To
not disadvantage the half-diameter example, we have chosen not to
analyse a more widely distributed array configuration in this study.
For stations  within the core, \prettyref{app:shared-processing-nodes}
discusses how $\Nst$ may be efficiently increased  by creating a
number `logical' stations, each with fewer elements, which share
a processing node and form a `physical' station. However, any requirement
to increase $\Nst$ needs to be traded with the signal processing
costs; in such a trade-off, the hardware prior to the station beamformer
is not a cost driver, but the imaging processor (and to some extent
the correlator) has the potential to be a large cost driver.

\subsection{Reducing the FoV requirement: defining a fixed beam--bandwidth product\label{sub:reducing-FoV}}

Other trade-offs are emerging through analysis of the \SKAi{} \DRM{}
(\DRMi{}). \DRMi{} captures the set of observations required to
achieve the \SKAi{} science goals and forms an `envelope' of technical
requirements for the telescope. One possible performance--cost trade-off
is to reduce the \SKAiLow{} signal processing capacity, defined by
the product of the bandwidth and the average number of station beams
formed. In this approach, the processing capacity only meets the beam-bandwidth
product required by the most demanding science application in \DRMi{}.
In contrast, the \SKAi{} \hlsd{} (HLSD) and the results presented
thus far assume sufficient signal processing capacity to concurrently
observe $\unit[20]{deg^{2}}$ of processed FoV over the entire 70--450\,MHz
band. 

To understand the cost advantages from such a trade-off, \prettyref{app:beam-bandwidth-trade}
considers the cost of a strawman design, where the representative
single and dual-band implementations are modified such that the signal
processing capacity is defined by the requirement to only observe
$\unit[20]{deg^{2}}$ across the 70--180\,MHz band, resulting in
a beam-bandwidth product of 4.8\,GHz. By comparison, the beam-bandwidth
capacity of our canonical single and dual-band implementations are
80\,GHz and 24\,GHz respectively, the latter being smaller because
fewer beams are required in the 180--450\,MHz frequency range to
form the $\unit[20]{deg^{2}}$ FoV.

Figures \ref{fig:cost-SKAlow-stations-smaller-low-FoV} and \ref{fig:total-cost-50dep-10site-smaller-stations-low-FoV}
in \prettyref{app:beam-bandwidth-trade} show that for such a strawman,
the cost of the sub-systems which scale with the number of station
beams are no longer significant (those being the station beamformer
and station--CPF transmission, as well as the correlator and imaging
processor). The costs which dominate are those which scale with the
number of antenna elements. Thus the dual-band implementation, with
twice the number of elements, is more expensive in all scenarios,
although the increase is less than the factor of two increase in cost
which one might naively expect for twice the number of elements. However,
this difference still makes the trade-off potentially significant.

\subsection{Intra-station signal transport and processing architecture considerations\label{sub:signal-architectures}}

The scalable high-level view of the system, which follows the elemental
signal path (\prettyref{sec:parametric-cost-modelling}), allows for
the realisation of  various signal transport and processing architectures.
\prettyref{sub:cost-data-sources} describes the architectures used
in this analysis, but a greater number of data transport and processing
architectures are conceivable, through the combination of:
\begin{itemize}
\item using different technologies to perform the sub-system function
\item rearranging the order of the sub-systems in the signal path
\item changing physical location the sub-system.
\end{itemize}
\prettyref{tab:transport-processing-architectures} shows the most
relevant options for intra-station signal transport and processing
architectures. Although comparing the cost-effectiveness of the different
architectures is beyond the scope of the present analysis, this section
discusses the performance and cost implications of some of these architectures,
focusing on the hierarchical beamforming and power supply aspects.
Some architectures are also discussed in \citet{FauAle10}, \citet{AAV11-Concept}
and \citet{FauVaa11-Deployment}. 

\begin{table}[!t]
\caption{Principal options for intra-station architectures.\textsuperscript{a}\label{tab:transport-processing-architectures}}

\centering{}\begin{threeparttable}%
\begin{tabular}{l>{\centering}p{0.3\textwidth}>{\centering}p{0.15\columnwidth}}
\toprule 
Sub-system & Physical location & Technology\tabularnewline
\midrule
Digitiser & receptor, tile or station & N/A\tabularnewline
Tile BF (optional) & tile or station & RF or digital\tabularnewline
Signal transport & receptor--tile and tile--station, or receptor--station & RF or digital\tabularnewline
Station beamformer & station node, or node serving multiple stations & digital\tabularnewline
\bottomrule
\end{tabular}\begin{tablenotes}
\small
\item[a] The options listed illustrate the range of signal transport and processing architectures. This does not imply that every architecture would meet all the \SKAiLow{} requirements, nor is every option listed.
\end{tablenotes}
\end{threeparttable}
\end{table}

\subsubsection{Digital hierarchical beamforming\label{sub:digital-hierarchical-beamforming}}

Some caution needs to be applied to the comparison between the analogue
tile and all-digital beamforming made in \prettyref{sub:RF-beamforming-cost-reduction}.
This is because alternative intra-station architectures using hierarchical,
or multi-stage, all-digital beamforming also have the potential to
reduce cost. For the all-digital architecture in \prettyref{fig:schematic-all-dig},
the digitised signal from every antenna element is transmitted to
the station beamformer block and both the tile and station stages
of beamforming are located in that block. 

\begin{figure}[!t]
\centering{}\includegraphics[width=1\textwidth]{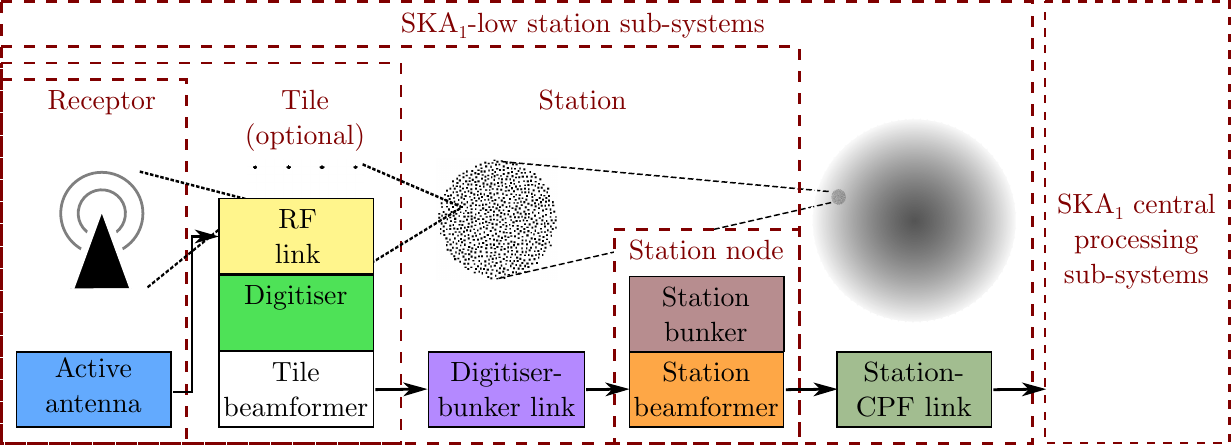}\caption{Schematic of the all-digital beamforming architecture, with a digital
tile beamforming block located at the tile.\label{fig:schematic-all-dig-tile-BF}}
\end{figure}

An alternative all-digital beamforming architecture is shown in \prettyref{fig:schematic-all-dig-tile-BF};
this more closely represents the all-digital architecture in \citet{FauVaa11-Deployment}.
The architecture performs the first stage of digital beamforming at
each tile to reduce the processed FoV earlier in the signal chain.
Only $\Nbt$\nomenclature[sNbt]{$\Nbt$}{\TNbt{}} digital tile beams
are transmitted to the station beamformer, reducing the total data
rate transmitted, hence digitiser--bunker link cost, by a factor of
$\Nbt/\Net$. Hierarchical beamforming also reduces the beamformer
processing load as described in \prettyref{app:hierarchical-beamforming},
although this applies regardless of how the hierarchical beamforming
processing is physically distributed. The \skadsCost{} cost estimate
for the station beamformer processing already includes a discount
from hierarchical beamforming. 

However, these savings may be offset or even exceeded by the extra
costs introduced by performing digital tile beamforming at distributed
locations in the station signal path, rather than just at the station
node. For example, the processing for distributed tile beamforming
could require extra power distribution infrastructure. The total cost
of controlled environment housings (including cooling and RFI shielding)
for each tile beamformer would probably be more expensive than implementing
both stages of beamforming within a larger controlled environment
housing at each station node. Also, upgrading the digital system is
easier if the processing is concentrated at a single location.

\subsubsection{Hierarchical beamforming performance\label{sub:hierarchical-beamforming-performance}}

Although hierarchical beamforming enables further design flexibility
for the intra-station architecture, it also has the potential to decrease
performance, regardless of whether the first-stage tile beamforming
is analogue or digital. As mentioned in \prettyref{sub:RF-beamforming-cost-reduction},
tile beamforming reduces the accessible FoV and observational flexibility
earlier in the signal path. If only a single tile beam is formed,
it precludes the flexibility of pointing station beams at independent
patches of the sky; the station beam pointing is restricted to within
the single tile beam. Hierarchical beamforming also introduces errors
in those station beams which are off-centre (i.e.~not pointing in
exactly the same direction as the tile beam), as discussed in \citet{FauAle10}.

\begin{figure}[!t]
\begin{centering}
\includegraphics[width=0.35\textwidth]{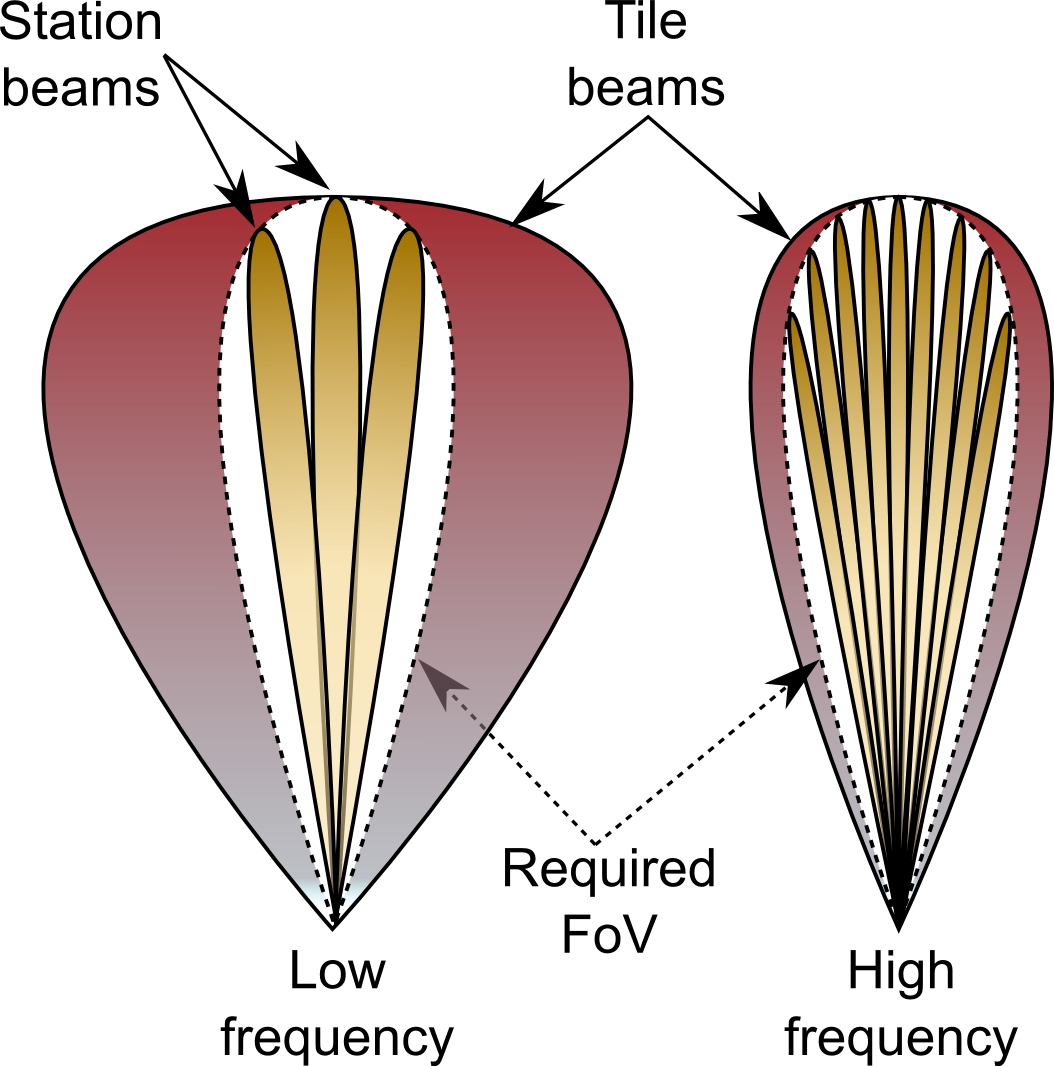}
\par\end{centering}

\caption{Schematic of the frequency-dependent relationship between the required
FoV (dotted white) and the tile and station beams.\label{fig:AA-beams-frequencies}}
\end{figure}

Tiles with more elements per tile further reduce the accessible FoV,
because of the larger tile diameter (assuming the inter-element spacing
is kept constant). If only one tile beam is formed\label{ass:one-tile-beam-formed},
then the accessible FoV is defined by $\Otile$, the FoV of that tile
beam. The required processed FoV,~$\Oreq$, is limited by the accessible
FoV, such that $\Oreq<\Otile$. As illustrated in \prettyref{fig:AA-beams-frequencies},
$\Otile$ is frequency-dependent, while $\Oreq$ is not. Because $\Otile$
is defined to be the FoV out to the half-power point of the beam,
equal sized FoVs ($\Otile=\Oreq$) would result in sensitivity loss
away from the centre of the tile beam. 

For example, consider a $4\times4$ element tile, with regularly (uniformly)
spaced elements. From \citet{Kra86}, the half-power beamwidth~$\HPBW$
at zenith for a linear array of $n$ elements with inter-element spacing
$\dee$ is approximately
\begin{equation}
\HPBW=0.88\sin^{-1}\left(\frac{\lambda}{n\dee}\right);
\end{equation}
away from zenith, $\HPBW$ is larger \citep{Mai95}. For $d=\unit[1.5]{m}$
at 450\,MHz, $\HPBW\approx\unit[5.6]{deg}$ and $\Otile\approx\unit[31]{deg^{2}}$,
while $\Oreq=\unit[20]{deg^{2}}$. Thus a single beam formed from
a 16 element tile does provide sufficient FoV at the highest frequency,
but there would be some sensitivity loss towards the edge of the tile
beam. This loss reduces at lower frequencies, given $\Otile\propto1/\nu^{2}$. 

Alternatively, more tile beams could be formed. However, this would
result in more signal paths to the station beamformer, and the cost
reduction would be less. Because this problem similarly applies to
digital tile beamforming, the full benefit of a reduction in data
transmission and signal processing cost is only realised if only one
digital tile beam is formed. For example, the all-digital architecture
in \citet{FauVaa11-Deployment} sends multiple tile beams from the
256 element tiles to the station beamformer, resulting in a total
data rate at the station beamformer which is nearly as high as transmitting
every digitised antenna element signal.

There is potential for both RF and digital tile beamforming approaches
to be implemented in different phases of the telescope. If designed
correctly, the analogue tile beamforming could be upgraded to all-digital
beamforming at a later date, once the cost and power consumption of
the digital components reduce so that such an upgrade is cost-effective.

\subsubsection{Example alternative architectures\label{sub:alternative-architectures-examples}}

Alternative intra-station architectures may prove to be more cost-effective
than those considered in this analysis, with implications for the
cost trade-offs. Technical factors also require consideration, such
as dissimilar power losses at different locations in the architecture,
and the deployability of different intra-station architectures. 

In the all-digital architecture (\prettyref{fig:schematic-all-dig}),
digitisation occurs relatively close to the antenna element and the
digitised signal is transmitted via optic fibre. Powering the active
antenna electronics, digitiser and digital optical transmission components
presents an extra cost, because they would require either an extensive
power distribution network which is appropriately sized to the peak
power load, or the installation of self-powered (solar power and energy
storage) solutions \citep[e.g.][]{Hal09,FauVaa11-Deployment}. Self-powered
antenna elements would remove the need to distribute power to the
electronics at every antenna element. However, to cater for the 180\,m
station diameter, such an architecture would require an increase to
the current maximum transmission distance of 50\,m for short-range,
high-speed digital optical transmission~\citep{FauVaa11-Deployment}. 

In contrast, the RF tile beamforming architecture (\prettyref{fig:schematic-RF-tile})
can deliver power to the active antenna element electronics and RF
tile beamformer from the station node via a copper-based RF link,
as is done for LOFAR \citep{Gun07-ADD}. Using the RF link for both
signal transport and power delivery presents a cost saving, because
the dedicated power distribution network only extends as far as the
station nodes, rather than to every antenna element. 

An alternative all-digital beamforming architecture, if technically
feasible, could put the digitisation at station node and use similar
RF links to power every antenna element and transport the signal to
the station node. The obvious additional cost is the extra RF links
required, but a separate power distribution network is not required.

\begin{figure}[!t]
\centering{}\includegraphics[width=1\textwidth]{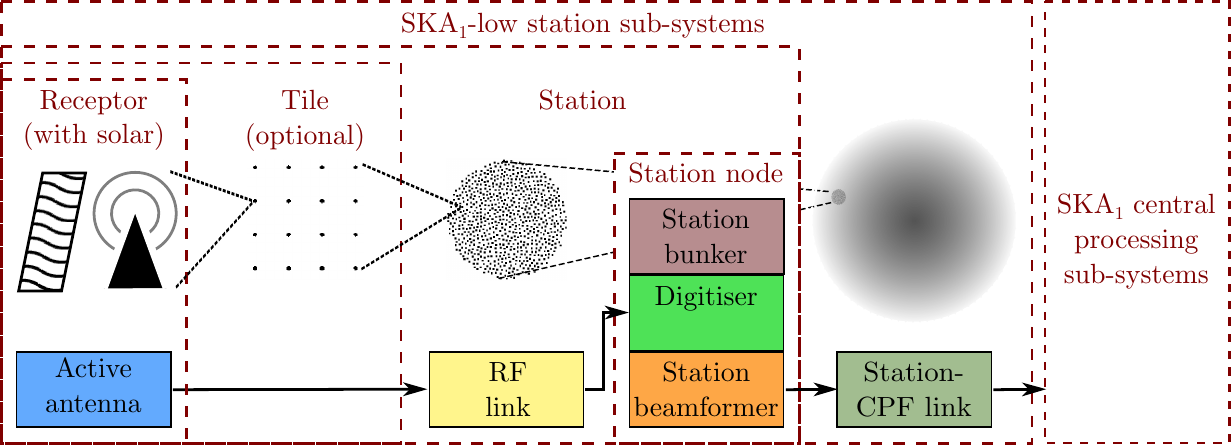}\caption{Schematic of a self-powered array, with the digital blocks located
at the station node.\label{fig:schematic-solar}}
\end{figure}

A different architecture could have only the analogue components at
the antenna element, and use \rof{} (RoF) technology to transmit
the analogue signals from each antenna element \citep{JusVaa11}.
These analogue signals would be transmitted to a node containing the
digitisers, channelisation and beamforming equipment, as shown in
\prettyref{fig:schematic-solar}. The node could serve one of more
stations; for transmission distances of 200--500\,m, such an architecture
would not be viable for analogue transmission over copper-based cable
\citep{Per11}. The fibre cables are also physically smaller than
copper-based cable, which may be important if there are tens of thousands
of cables entering one node. The life-cycle cost benefits of these
self-powered architectures with fibre links include simplified deployment
from fewer connections (than the all-digital architecture with distributed
power), potential for RFI reduction, increased resilience against
lightning strikes and increased upgrade flexibility because all the
received data arrives at the node \citep{FauVaa11-Deployment}.

\section{Further work\label{sec:further-work}}

In the absence of all the necessary cost information to make a complete
cost estimate, this first-order analysis provides a simple comparison
of the cost differences between representative single and dual-band
implementations. Along with obtaining new cost information, there
is scope for continued expert attention to refine the existing cost
estimates and better understand the uncertainties. Cost refinement
means improving the accuracy and precision of the cost, using \refClass{}
costing from other projects or studies, or conducting new bottom-up
cost studies. Prime candidates for cost refinement are:
\begin{itemize}
\item active antenna elements, for both the single-band (6.5:1) and dual-band
(2.5:1) elements
\item antenna deployment and site preparation
\item central processing sub-systems.
\end{itemize}
There will always be a level of uncertainty in parametric studies
because the models use fairly simple scaling relationships for the
trade-off analysis. This uncertainty can be mitigated by solidifying
the \SKAiLow{} performance requirements, and the specific trade-offs
required.

The use of our parametric model need not be limited to comparing single
and dual-band arrays. \prettyref{sec:other-considerations} illustrates
just a few of the trade-offs which can be explored; other top-level
parameters within our model which could be varied include:
\begin{itemize}
\item inter-element spacing
\item antenna element gain
\item the frequency split point of the dual-band array
\item number of elements per station (to vary $\AonT$ while keeping the
number of stations fixed).
\end{itemize}
The optimal number of stations and elements per station, for constant
sensitivity, is an open question, but requirements other than cost
must be considered. For example, the number and geographical placement
of stations to achieve adequate \uvc{} and telescope calibration
will also affect this optimisation. Such an optimisation requires
a better understanding of the relative life-cycle costs between the
\SKAiLow{} station digital sub-systems and the central processing
sub-systems, in conjunction with refined performance requirements
for station sensitivity and processed FoV (\prettyref{app:station-performance-further-work}).

We mention in \prettyref{sec:introduction} the close link between
the telescope performance, cost and risk, and the science requirements.
Our parametric model is sufficiently robust to consider differential
cost trends between implementations, but is less suitable for determining
absolute costs; calculating the monetary cost of changing a science
requirement requires care.

\section{Conclusions\label{sec:conclusions}}

We have developed a scalable parametric model to compare the cost
of implementing \SKAiLow{} as a single or dual-band \aa{}, considering
the cost impact on both the station hardware costs and the broader
telescope system costs. Perhaps somewhat surprisingly, despite the
dual-band array having twice the number of antenna elements, neither
the representative single or dual-band implementations are conclusively
cheaper.

The cost difference between the single and dual-band implementations
is, in essence, a comparison between the number and cost of the signal
paths prior to station beamforming, and the downstream signal processing
costs. The dual-band implementation has twice as many signal paths,
but achieves a given field of view using fewer station beams, thereby
reducing the downstream processing load. As a result, the dual-band
implementation is more sensitive to changes in costs that scale with
the number of signal paths, such as those of active antenna element
hardware and deployment. Conversely, the single-band implementation
is more sensitive to changes in the cost of signal processing sub-systems
such as the station beamformer, correlator, and imaging and non-imaging
processors. The cost difference between the single and dual-band implementations
depends on the fractional cost of each of these groupings.

A particularly important \SKAlow{} parameter requiring further scientific
consideration is the product of the processed FoV and bandwidth---the
beam-bandwidth product, discussed in \prettyref{sub:reducing-FoV}.
If the beam-bandwidth capacity of the processing can be significantly
reduced, the costs which then dominate are those that scale with the
number of antenna elements. This increases the cost of the dual-band
implementation relative to the single-band implementation, but the
difference is much less than the factor of two increase which might
naively be expected from an array with twice the number of elements.

We find that implementing a first stage of RF (analogue) tile beamforming
prior to the digital station beamformer enables a potentially significant
reduction in station hardware costs and power demand, the effect of
which can outweigh any difference between single and dual-band implementations.
Furthermore, if 90\,m diameter stations are considered instead of
180\,m, the correlator and imaging processor costs become cost drivers,
dominating the station hardware costs.

To establish whether the single or dual-band array is the most cost-effective
\SKAiLow{} implementation, improved cost information and further
optimisation of the putative telescope designs are required. In terms
of specific studies likely to assist in system design choices, we
note that central processing, antenna deployment and site preparation
costs are potentially significant cost drivers which have so far not
had sufficient attention.

\section*{Acknowledgements}

We thank Jan Geralt Bij de Vaate, Andrew Faulkner, Aziz Jiwani, and
ICRAR, ASTRON, SPDO and University of Cambridge colleagues for the
discussions surrounding this topic.

\pagebreak{}

\appendix
\addappheadtotoc

\addtocontents{toc}{\protect\setcounter{tocdepth}{0}}

\let\stdsection\section
\def\section*#1{\stdsection{#1}}
\printnomenclature{}
\let\section\stdsection

\section{Summary of assumptions \label{app:assumptions}}

For clarity, the key assumptions made for this analysis are summarised
here and page references listed.
\begin{itemize}
\item Performance-related assumptions:

\begin{itemize}
\item signals are dual-polarisation ($\Npol=2$), or full-Stokes \prettyref{ass:dual-pol}
\item $\unit[20]{deg^{2}}$ processed field of view is observed concurrently
across the 70--450\,MHz band \prettyref{ass:concurrent-FoV}
\item antenna element gain is the same for the single and dual-band (low
and high) antenna elements to ensure that the first-order station
$\AonT$ estimates are comparable \prettyref{ass:antenna-element-gain}
\item the intra-station element layout is an irregular layout of uniform
element distribution \prettyref{ass:intra-station-layout}
\item only one tile beam is formed for each tile \prettyref{ass:one-tile-beam-formed}.
\end{itemize}
\item Cost-related assumptions:

\begin{itemize}
\item the \lofarCost{} cost estimate is for 50\,MHz output bandwidth with
2007 technology, so by 2016 newer technology will allow for the processing
of the full 380\,MHz bandwidth for the same cost \prettyref{ass:LOFAR-50MHz-BW}
\item the estimated costs do not represent the total cost of building the
telescope, because costs are excluded when it can be shown that they
remain constant between the single and dual-band implementations \prettyref{ass:costs-not-price-to-build}
\item the low and high-band cores are separate \prettyref{ass:low-high-cores}
\item low and high-band stations are co-located beyond the core, and the
trenching and cables for the data transmission and power to these
stations are shared \prettyref{ass:trenching-cabling-co-location}
\item infrastructure common to both stations, such as housing for the station
processing node, is not shared \prettyref{ass:no-shared-station-infrastructure}.
\end{itemize}
\item Signal processing assumptions:

\begin{itemize}
\item a time to frequency domain transformation and cross-correlation `FX'
correlator is used \prettyref{ass:FX-correlator}
\item the number of coarse frequency channels and their channel width is
constant \prettyref{ass:coarse-channel-width}
\item the station beamformer cost scaling is the same for both frequency
and time domain beamforming (i.e. the coarse filterbank does not dominate
the beamformer cost) \prettyref{ass:freq-time-BF-scaling}
\item the \SKAiLow{} correlator frequency resolution is defined by the
HLSD ($\chBW=\unit[1]{kHz}$) rather than $\Dst$ \prettyref{ass:channel-width}
\item the \cpf{} sub-systems operate on a `per beam' basis \prettyref{ass:central-processing}
\item the imaging cost is dominated by the data buffer, rather than the
processing \prettyref{ass:imaging-cost}. 
\item only the high-band core of the dual-band array is used in the \nip{},
and the processing for the AAs, not the dishes, dominate the cost
\prettyref{ass:NIP}.
\end{itemize}
\end{itemize}

\section{Filterbank and cross-correlation architecture\label{app:FX-architecture}}

The models assume a time to frequency domain transformation and cross-correlation
`FX' correlator is used. This is the most cost-effective architecture
for the SKA, as opposed to other correlator topologies such as `XF'.
The FX correlator architecture is discussed in \citet{Bun00}; the
architecture is more cost-effective with an increasing number of correlatable
inputs (stations in this document). An FX architecture also allows
signal processing actions, such as beamforming and RFI excision, to
be efficiently performed \citep{HalSch08}. 

In an FX correlator architecture, the signal from each element or
tile input is filtered into frequency channels (`F') and for each
channel, the input signals are cross-correlated (`X'). \citet{Bun03}
shows that the filtering option can be efficiently implemented with
a polyphase filterbank (PFB), such that the data rate of the output
signal is equivalent to the input data rate. Although computationally
more expensive, cascaded or multi-stage filterbanks allow for more
efficient implementation on discrete processing units, such as ASICs
and FPGAs, without the use of external RAM and the associated memory
bandwidth costs. Oversampling is also required at all stages except
the last to maintain performance where the channels overlap. 

Cascaded filterbanks allow flexibility in the implementation architecture,
where each filterbank stage is appropriately located to suit the architecture.
The all-digital beamforming architecture in \citet{AAV11-Concept}
integrates the first-stage coarse channel filterbank (CFB) between
the digitisation and the tile beamforming blocks shown in \prettyref{fig:schematic-all-dig-tile-BF}
(p \pageref{fig:schematic-all-dig-tile-BF}). Channels outside processed
bandwidth $\BW=\fmax-\fmin$\nomenclature[s~deltanu]{$\BW$}{\TBW{}}
are discarded after the coarse filterbank, hence are not transported
to the station beamformer. Although the data transport requirements
are reduced, there are additional infrastructure and power supply
implications in having the filterbank at the digitiser. 

For simplicity, this analysis assumes that first-stage coarse channel
filterbank (CFB) is implemented at the station bunker, prior to the
beamforming; its cost is implicitly included in the station beamformer
cost (\prettyref{app:station-beamformer}).\label{ass:CFB-at-station}
The second-stage fine channel filterbank (FFB) is located at the correlator
(\prettyref{app:fine-channelisation-correlation}). Because any scalable
digital signal processing description (e.g \citealp{Bun10}) requires
a specific architecture with implicit assumptions, this first-order
analysis keeps the number of coarse frequency channels and the channel
width constant.\label{ass:coarse-channel-width}

\section{Parametric models and costs for  \texorpdfstring{\SKAiLow{}}{SKA1-low} stations\label{app:models-SKAlow}}

The parametric equations for the station sub-system blocks consist
of fixed and variable units costs, $\Cfix{}$ and $\Cvar{}$ respectively.
The variable costs scale with one or more parameters. The total cost
of a particular block in the system is the product of quantity and
cost. Table \ref{tab:blocks-summary} summarises the quantity and
the cost scaling of the blocks in this analysis (see \prettyref{sec:parametric-cost-modelling}).
The detailed block descriptions below provide further justification
and references for the cost scaling. \prettyref{tab:blocks-cost}
lists the fixed and variable unit costs of each block. \nomenclature[sCfix]{$\Cfix{}$}{\TCfix{}}\nomenclature[sCvar]{$\Cvar{}$}{\TCvar{}}

\begin{table}[!t]
\caption{Summary of blocks and scaling for \SKAiLow{} sub-systems.\label{tab:blocks-summary}}

\begin{tabular}{>{\raggedright}p{0.2\textwidth}>{\raggedright}p{0.22\textwidth}>{\raggedright}p{0.25\textwidth}>{\raggedright}p{0.28\textwidth}}
\toprule 
Block name & Quantity in \SKAi{} & Parametric equation & Block coverage\tabularnewline
\midrule
Active antenna element & $\Nst\Nest$  & $\Cfix{}$ & mechanical element, LNA, gain and filter, housing and ground plane\tabularnewline
Analogue (RF) tile beamformer\textsuperscript{a}  & $\Nst\Ntst$ & $\Nbt\Net\Cvar{}$ & hardware\tabularnewline
Element/tile--digitiser RF link & RF: $\Nst\Ntst\Nbt$

Dig: $\Nst\Nest$ & RF: $\Cfix{RF}$

Dig: $\Cfix{Dig}$ & RF cable\tabularnewline
Digitiser & RF: $\Nst\Ntst\Nbt$

Dig: $\Nst\Nest$ & $\Cfix{}$ & digitiser\tabularnewline
Digitiser--bunker link \textsuperscript{b} & Dig: $\Nst\Nest$ & $\Cvar 1\Rdig+\Cvar 2\overline{L_{{\rm e-st}}}$ & $\Cvar 1$: electronics, cable connectors

$\Cvar 2$: fibre cable\tabularnewline
Station beamformer\textsuperscript{}\textsuperscript{c} & $\Nst$ & RF: $\Nbstavg\Ntst\Cvar{}$

Dig: $\Nbstavg\Nest\Cvar{}$ & coarse filterbank, station beamformer\tabularnewline
Station infrastructure (bunker) & $\Nst$ & RF: $\Cfix{}+\Agst\Cvar 1+\Nbstavg\Ntst\Cvar 2$

Dig: $\Cfix{}+\Agst\Cvar 1+\Nbstavg\Nest\Cvar 2$ & $\Cfix{}$: building etc.

$\Cvar 1$: preparation, trenching etc.

$\Cvar 2$: environmental conditioning, rack space etc. \tabularnewline
Station--CPF link transmission & $\Nst$ & $\Rst\Cvar{}$ & fibre transmission\tabularnewline
\bottomrule
\end{tabular}

\textsuperscript{a}Optional block. If analogue tile beamforming is
included, subsequent quantities and costs are denoted `RF'. If not,
the system is all-digital beamforming, denoted by `Dig'.\\
\textsuperscript{b}Optional block. Assumes no digital beamforming
at the tile. See \prettyref{sub:signal-architectures} for alternative
architectures.

\textsuperscript{c}Approximate cost scaling, see \prettyref{app:beamforming-computational-cost}.
\end{table}

\begin{table}[!t]
\centering{}\caption{\SKAiLow{} sub-system unit costs in \euro~(2007). \label{tab:blocks-cost}}
\begin{threeparttable}%
\begin{tabular}{>{\raggedright}p{0.25\textwidth}>{\raggedright}p{0.25\textwidth}>{\raggedright}p{0.2\textwidth}>{\raggedright}p{0.12\textwidth}}
\toprule 
Block name & \SkadsCost{} cost estimate\textsuperscript{a} & \LofarCost{} cost estimate\textsuperscript{b} & Block unit\textsuperscript{c}\tabularnewline
\midrule
Active antenna element & $\Cfix{}$: 75 & $\Cfix{}$: 172 & per element\tabularnewline
RF tile beamformer & $\Cvar{}$: 7.0 per output beam & $\Cvar{}$: same as low & per element\tabularnewline
Element/tile--digitiser RF link & $\Cfix{RF}$: 81

$\Cfix{Dig}$: 18 & $\Cfix{RF}$: same as low

$\Cfix{Dig}$: same as low & per signal\tabularnewline
Digitiser & $\Cfix{}$: 12 & $\Cfix{}$: 80 & per signal\tabularnewline
Digitiser--bunker link & $\Cvar 1$: $1\,{\rm Gbps^{-1}}$

$\Cvar 2$: $0.01\,{\rm Gbps^{-1}}{\rm m^{-1}}$ & $\Cvar 1$: same as low

$\Cvar 2$: same as low & per link\tabularnewline
Station beamformer & $\Cvar{}:$ 0.1 per input per output beam & $\Cvar{}:$ 2.6 per input per output beam & per station\tabularnewline
Station infrastructure & $\Cfix{}$: $28\,{\rm k}$

$\Cvar 1$: 0

$\Cvar 2$: 0.06 per input per output beam & $\Cfix{}$: $74\,{\rm k}$

$\Cvar 1$: same as low

$\Cvar 2$: 2.6 per input per output beam & per station\tabularnewline
Station--CPF link transmission & $\Cvar{}$: $\unit[100]{Gbps^{-1}}$ & $\Cvar{}$: same as low & per link\tabularnewline
\bottomrule
\end{tabular}\begin{tablenotes}
\small
\item[a] All-digital beamforming,  and technology advances. Most costs extrapolated from Table 3 of \citet{FauVaa11-Deployment}.
\item[b] LOFAR estimate for 50 MHz output bandwidth and analogue (RF) tile beamforming. Most costs extrapolated from Table 4 of \citet{FauVaa11-Deployment}. The availability of only one cost estimate is indicated by `same as low'.
\item[c] All elements, beams, inputs and signals are dual polarisation.
\end{tablenotes}
\end{threeparttable}
\end{table}

These station cost estimates are for the sub-system hardware costs,
where a sub-system is generally described by one of the scalable blocks.
The station cost estimates include `sub-system infrastructure',
such as the housing for the signal processing units. But the sub-system
hardware costs themselves are not just the procurable components (the
physical hardware); they also include costs such as \nre{}, assembly
and  integration and testing for the sub-system. \citet{FauVaa11-Deployment}
details the cost coverage for the AA CoDR estimates; in general, only
the component costs are accounted for. Note that some components may
implicitly include non-component hardware costs in the component purchase
price. For example, the purchase price of a digitiser board would
usually include assembly, integration and testing prior to delivery.

\subsection{Active antenna element\label{app:active-antenna-element}}

The active antenna element describes an integrated system, which includes
the mechanical element, LNA, gain and filter, housing and ground plane.
We only consider a consolidated unit cost for the active antenna element,
because its constituent components are specified and costed for a
particular design. For example, the mechanical antenna element is
matched to the LNA to minimise the receiver noise across the frequency
band \citep{ArdBre09}, and the ground plane is designed for a given
antenna element. Additionally, the active antenna element design should
minimise manufacturing, transportation, deployment and operations
costs, as discussed in \citet{FauVaa11-Deployment}. 

Consolidating the active antenna element cost does conceal potential
trade-offs within that design space; a trade-off of topical interest
is the cost of providing a ground plane for the element.  However,
a simple analysis in \prettyref{app:ground-plane} finds that there
is not currently the justification for costing the ground plane separately
to the rest of the active antenna element, because the cost of ground
plane is more closely linked to the type of antenna element than the
areal cost, and the ground plane cost is not significant in the broader
\SKAiLow{} context, given the uncertainties of the present first-order
analysis.

\subsubsection{Active antenna element costs\label{app:active-antenna-element-costs}}

The cost data sources in \citet{FauVaa11-Deployment} provide a consolidated
cost for the active antenna element. The cost per dual polarisation
active antenna element is \euro75 for the \skadsCost{} estimate,
the cost being taken from existing arrays. The \lofarCost{} estimate
for the same system is \euro179 and is a direct transfer from the
cost of the LOFAR high band (120--240\,MHz) antennas. For the purposes
of our analysis, the antenna element's RF beamformer cost of \euro7
per element is subtracted from this cost and listed separately in
\prettyref{tab:blocks-cost}, resulting in a \lofarCost{} element
cost of~\euro172.

A comparison of the representative single and dual-band implementations
clearly requires estimation of the low and high-band active antenna
element costs. Factors to consider are:
\begin{itemize}
\item The fractional bandwidth of both the low and high-band elements is
approximately 2.5:1, as opposed to 6.5:1 for the single-band element.
\item The low-band elements, with a maximum frequency $\fmax=\unit[180]{MHz}$,
will require less exacting manufacturing standards than the single
and high-band elements.
\item The high-band elements, with a minimum frequency $\fmin=\unit[180]{MHz}$,
are physically smaller than the single and low-band elements.
\item The average inter-element spacing for the high-band array is 0.75\,m,
compared with 1.5\,m for the single and low-band arrays.
\end{itemize}
To provide a first-order estimate of the influence of these factors
on the low and high-band active antenna element costs, multipliers
(discounts) are applied to the single-band element costs (\euro75
and \euro172 for the \skadsCost{} and \lofarCost{} estimates respectively).
The cost of the active antenna element is assumed to be split 2:1
between the physical components and the electronics (LNAs, filters
etc). The cost of the low-band physical components is estimated at
80\,\% of the single-band cost. The high-band $\fmin$ is about 2.5
times the single-band $\fmin$, hence the cost of the physical components
is estimated at 40\,\% of the single-band cost. The electronics for
each band in the dual-band array are assumed to be only marginally
cheaper, at 90\,\% of the single-band cost. The multipliers are thus
calculated as 
\begin{align}
\text{Low-band} & =\frac{2\times0.8+0.9}{3}\\
 & =0.83
\end{align}
and 
\begin{align}
\text{High-band} & =\frac{2\times0.4+0.9}{3}\\
 & =0.57.
\end{align}
These multipliers are first-order estimates; a detailed analysis,
including the applicability to more directive antennas, is an important
investigation for future SKA studies.

\subsubsection{Ground plane costs\label{app:ground-plane}}

The bottom-up SKADS and reference class LOFAR active antenna element
estimates include the cost of the ground plane. However, this section
makes a simple analysis of ground plane costs to clarify its contribution
to the consolidated active antenna element cost. The cost of a wire
mesh ground plane will vary with the area of the mesh and the size
of the openings in the mesh. One cost estimation method, applicable
to both the single and dual-band implementations, is to calculate
the total length of wire $l_{{\rm wire}}$ used to manufacture the
mesh and assume that, for large quantities, the cost of the ground
plane is linearly proportional to the cost of the wire. For a square
mesh ground screen, $l_{{\rm wire}}=2/l_{{\rm opening}}$ per~m\textsuperscript{2}
of mesh, where $l_{{\rm opening}}$ is the opening width (distance
between wires). To ensure a radio mirror surface, a rule-of-thumb
is that $l_{{\rm opening}}\le0.1\lmin$ , where $\lmin$ is the minimum
wavelength to be observed. The ground plane area for an antenna element
will depend on the inter-element spacing and the intra-station element
layout. For an irregular layout of approximately uniform element distribution,
the ground plane area for each element is approximately $\deeavg^{2}$,
where $\deeavg$ is the average inter-element spacing. Thus, the
total length of wire per antenna element is $l_{{\rm wire}}=2\deeavg^{2}/l_{{\rm opening}}$.

\prettyref{tab:ground-plane-specs} shows the ground plane specifications
and cost for the representative single and dual-band implementations.
A ground plane which reflects all frequencies in the band is shown,
as is a ground plane suitable for antenna elements such as the more
directional log-periodic dipole (LPD), where the higher frequency
portion of the element uses the element structure as the ground plane
\citep{BraCap06}. In that case, the ground plane openings are sized
only for the lower frequencies. Example ground plane costs (per~m\textsuperscript{2}
and per antenna element) are given in \prettyref{tab:ground-plane-specs},
based on the total length and cost of the wire in galvanised steel
mesh. For mesh with $\unit[50\times50]{mm}$ openings, as used for
the MWA, $l_{{\rm wire}}=\unit[40]{m\, per\, m^{2}}$ of mesh. Extrapolating
actual MWA costs, this equates to a wire cost of \euro0.15\,m\textsuperscript{-1}. 

\begin{table}[!t]
\caption{Ground plane specifications and cost, for the single and dual-band
implementations. \label{tab:ground-plane-specs} }

\renewcommand{\multirowsetup}{\centering} 

\begin{centering}
\begin{threeparttable}%
\begin{tabular}{@{}ll>{\raggedright}p{0.05\textwidth}>{\raggedright}p{0.08\textwidth}>{\centering}p{0.08\textwidth}>{\centering}p{0.12\textwidth}>{\centering}p{0.1\textwidth}>{\centering}p{0.09\textwidth}>{\centering}p{0.12\textwidth}}
\cmidrule{2-9} 
 & Band & $\deeavg$ (m) & Antenna element $\fmax$ (MHz) & Suitable antenna types\textsubscript{} & Ground plane opening dimensions (mm)\textsuperscript{a} & Ground plane usable $\fmax$ (MHz)\textsuperscript{b} & Example cost per~m\textsuperscript{2} (\euro\,2007)  & Example cost per antenna element (\euro\,2007)\textsuperscript{c} \tabularnewline
\cmidrule{2-9} 
 & \multirow{2}{*}{Single} & \multirow{2}{0.05\textwidth}{1.5} & \multirow{2}{0.08\textwidth}{450} & all & $50\times50$ & 600 & 6.0 & 13.5\tabularnewline
 &  &  &  & LPD & $200\times200$\textsuperscript{d} & 150 & 1.5 & 3.4\tabularnewline
\addlinespace[6pt]
 & \multirow{2}{*}{Low} & \multirow{2}{0.05\textwidth}{1.5} & \multirow{2}{0.08\textwidth}{180} & all & $150\times150$ & 200 & 2.0 & 4.5\tabularnewline
 &  &  &  & LPD & $200\times200$\textsuperscript{d} & 150\textsuperscript{} & 1.5 & 3.4\tabularnewline
\addlinespace[6pt]
 & \multirow{2}{*}{High} & \multirow{2}{0.05\textwidth}{0.75} & \multirow{2}{0.08\textwidth}{450} & all & $50\times50$ & 600 & 6.0 & 3.4\tabularnewline
 &  &  &  & LPD & $100\times100$\textsuperscript{d} & 300 & 3.0 & 1.7\tabularnewline
\cmidrule{2-9} 
\end{tabular}\begin{tablenotes}
\small
\item[a] Rounded to multiples of 25\,mm. 
\item[b] Calculated assuming $l_{{\rm opening}}\le0.1\lmin$.
\item[c] For a ground plane of area $\deeavg^{2}$. Wire (\euro0.15\,m\textsuperscript{-1}) costs assumed to represent the ground plane cost.
\item[d] Sizes and usable $\fmax$ are an example only and are not based on a particular antenna design.
\end{tablenotes}
\end{threeparttable}
\par\end{centering}

\end{table}

In \prettyref{tab:ground-plane-specs}, the type of antenna element
is a dominant factor in determining the ground plane costs. This effect
can be seen in the cost per~m\textsuperscript{2} column, where there
is a factor of 4 variation. This means that rather than being costed
as a distinct component, the ground plane should be considered in
conjunction with the antenna element design and included in the consolidated
active antenna element cost. Fortunately, the ground plane cost is
not significant in the broader context of \SKAiLow{} station hardware
and system variable costs. Even the largest cost difference, \euro10.1
per element between the two single-band ground planes, only equates
to a \euro113\,k cost per station. In the context of station hardware
costing at least \euro1.2\,M (see \prettyref{fig:Cost-SKAlow-sub-systems-err}),
this is less than 10\,\% of the hardware cost and is much smaller
than the uncertainties described in \prettyref{sub:risk-uncertainty}.

An additional factor to consider is that the wire cost, which has
been used as a proxy for ground plane cost, will depend on the wire
diameter. The following practical requirements will influence the
wire diameter:
\begin{itemize}
\item Rigidity: ensures a planar surface, within some level of tolerance.

\item Deployability: is the ground plane to be deployed as sheets of mesh
(larger diameter wire) or longer rolls of mesh (smaller diameter wire)?
\item Durability: will the mesh entirely cover the station area, such that
it needs to be durable enough to be walked on to enable hardware maintenance? 
\end{itemize}
The wire diameter used for the MWA mesh sheets is approximately 3\,mm,
but if the requirements are less stringent and the wire diameter can
be smaller, then the ground plane costs would further reduce.

\subsection{Optional: RF tile beamformer}

The RF tile beamformer is assumed to be a part of the active antenna
element system. Only one tile beam is formed ($\Nbt=1$) for the comparisons
made in this document. Multiple independent FoVs require the formation
of multiple tile beams.

\subsection{Element/tile--digitiser RF links}

The links listed in \citet{FauVaa11-Deployment} are CAT-7 for the
all-digital system and co-axial cable for the RF tile beamformed system,
although in principle, CAT-7 or co-axial cable could be used in either
system. The co-axial cable for the RF tile beamformed system is more
expensive because of the longer cable lengths required for that architecture.

\subsection{Digitiser}

The digitiser sample rate $\Rsample$ and number of bits $\NbitDig$
are fixed at 1\,GS/s and 8 bits respectively. The digitiser over-samples
the data by having a sample rate larger than the maximum frequency.\nomenclature[sRsample]{$\Rsample$}{\TRsample{}}\nomenclature[sNbitDig]{$\NbitDig$}{\TNbitDig{}}

\subsection{Optional: digitiser--bunker links\label{app:tile-station-links}}

The data from the digitiser is transmitted over fibre to the station
processing. This link assumes digitisation occurs at or near the tile
or element and the station processing is near the centre of the station.
Table~3 of \citet{FauVaa11-Deployment} describes a unit cost of
\euro152 for the short optical fibre link. This has the capacity
of 120\,Gbps per fibre (12$\times$10\,Gbps channels). To fully
utilise this capacity, a few digitised element or tile beam signals
could be transmitted on each link. The cost is composed of transmit
(at antenna) and receive (at station) units, connectors and the fibre.
A simplistic cost breakdown is a cost of \euro120 for the electronics
and connectors and $\euro1.0\,{\rm m^{-1}}$ for the fibre, which
approximates to $\Cvar 1=\euro1\,{\rm Gbps^{-1}}$ and $\Cvar 2=\euro0.01\,{\rm {\rm Gbps^{-1}}{\rm m^{-1}}}$
respectively.

The parameter $\Lestavg$ is the average link length between the element
or tile and the station processing.  A simplified calculation (average
radius to a circle centre) applies to an irregular layout of uniform
element distribution, such that $\Lestavg\approx\Dst/3$.  \nomenclature[sLestavg]{$\Lestavg$}{\TLestavg{}}

The data rate out of the digitiser $\Rdig$is given by
\begin{equation}
\Rdig=\Npol\NbitDig R_{{\rm sample}}OH
\end{equation}
where $OH$ is the overhead, assumed to be 1.25 for digital encoding.
\nomenclature[sRdig]{$\Rdig$}{\TRdig{}}

\subsection{Station beamformer (including coarse channel filterbank) \label{app:station-beamformer}}

Station beamforming is required for the SKA to reduce the number of
inputs to the correlator. Digital beamforming can be done in the time
domain, or in the frequency domain on the channelised signal. The
computational cost of the frequency and time domain beamforming approaches
is discussed in \citet{BarMil11} and \citet{KhlZar10}, where computation
costs are expressed as a function of the number of beams, input antennas
and frequency channels, and other costs to implement a time delay
(where necessary) and the FFT. These latter costs are architecture
specific, as is the cost scaling with the number of channels. \citet{JonZar11-Station-BF}
discuss beamforming architectures in more detail.

The station beamformer cost is approximated using \prettyref{eq:cost-BF}
below, and is calculated as cost per input per output beam. This cost
is extrapolated from \citet{FauVaa11-Deployment}; 11\,264 inputs
and 160 output beams (averaged over the band) is assumed to make the
extrapolation. The \skadsCost{} station beamformer cost already takes
into account the processing discount from the two-stage beamforming.
The \lofarCost{} cost in Table 4 of \citet{AAV11-Concept} is for
only 50 MHz bandwidth, however it is also for 2007 technology. We
assume that newer technology will allow for the beamforming of the
full 380\,MHz bandwidth for the same cost.

\subsubsection{Computational cost of frequency and time domain beamforming\label{app:beamforming-computational-cost}}

From \citet{BarMil11} and \citet{KhlZar10}, for a given architecture
and number of channels, the frequency and time domain station beamformer
processing load can be respectively simplified to 
\begin{equation}
P_{{\rm BF[\nu]}}\propto\Nin(K_{{\rm CFB}}+K_{{\rm BF[\nu]}}\Nbst)
\end{equation}
and
\begin{equation}
P_{{\rm BF[t]}}\propto\Nbst(K_{{\rm BF[t]}}\Nin+K_{{\rm CFB}}),
\end{equation}
where $\Nin$ is the number of elements or tiles being beamformed,
$\Nbst$ is the number of station beams formed, the constant $K_{C{\rm FB}}$
is the coarse filterbank cost per frequency channel and $K_{{\rm BF[\nu]}}$
and $K_{{\rm BF[t]}}$ are the beamforming costs per channel (frequency
domain) or antenna (time domain). 

If the CFB cost does not dominate (i.e. $\Nbst\gg K_{{\rm CFB}}/K_{{\rm BF[\nu]}}$
and $\Nin\gg K_{{\rm CFB}}/K_{{\rm BF[t]}}$), the cost scaling is
the same for both frequency and time domain beamforming.\label{ass:freq-time-BF-scaling}\nomenclature[sNin]{$\Nin$}{\TNin{}}
The processing cost of the station beamformer is thus approximated
by 
\begin{equation}
P_{{\rm BF}}\propto\Nin\Nbst.\label{eq:cost-BF}
\end{equation}
In this analysis, $\Nin$ is either $\Nest$ or $\Ntst$ and $\Nbst$
is $\Nbstavg$.

This approximation is used to derive the unit costs from the cost
data sources. In deriving the station beamformer cost from the \lofarCost{}
estimate, a factor of 16 fewer inputs (equal to the number of elements
per tile) is used relative to the \skadsCost{} station beamformer
cost estimate, which is calculated from the aggregate of the first-stage
and station processing costs. This results in a factor of 26 difference
in station beam unit cost (\prettyref{tab:blocks-cost}). The same
scaling approximation is applied to 80\,\% of the station bunker
cost (see \prettyref{app:station-infrastructure-bunker}), hence the
factor of 43 difference in the variable unit cost for that sub-system.

\subsubsection{Hierarchical beamforming\label{app:hierarchical-beamforming}}

Hierarchical digital beamforming reduces the data transport and processing
load on the system. A simple example of this is shown here. \citet{FauAle10}
presents a two-stage digital beamforming approach and this is reflected
in Table 3 of \citet{FauVaa11-Deployment}, where the first stage
consists of a tile of 256 antenna elements as inputs ($\Ninfn{tile}=256$).
There are 44 tiles in a station ($\Ntst=44)$\nomenclature[sNtst]{$\Ntst$}{\TNtst{}},
so one beam from each of the 44 tiles (all pointing in the same direction)
are input into the second-stage (station) beamformer ($\Ninfn{stn}=$44).
The total station processing load to form $\Nbst$ station beams (from
\prettyref{eq:cost-BF}) is approximately:
\begin{equation}
P_{{\rm BF}}\propto\Ntst\Ninfn{tile}\Nbt+\Ninfn{stn}\Nbst.
\end{equation}
Given $\Nest=\Ntst\Ninfn{tile}$ and $\Ninfn{stn}=\Ntst$, this becomes
\begin{equation}
P_{{\rm BF}}\propto\Nest\Nbt+\Ntst\Nbst.\label{eq:two-stage-bf}
\end{equation}
For this example, $P_{{\rm BF}}\propto11\,264\Nbt+44\Nbst.$ In comparison,
for a single stage of beamforming, where $\Nin=11\,264$, $P_{{\rm BF}}\propto11\,264\Nbst.$
Hence when $\Nbst>1$, the two-stage beamforming reduces processing
costs. However, caution should be used to ensure that the assumptions
for \prettyref{eq:cost-BF} still hold.

\subsubsection{Trading $\Nest$ for $\Nst$\label{app:beamforming-Nst-Nest}}

From \prettyref{eq:cost-BF}, the total processing cost for $\Nst$
single-stage station beamformers is approximately
\begin{equation}
P_{{\rm BF-total}}\propto\Nst\Nest\Nbstavg.
\end{equation}
Given $\Nst\propto1/(\Nest\FFst\deeavg^{2})$ for constant $\Aearr$
(\prettyref{eq:Nst-Nest-FF-trade}) and $\overline{\Nbst}\propto\Nest\deeavg^{2}$
(\prettyref{eq:Nbst-avg-prop-Nest-deeavg}), 
\begin{equation}
P_{{\rm BF-total}}\propto\frac{\Nest}{\FFst},
\end{equation}
where $\FFst$ is the frequency-dependent station filling factor (\prettyref{eq:FFst-detailed}).
If $\deeavg$ and the antenna element gain $\GGe$ do not change,
then the function $\FFst$ remains constant and 
\begin{equation}
P_{{\rm BF-total}}\propto\Nest.\label{eq:cost-BF-total}
\end{equation}
As long as the assumptions for the approximation (\prettyref{eq:cost-BF})
still hold, the scaling relationship can also be applied to hierarchical
beamforming. 

As shown in \prettyref{app:constant-FoV-dual-band}, the ratio of
the average number of beams between single and dual-band implementations
$\NbstavgRatio$ is independent of $\Nst$ for constant $\Aearr$.
Because $P_{{\rm BF-total}}\propto\Nbstavg$, the beamformer processing
cost ratio between the single and dual-band implementations is independent
of station diameter, although the absolute (euro) cost difference
is less for the smaller stations.

\subsection{Station infrastructure (bunker)\label{app:station-infrastructure-bunker}}

Within a station array, it is assumed that the antenna elements are
closely packed, hence cables (power and fibre) would be laid as part
of the station construction, rather than individual trenches being
dug. Thus these costs scale with area. However, this areal infrastructure
cost is difficult to estimate with much accuracy until after the site
selection. There are also costs for a controlled environment housing
at each station for the processing hardware. This cost would increase
linearly with the amount of processing, although there will be a fixed
cost for the housing. These costs are extrapolated from \citet{FauVaa11-Deployment}
to obtain a cost per input signal per output beam and a fixed cost.
We estimate the zeroth-order breakdown of costs between the variable
and fixed costs to be 80\% and 20\% respectively.

\subsection{Station--CPF link transmission\label{app:station-CPF-link-transmission}}

The output data from the station beamformer is transmitted over fibre
to the central processing facility.  This analysis only considers
the transmission costs; the per metre trenching and cabling cost is
ignored because the layout (configuration) of the \SKAiLow{} stations
do not change significantly between representative systems. The layouts
shown in \citet{BolMil11-Config} have a compact core and spiral arms.
The spiral arms may require trenching for the data links, but for
the compact core this can be absorbed into the areal infrastructure
cost. For the HLSD, 97\% of the network infrastructure costs are in
the trench network \citep{McC11-STaN-infrastructure}. This means
that although a higher data rate may require more strands of fibre,
the cost of the fibre cable is not significant, being less than 3\%. 

For the transmission costs, there are technology steps, where more
expensive transmission equipment is required for longer distance.
 An estimate of dense wavelength division multiplexing (DWDM) transmission
system costs per channel are derived from the SKA Design Studies costing
work in SKACost \citep{BolAle09}. Although there is some variation
with distance and the discrete cost steps of DWDM transmission, for
this first order costing it is sufficient to estimate an average cost
of \euro1\,k per 10 Gbps channel for all links, or \euro100 per
Gbps.

The total data rate out of a station beamformer is given by 
\begin{equation}
\Rst=\Nbstavg\Npol\NbitCFB\Delta\nu\times OS\times OH\times{\rm Nyq.,}
\end{equation}
where $\Nbstavg$ is the average number of station beams formed over
the band. The number of bits out of the coarse channel filter-bank
$\NbitCFB=4$, oversampling $OS=1.1$, digital encoding overhead $OH=1.25$
and the data is Nyquist sampled: ${\rm Nyq.}=2$. \nomenclature[sRst]{$\Rst$}{\TRst{}}\nomenclature[sNbitCFB]{$\NbitCFB$}{\TNbitCFB{}}

\subsection{Dual-band station costs\label{app:dual-band-costs}}

The dual-band implementation is costed as a separate a low-band (70--180\,MHz)
and high-band (180--450\,MHz) array. For each station block (sub-system),
its cost will either remain the same, or be some fraction of the single-band
cost, depending on how each block is modelled. This `cost multiplier'
is then applied to the costs in \prettyref{tab:blocks-cost}. \prettyref{tab:blocks-cost-dual}
shows the cost multipliers used for each block in the low and high-band
arrays. The cost multipliers chosen are reasonable approximations;
this is an area that requires expert attention to verify these numbers.

\begin{table}[!t]
\begin{centering}
\caption{Cost multiplier estimates for the dual-band array, where each multiplier
is a fraction of the single-band unit cost shown in \prettyref{tab:blocks-cost}.
\label{tab:blocks-cost-dual}}
\begin{threeparttable}%
\begin{tabular}{>{\raggedright}p{0.2\textwidth}>{\raggedright}p{0.15\textwidth}>{\raggedright}p{0.15\textwidth}>{\raggedright}p{0.4\textwidth}}
\toprule 
Block name & Low-band & High-band & Comments\tabularnewline
\midrule
Active antenna element\textsuperscript{a} & $\Cfix{}$: $0.83$ & $\Cfix{}$: $0.57$ & See \prettyref{app:active-antenna-element}.\tabularnewline
Optional: RF tile beamformer & $\Cvar{}$: 1 & $\Cvar{}$: 1 & \tabularnewline
Element/tile--digitiser RF link & $\Cfix{RF}$: 1

$\Cfix{Dig}$: 1 & $\Cfix{RF}$: 0.3

$\Cfix{Dig}$: 0.3 & Closer spacing means smaller tiles and stations.\tabularnewline
Digitiser\textsuperscript{a} & $\Cfix{}$: 0.4 & $\Cfix{}$: 1 & Sampling speed is set by $\fmax$ of each band (0.4\,GS/s for the
low-band). Assumes that cost is linearly proportional to sampling
speed.\tabularnewline
Digitiser--bunker link  & $\Cvar 1$: 1

$\Cvar 2$: 1 & $\Cvar 1$: 1

$\Cvar 2$: 1 & $\Cvar 1$: Per unit of data.

$\Cvar 2$: Per unit of data per unit length.\tabularnewline
Station beamformer\textsuperscript{a} & $\Cvar{}:$ 0.29 & $\Cvar{}:$ 0.71 & Assumes cost is linearly proportional to processed bandwidth.\tabularnewline
Station infrastructure\textsuperscript{a} & $\Cfix{}$: 1

$\Cvar 1$: 1

$\Cvar 2$: 0.29 & $\Cfix{}$: 1

$\Cvar 1$: 1

$\Cvar 2$: 0.71 & Processing infrastructure is the same fraction as the station beamformer.\tabularnewline
Station--CPF link transmission & $\Cvar{}$: 1 & $\Cvar{}$: 1 & Per unit of data.\tabularnewline
\bottomrule
\end{tabular}\begin{tablenotes}
\small
\item[a] These values are for a dual-band split frequency of 180\,MHz; they will vary for other split frequencies and overlapping bands.
\end{tablenotes}
\end{threeparttable}
\par\end{centering}

\end{table}

\section{Constant FoV as a function of frequency\label{app:constant-FoV}}

\subsection{Single-band implementation}

If a constant FoV as a function of frequency is a requirement, then
determining the average number of beams required over the receptor
bandwidth can simplify calculations \citep{AleBre09}. The number
of beams as a function of frequency can be given by
\begin{equation}
\Nbst(\nu)=\Nbst(\ftarget)\text{\ensuremath{\left(\frac{\nu}{\ftarget}\right)}}^{2},\label{eq:Nbst(freq)}
\end{equation}
where $\Nbst(\ftarget)$\nomenclature[sNbst]{$\Nbst$}{\TNbst} is
the number of dual-polarisation beams required at frequency $\ftarget$,
calculated as 
\begin{equation}
\Nbst(\nu_{0})=\Oreq/\Ost(\nu_{0}).\label{eq:Nbst-FoV}
\end{equation}
Integrating \prettyref{eq:Nbst(freq)} over the processed bandwidth
$\BW=\fmax-\fmin$ gives the number of beams of unit bandwidth:
\begin{align}
N_{{\rm b-st\_Hz}} & =\int_{\fmin}^{\fmax}\Nbst(\ftarget)\text{\ensuremath{\left(\frac{\nu}{\ftarget}\right)}}^{2}dv\nonumber \\
 & =\frac{\Nbst(\ftarget)}{\ftarget^{2}}\left[\frac{\nu^{3}}{3}\right]_{\fmin}^{\fmax}\nonumber \\
 & =\frac{\Nbst(\ftarget)(\fmax^{3}-\fmin^{3})}{3\ftarget^{2}}.
\end{align}
The average number of dual-polarisation beams over the band is 
\begin{equation}
\overline{\Nbst}=\frac{N_{{\rm b-st\_Hz}}}{\fmax-\fmin}.\label{eq:Nbst-avg}
\end{equation}

Also, given $\Ost\propto\Dst{}^{-2}$ and $\Nbst\propto\Ost^{-1}$,
substitution into \prettyref{eq:Nbst-avg} shows that 
\begin{equation}
\overline{\Nbst}\propto\Dst^{2}.\label{eq:Nbst-avg-prop-Dst}
\end{equation}

An actual implementation requires that a discrete number of beams
be formed at each frequency channel. This introduces some error, because
enough beams need to be formed for every frequency channel such that
the FoV requirement is always met; i.e. $\Nbst(\nu)\Ost(\nu)\ge\Oreq$.
An actual calculation requires a summation to replace the integral,
where the ceiling $\left\lceil \Nbst(\nu)\right\rceil $ is taken
for each frequency channel (strictly speaking, $\Nbst(\nu)$ should
to be calculated at the maximum frequency of each channel, not the
centre frequency). Although the error is larger at the lower frequencies
where $\Ost$ is larger, even at 70\,MHz the error is <10\,\% for
$\Oreq=\unit[20]{deg^{2}}$ (higher error for smaller FoV).

\subsection{Dual-band implementation\label{app:constant-FoV-dual-band}}

The dual-band implementation is more complex. \prettyref{eq:Nbst-avg}
can be applied separately to each band in the dual-band array, resulting
in the average number of beams in each of the low ($\mbox{\ensuremath{\NbstavgLow}}$)
and high ($\NbstavgHigh$) bands. The average number of beams over
the full band is given by 
\begin{equation}
\NbstavgDual=\frac{(\fsplit-\fmin)\NbstavgLow+(\fmax-\fsplit)\NbstavgHigh}{\fmax-\fmin},
\end{equation}
 where $\fsplit$ is the frequency split between the bands\nomenclature[s~nusplit]{$\fsplit$}{\Tfsplit}
and $\fmin$ and $\fmax$ are the minimum and maximum frequencies
of the dual-band implementation (i.e. 70 and 450\,MHz). From \prettyref{eq:Nbst-avg},
\begin{equation}
\NbstavgLow=\frac{\NbstLow(\ftarget)(\fsplit^{3}-\fmin^{3})}{3\ftarget^{2}(\fsplit-\fmin)},
\end{equation}
and an equivalent substitution can be made for $\NbstavgHigh$. 

The ratio of the average number of beams across the band between the
dual and single-band implementations, given by $\NbstavgRatio=\NbstavgDual/\NbstavgSingle$,
is a useful metric to compare the data rate from the station and also
the processing costs in the \cpf{}. Substitution gives
\begin{equation}
\NbstavgRatio=\frac{\NbstLow(\ftarget)(\fsplit^{3}-\fmin^{3})+\NbstHigh(\ftarget)(\fmax^{3}-\fsplit^{3})}{\NbstSingle(\ftarget)(\fmax^{3}-\fmin^{3})},\label{eq:Nbst-avg-ratio}
\end{equation}
or 
\begin{equation}
{\rm \NbstavgRatio=\frac{\DstLow^{2}(\fsplit^{3}-\fmin^{3})+\DstHigh^{2}(\fmax^{3}-\fsplit^{3})}{\DstSingle^{2}(\fmax^{3}-\fmin^{3})}},\label{eq:Nbst-avg-ratio-diameter}
\end{equation}
where the L, H and S sub-scripts indicate low, high and single-band
arrays respectively. Thus for a constant FoV across the processed
bandwidth $\BW$, the ratio depends on both the station diameter of
each band and the frequency split between the low and high-band arrays.\nomenclature[sNbstavgLow]{$\NbstavgLow$}{\TNbstavgLow}\nomenclature[sNbstavgHigh]{$\NbstavgHigh$}{\TNbstavgHigh}

\subsection{Trading \texorpdfstring{$\Nest$}{Ne/st} for \texorpdfstring{$\Nst$}{Nst}\label{app:constant-FoV-Nest-Nst}}

As shown in \prettyref{app:trade-Dst-Nst}, the station diameter is
both a function of the number of elements in the station and the average
inter-element spacing. To make a distinction between these effects,
Equations \ref{eq:Nbst-avg-prop-Dst} and \ref{eq:Nbst-avg-ratio-diameter}
requires the substitution of $\Dst\propto\sqrt{\Nest}\deeavg$ (\prettyref{eq:Dst-prop-Nest-deeavg}),
such that
\begin{equation}
\overline{\Nbst}\propto\Nest\deeavg^{2}\label{eq:Nbst-avg-prop-Nest-deeavg}
\end{equation}
 and
\begin{equation}
\NbstavgRatio=\frac{\NestLow\deeavgLow^{2}(\fsplit^{3}-\fmin^{3})+\NestHigh\deeavgHigh^{2}(\fmax^{3}-\fsplit^{3})}{\NestSingle\deeavgSingle^{2}(\fmax^{3}-\fmin^{3})}.\label{eq:Nbst-avg-ratio-Nest-deeavg}
\end{equation}
For the representative implementations considered in this analysis,
$\NestLow=\NestHigh=\NestSingle$, thus $\Nest$ is constant as a
function of frequency, resulting in 
\begin{equation}
\NbstavgRatio=\frac{\deeavgLow^{2}(\fsplit^{3}-\fmin^{3})+\deeavgHigh^{2}(\fmax^{3}-\fsplit^{3})}{\deeavgSingle^{2}(\fmax^{3}-\fmin^{3})}.\label{eq:Nbst-avg-ratio-deeavg}
\end{equation}
So if $\Dst$ is traded for $\Nst$ (for fixed $\Aearr$), this ratio
still holds as long as the same trade is made for all bands. For example,
if $\Dst$ is halved, then $\Nest$ decreases by a factor of 4. But
given $\NestLow=\NestHigh=\NestSingle$, $\NbstavgRatio$ does not
change.

\section{Parametric models and costs for other SKA sub-systems\label{app:models-SKA-other}}

The presentation of the models and costs in this section are, in general,
similar to \prettyref{app:models-SKAlow}. However, only zeroth-order
cost estimates are given, with the purpose of illustrating system-level
costs (\prettyref{sec:variable-cost-system-implications}). \prettyref{tab:blocks-summary-other-variable}
summarises the quantity and the cost scaling of the blocks which are
not \SKAiLow{} sub-systems, while the detailed block descriptions
provide further justification and references for the cost scaling.
\prettyref{tab:other-variable-costs} lists the fixed and variable
unit costs of each block.

\subsection{Site-related costs\label{app:models-SKA-site-related}}

Although site-related costs are not publicly available, a comparison
of single and dual-band implementations can be made by only considering
the costs that vary between implementations, while excluding costs
which remain fixed. For example, if an activity such as site preparation
is being undertaken, the fixed costs of that activity could include
contractor mobilisation and demobilisation, whereas the variable costs
are one or more pro rata (such as per hour or per m\textsuperscript{2})
costs of undertaking the activity.

Some site-related costs are independent of whether the implementation
is single or dual-band. The first-order array infrastructure costs
discussed in the HLSD, such as road networks and power and fibre
reticulation, will generally be independent of the number of stations
or their exact location and are not considered here. In the core region,
the spatial density of stations will be high enough that any infrastructure
work can be incorporated into the site preparation costs. For the
inner and mid region stations, located beyond the core, the stations
are placed in groups (clusters) on spiral arms \citep{BolMil11-Config}.
The array infrastructure requirements to connect cluster will be similar
between the single and dual-band implementations, regardless of the
number of stations at each cluster. For the central processing facility
buildings, there may be some cost scaling with the amount of processing.

\subsubsection{Antenna element deployment}

\citet{FauVaa11-Deployment} costs deployment of the antenna elements
at less than \euro50 per element. This excludes the deployment and
connection of the rest of the infrastructure, such as tile or station
processing nodes. To show the sensitivity of the single versus dual-band
comparison to changes in this cost, two deployment costs are considered:
\euro50 and \euro100 per element.

\subsubsection{Site preparation cost}

There is no published data on site preparation costs, and it is likely
to be highly dependent on what activities the site preparation involves.
For example, is it a simple land clearing activity, or are earthworks
and trenching required? To show the sensitivity of the single versus
dual-band comparison to changes in this cost, two areal site preparation
costs are considered: \euro$\unit[10]{m^{-2}}$ and \euro$\unit[100]{m^{-2}}$.

\begin{table}[!t]
\centering{}\caption{Summary of blocks and scaling for other sub-systems relevant to \SKAiLow{}.\label{tab:blocks-summary-other-variable}}
\begin{tabular}{>{\raggedright}p{0.25\textwidth}>{\raggedright}p{0.2\textwidth}ll}
\toprule 
Block name & Quantity in \SKAi{} & Parametric equation & Block coverage\tabularnewline
\midrule
Antenna element deployment & $\Nst\Nest$ & $\Cfix{}$ & deployment\tabularnewline
Site preparation  & $\Nst\Nest$ & $\deeavg^{2}$$\Cvar{}$ & site preparation\tabularnewline
Correlator & 1 & $\Nbstavg\Nst^{\alpha}\Cvar{};$ $1<\alpha<2$ & see \prettyref{app:fine-channelisation-correlation}\tabularnewline
Correlator--computing data transport & 1 & $\Rcorrout\Cvar{}$ & fibre transmission\tabularnewline
Imaging processor & 1 & $\Rcorrout\Cvar{}$ & \tabularnewline
\Nip{} & 1 & see \prettyref{app:NIP} & \tabularnewline
\bottomrule
\end{tabular}
\end{table}
\begin{table}[!t]
\centering{}\caption{Unit costs for other sub-systems relevant to \SKAiLow{}~(\euro2007).
\label{tab:other-variable-costs}}
\begin{tabular}{>{\raggedright}p{0.25\textwidth}>{\raggedright}p{0.25\textwidth}>{\raggedright}p{0.2\textwidth}>{\raggedright}p{0.15\textwidth}}
\toprule 
Block name & Cost estimate A  & Cost estimate B & Block unit\tabularnewline
\midrule
Antenna element deployment & $\Cfix{}$: 50 & $\Cfix{}$: 100 & per element\tabularnewline
Site preparation  & $\Cvar{}$: $\unit[10]{m^{-2}}$  & $\Cvar{}$: $\unit[100]{m^{-2}}$  & per element\tabularnewline
Correlator & $\Cvar{}$: 40\,k per input beam  & $\Cvar{}$: same as A & per correlator\tabularnewline
Correlator--computing data transport & $\Cvar{}$: $\unit[200]{Gbps^{-1}}$ & $\Cvar{}$: same as A & \tabularnewline
Imaging processor & 20\,M for single-band implementation & same as A & N/A\tabularnewline
\Nip{} & 30\,M for single-band implementation & same as A & N/A\tabularnewline
\bottomrule
\end{tabular}
\end{table}

\subsection{Central processing facility sub-systems\label{app:central-processing}}

\subsubsection{Correlator frequency resolution and integration time requirements\label{app:correlator-frequency-resolution-integration}}

Two parameters which are are relevant to the correlator and image
processor design are the correlator frequency resolution (channel
width) $\chBW$ and integration time $\Dt$; these need to be small
enough to respectively keep the radial and circumferential smearing
below some acceptable threshold \citep{ThoMor01}. The frequency
resolution and integration time required is inversely proportional
to antenna or station beamwidth \citep{TurFau11-HLSD}. If the maximum
baseline (distance between antenna pairs) and station beam taper $\Kst$
are constant, then 
\begin{equation}
\chBW\propto\Dst\label{eq:frequency-resolution-proportionality}
\end{equation}
\nomenclature[s~Deltatch]{$\chBW$}{\TchBW{}}\nomenclature[s~Deltat]{$\Dt$}{\TDt{}}and
\begin{equation}
\Dt\propto\Dst.\label{eq:integration-time-proportionality}
\end{equation}
However, \SKAiLow{} has a more stringent requirement on frequency
resolution which makes it independent of station size. The HLSD (Table
4) specifies a scientifically derived requirement of $\chBW=\unit[1]{kHz}$
for AAs, while \citet{TurFau11-HLSD} calculates $\chBW=\unit[590]{kHz}$
to meet the 2\% smearing requirement with a 180\,m diameter station.
Given the much more stringent specification in the HLSD, any change
in $\Dst$ for SKA-low will not affect $\chBW$.\label{ass:channel-width}

\subsubsection{Fine channelisation and correlation\label{app:fine-channelisation-correlation}}

The `FX' correlator cost scalings can be understood by analysing
the data streams flowing through the fine filterbank (channelisation)
and cross-correlation sub-systems, described in \prettyref{app:FX-architecture}.
The station beamformer outputs coarsely channelised station beams,
where each channel has width $\chBWcoarse$. A single data stream
from a station will contain one coarse channel from one station beam.
Each data stream is input into a fine filterbank (FFB) and split into
channels of smaller frequency resolution. The total processing cost
of the FFBs is thus 
\begin{equation}
P_{{\rm FFB}}=\NchCoarse\Nbstavg\Nst K_{{\rm FFB}},\label{eq:FFB-processing}
\end{equation}
where $\NchCoarse$ is the number of channels per coarse filterbank
($\NchCoarse=\BW/\chBWcoarse$) and $K_{{\rm FFB}}$ is the FFB processing
cost for a single data stream. 

The value of $K_{{\rm FFB}}$ is not easily determinable, because
the filterbank processing architecture is designed so that, for each
filterbank, the data flow, mathematical operations and memory usage
are optimised for some output frequency resolution \citep[e.g.][]{Bun10,BarMil11}.
However, $K_{{\rm FFB}}$ is constant in the present analysis, because
the FFB output frequency resolution (equal to $\chBW$) is fixed by
the scientific requirements described above. Additionally, $\chBWcoarse$
is held fixed in the present analysis; \citet{AAV11-Concept} specifies
$\chBWcoarse=\unit[0.25]{MHz}$. Therefore a re-evaluation of \prettyref{eq:FFB-processing}
gives
\begin{equation}
P_{{\rm FFB}}\propto\BW\Nbstavg\Nst.
\end{equation}

Once the data streams are split into fine channels, equivalent data
streams (i.e. m\textsuperscript{th} fine channel of the n\textsuperscript{th}
beam) from each antenna pair ($\sim\Nst^{2}/2$ pairs) are cross-correlated,
at a correlation rate equal to the sample rate $\chBW$ \citep{Bun00}.
Because there are $\Nch=\BW/\chBW$\nomenclature[sNch]{$\Nch$}{\TNch{}}
fine channels to be correlated, and $\chBW$ is constant, the correlation
processing cost is approximately
\begin{equation}
P_{{\rm X}}\propto\BW\Nbstavg\Nst^{2}.\label{eq:X-processing}
\end{equation}

Additional to the `F' and `X' computation hardware, there are
also processing costs in the form of memory buffers and the inter-connects
(corner turn) between the filterbanks and correlation devices \citep{TurFau11-HLSD}.
These costs are design dependent, and one or more of these costs may
dominate the total correlator processing hardware cost. But it is
reasonable to expect that the total cost will scale as
\begin{equation}
\Pcorr\propto\BW\Nbstavg\Nst^{\alpha},\label{eq:correlator-processing-Nst}
\end{equation}
where $1<\alpha<2$ depending on the design and technologies used.
If $\Nst$ is held constant, 
\begin{equation}
\Pcorr\propto\BW\Nbstavg.\label{eq:correlator-processing-simple}
\end{equation}

\subsubsection*{Dual-band implementation}

The correlator processing  for the dual-band implementation simply
requires the substitution of $\NbstavgDual$ (\prettyref{app:constant-FoV-dual-band})
into \prettyref{eq:correlator-processing-simple}. The correlator
processing ratio between the dual and single-band implementations
is therefore 
\begin{equation}
\Pcorr{\rm (dual:single)=\NbstavgRatio}.\label{eq:correlator-ratio}
\end{equation}

\subsubsection*{Trading $\Nest$ for $\Nst$}

For constant $\Aearr$, Equations \ref{eq:Nbst-avg-prop-Nest-deeavg}
and \ref{eq:Nst-Nest-FF-trade} can be substituted into \prettyref{eq:correlator-processing-Nst}.
Assuming $\deeavg$ and the antenna element gain $\GGe$ remain constant,
\begin{equation}
\Pcorr\propto\BW\Nest^{1-\alpha},\label{eq:correlator-processing-Nest}
\end{equation}
where $1<\alpha<2$. 

Because $\Nst$ changes equally for both the single and dual-band
implementations, \prettyref{eq:correlator-ratio} still holds true.
Although the cost of the correlator processing increases with $\Nst$,
the relative cost is independent of the $\Nst$ vs. $\Nest$ trade-off.

\subsubsection*{Fine channelisation and correlation costs}

There are a range of correlator cost estimates (<\euro1\,M to nearly
\euro100\,M) for \SKAi{} in \citet{Tur11-Cost}, representing different
architectures, technologies and options for flexibility. The correlation
of \SKAiLow{} stations, rather than the dishes, represent most of
the cost. The cost estimates generally only encompass the processing
units and data inter-connects; doubling the cost accounts for all
the accessory hardware required to support these processing units.
Costs such as \nre{}, which are generally not included in the estimates,
can be considered a fixed cost.  A zeroth-order estimate of the cost
of the parts of the \SKAiLow{} correlator which scale as the number
of input beams is \euro20\,M; this equates to a GPU-class correlator.
This estimate is specified for 480 station beams ($\BW=\unit[380]{MHz}$),
which is approximately \euro40\,k per station beam.

\subsubsection{Correlator--computing data transport\label{app:correlator-computing-data-rate}}

Each datum produced from the correlator (\prettyref{eq:X-processing})
is integrated for time $\Dt$. Thus the data rate out of the correlator
is:
\begin{equation}
\Rcorrout\propto\frac{\Nch\Nbstavg\Nst^{2}}{\Dt}.\label{eq:correlator-data-out}
\end{equation}
For a fixed processed FoV $\Ostproc$\nomenclature[s~Oproc]{$\Ostproc$}{\TOstproc{}},
the data rate out of the correlator $\Rcorrout$\nomenclature[sRcorrout]{$\Rcorrout$}{\TRcorrout{}}
is independent of station diameter when the frequency resolution and
integration time are set by the maximum smearing requirements (Equations
\ref{eq:frequency-resolution-proportionality} and \ref{eq:integration-time-proportionality}).
In that case,
\begin{align}
\Rcorrout & \propto\frac{\BW\Nbstavg\Nst^{2}}{\Dst^{2}}\\
 & \propto\BW\Ostproc\Nst^{2},
\end{align}
given $\Ostproc\propto\Nbstavg\Ost\propto\Nbstavg\Dst^{-2}$. This
result is independent of station diameter. 

However, the maximum smearing requirements do not set the frequency
resolution for \SKAiLow{}. The more stringent requirement on frequency
resolution discussed earlier means that $\chBW$, hence $\Nch$, does
not vary with station diameter. Thus a re-evaluation of \prettyref{eq:correlator-data-out}
gives
\begin{eqnarray}
\Rcorrout & \propto & \frac{\BW\Nbstavg\Nst^{2}}{\Dst}.\label{eq:correlator-data-out-specific-single}
\end{eqnarray}
This is consistent with the more detailed analysis in \citet{McC11-STaN-HLD}.

\subsubsection*{Dual-band implementation}

For the dual-band implementation, $\Rcorrout$ is calculated separately
for the low and high band station diameters and summed to achieve
a data rate for the full bandwidth. Thus
\begin{equation}
\Rcorrout\propto\text{\ensuremath{\left(\frac{(\fsplit-\fmin)\NbstavgLow\NstLow^{2}}{\DstLow}+\frac{(\fmax-\fsplit)\NbstavgHigh\NstHigh^{2}}{\DstHigh}\right)}}.\label{eq:correlator-data-out-specific-dual}
\end{equation}

Given $\NstLow=\NstHigh=\NstSingle$ and $\Nbstavg\propto\Dst^{2}$
(\prettyref{eq:Nbst-avg-prop-Dst}), the ratio $\RcorroutRatio$ can
be simplified to
\begin{equation}
\RcorroutRatio=\frac{\DstLow(\fsplit^{3}-\fmin^{3})+\DstHigh(\fmax^{3}-\fsplit^{3})}{\DstSingle(\fmax^{3}-\fmin^{3})}.\label{eq:correlator-data-out-specific-ratio}
\end{equation}

\subsubsection*{Trading $\Nest$ for $\Nst$}

For constant $\Aearr$, Equations \ref{eq:Nbst-avg-prop-Nest-deeavg},
\ref{eq:Nst-Nest-FF-trade} and \ref{eq:Dst-prop-Nest-deeavg} can
be substituted into \prettyref{eq:correlator-data-out-specific-single}.
Assuming $\deeavg$ and the antenna element gain $\GGe$ remain constant,
\begin{equation}
\Rcorrout\propto\frac{\BW}{\Nest^{3/2}}.\label{eq:correlator-data-out-Nest}
\end{equation}

To account for the trade between $\Nest$ and $\Nst$ for fixed $\Aearr$,
as well as the different inter-element spacing between the low and
high-band arrays,  $\Dst\propto\sqrt{\Nest}\deeavg$ (\prettyref{eq:Dst-prop-Nest-deeavg})
can be substituted into \prettyref{eq:correlator-data-out-specific-ratio}.
Because $\NestSingle=\NestLow=\NestHigh$,
\begin{equation}
\RcorroutRatio=\frac{\deeavgLow(\fsplit^{3}-\fmin^{3})+\deeavgHigh(\fmax^{3}-\fsplit^{3})}{\deeavgSingle(\fmax^{3}-\fmin^{3})}.
\end{equation}

Although the correlator output data rate increases with $\Nst$ (Equations
\ref{eq:correlator-data-out-specific-single} and \ref{eq:correlator-data-out-specific-dual}),
the relative data rate between the single and dual-band implementations
is independent of the $\Nst$ vs. $\Nest$ trade.

\subsubsection*{Correlator--computing data transport cost}

For \SKAiLow{}, the HLSD lists an average data rate of $\unit[332\e 9]{byte\, s^{-1}}$
(2.66\,Tbps) from the correlator to the computer; a factor of 1.25
encoding overhead brings this to 3.32\,Tbps. Taking an approach similar
to the station-CPF link (\prettyref{app:station-CPF-link-transmission}),
only the correlator--computing data transmission is costed; the trenching
and cabling cost is ignored. Assuming that the computing is off-site
at a nearby city or other suitable location, a cost-effective option
is to multiplex the signals onto fibre using DWDM technology \citep[e.g. Figure 26 of][]{BolFau09}. 

The DWDM transmission can be split into a data transmit--receive cost
and a signal amplification cost. \citet{McC10} costs the transmitter--receiver
pair at \euro2\,k per 10\,Gbps channel and the optical amplifier
and dispersion compensator at \euro10\,k per 16-channel unit. The
optical signal amplification is required every 80\,km. If three amplification
units are required, this results in a conservative estimate of \euro4\,k
per 10\,Gbps channel, or \euro400 per Gbps for the data transmission.
This is higher than the average cost of the station--CPF transmission,
because of the greater distances. The correlator--computing transmission
cost for the HLSD is then approximately \euro1.3\,M.

\subsubsection{Imaging \label{app:imaging}}

The SKA post-correlator processing requirements and algorithms, and
their effect on computational cost, is an area of active research
\citep[e.g.][]{Ale11}. The `imaging' sub-system encompasses the
processing of visibilities from the correlator into imaged data products,
as outlined in \citet{AleBre09}. Considered here are the `gridding'
operations on the visibility data which are the main computational
cost \citep{Cor04,AleBre09}. Many of the imaging operations act
on the data from the correlator, thus buffering of these data is required
\citep{FauAle10}. Assuming that the imaging cost is dominated by
the data buffer, rather than the operations cost of the imaging algorithms,
then the data rate out of the correlator (\prettyref{app:correlator-computing-data-rate})
can be used as an indicator of cost \citep{AleBre09}.\label{ass:imaging-cost}

If the processing cost is considered, then there is a contribution
from the data volume, but also a cost for correcting for non-coplanar
baselines in wide-field images. For continuum imaging, which requires
high dynamic range in the presence of confusing sources, \citet{PerCla03}
and \citet{Cor04} establish scaling relationships for an array of
\spfeed{} dishes. \citet{Cor05} extends this analysis to multi-beam
systems such as aperture arrays and dishes equipped with \paf{}s.
The processing cost and its scaling as a function of dish or station
diameter varies: the data rate out of the correlator is one factor;
the other is the cost per visibility to correct for non-coplanar baselines.
That cost depends on whether the correction can be done by separately
imaging each independently pointed station beam, imaging the entire
processed FoV at once, or by using some other algorithm. Modelling
the cost of this processing and its scaling relationship with station
diameter is beyond the scope of this work.

\subsubsection*{Imaging processor cost}

The hardware implementation for SKA computing is currently ill-defined,
but a simple estimate can be derived from the SKA budget. \citet{GarCor10}
budget \euro{}350\,M for \SKAiLow{} capital investment, which includes
a ``significant element of contingency''. We use \euro20\,M as
a zeroth-order estimate of the processing hardware and related infrastructure
for the imaging aspect of the computing. Software is the other major
computing cost, but estimating any cost difference between single
and dual-band is beyond the scope of this work.

\subsubsection{Non-imaging processing\label{app:NIP}\label{ass:NIP}}

The main sub-systems of the \niping{} are a central beamformer, and
pulsar searching and timing on the beams formed, as described in \citet{TurFau11-HLSD}.
Each beam is a phased or `tied' array beam, formed using some or
all the stations in the array as inputs to the beamformer. (In contrast,
station beams are formed from the antenna elements or tiles in the
station). 

The pulsar survey costs are assumed to be dominant, compared to the
pulsar timing costs \citep{Tur11-Cost}. Also, although the central
beamformer may be a combined with the correlator sub-system \citep[e.g.][]{TurFau11-HLSD},
the processing for the pulsar survey is likely to be a significantly
larger cost. The search for pulsars is conducted on a `per beam'
basis and can be computationally expensive, especially if searching
for binary pulsars using acceleration searches. For example, \citet{SmiKra09}
calculate the number of computational operations required for the
acceleration search with SKA Phase~2, for a fiducial set of search
parameters; the number of operations for the acceleration search are
two orders of magnitude greater than for the array beamforming. However,
determining the algorithm for optimal processing loads and data rates
for the SKA requires further investigation \citep{TurFau11-HLSD}.

Regardless of the algorithm, the general processing trend can be analysed
because the search is conducted on each beam. The number of array
beams $\Nbarr$ required to survey the sky depends on the FoV of each
beam. That in turn depends on the frequency of observation, and the
diameter of the array from which the beams are formed: $\Nbarr\propto(\nu\Darr)^{2}$.\nomenclature[sNbarr]{$\Nbarr$}{\TNbarr}\nomenclature[sDarr]{$\Darr$}{\TDarr{}}
Although using more stations increases sensitivity, it also increases
$\Darr$. To limit the computational requirements, \citet{SmiSta11-SKA1}
suggest that only stations in the 1\,km diameter core be used to
form the processed FoV. 

The single-band implementation (\prettyref{fig:single-station}) has
a 1\,km diameter core. However, for the high band of our representative
dual-band implementation, the smaller inter-element spacing means
that the core is only 0.5\,km in diameter (\prettyref{fig:dual-station}).
This results in the required FoV being met with factor of four fewer
core array beams. The cost of beamforming is approximately linearly
proportional to the number of beams (\prettyref{app:beamforming-computational-cost})
and the subsequent processing to search for pulsars is conducted on
the per beam basis, hence the processing cost for the dual-band implementation
is 25\% of the single-band implementation.

\label{ass:NIP-AA-dishes}To realise the factor of 4 cost reduction,
the 0.5\,km diameter high-band core must be physically separate to
the low-band core, as shown in \prettyref{fig:dual-station}. Also,
the processing capacity (hence cost) of the \nip{} is assumed to
be specified by the AA pulsar survey, rather than the dish pulsar
survey. But if the requirements for the dish pulsar survey set the
processing capacity, a factor of 4 increase in AA pulsar survey performance
is still achievable, despite no cost reduction being realised. This
is because the smaller diameter of the high-band core results in the
FoV of each core array beam increasing by a factor of 4, meaning fewer
beams are searched to survey the same sized area of sky.

\citet{TurFau11-HLSD} assume that the 35 \SKAiLow{} stations (180\,m
diameter) in the central 5\,km of the array are used for pulsar searches.
In that case, more compact stations will not necessarily decrease
this 5\,km diameter, hence the dual-band array provides no extra
benefit. However, using all the stations within the 5\,km is not
cost-effective. As Table 3 of \citet{ColCla11} shows, outside the
densely packed core of the \SKAiLow{} array, the cost-effectiveness
of high time resolution searches is significantly reduced; the extra
sensitivity gained from including more stations is insufficient to
offset the many more array beams which must be formed to meet the
required FoV.

Another aspect to non-imaging processing is searches for `fast transients',
which are highly energetic, single-pulse events. Like pulsars, these
searches are generally conducted on a per beam basis, where for \SKAi{},
the beams would be either core array beams or station beams, depending
on the search strategy \citep[see][and references therein]{ColCla11}.
Thus the fast transients processing cost for the high-band array is
25\% of the single-band implementation. If searches are done with
the low-band array, the cost is equal to the single-band implementation.

\subsubsection*{\Nip{} cost}

\citet{Tur11-Cost} provides a cost of \euro28\,m for a pulsar
search concept description by \citet{KniHor11-Pulsar}. This cost
is for processing hardware including server cases. Adding in racks
and power distribution (which scale approximately linearly with processing),
we round the cost up to \euro30\,m. The concept description does
not specify whether processing hardware searches the $\unit[1.25]{deg^{2}}$
FoV or some subset of that. However, given the search is done on
a per beam basis, this is not important because the relative cost
applies regardless.

\subsection{Power demand\label{app:power-demand}}

The power demand estimate is sourced from the bottom-up SKADS power
budget in \citet{FauVaa11-Deployment}. The power budget is for the
all-digital station architecture, but extrapolation to the RF tile
beamforming architecture is done by including a power cost for the
RF beamformer. The scaling relationships for power demand are shown
in \prettyref{tab:blocks-power-scaling} and the unit costs in \prettyref{tab:blocks-power-cost}.

\begin{table}[!t]
\caption{Summary of blocks and scaling for power demand in \SKAiLow{} station
sub-systems (RF first-stage beamforming is optional).\label{tab:blocks-power-scaling}}
\begin{threeparttable}%
\begin{tabular}{>{\raggedright}p{0.2\textwidth}>{\raggedright}p{0.22\textwidth}>{\raggedright}p{0.21\textwidth}>{\raggedright}p{0.27\textwidth}}
\toprule 
Block name\textsuperscript{} & Quantity in \SKAi{} & Power scaling & Block coverage\tabularnewline
\midrule
Active antenna element & $\Nst\Nest$  & $\Cfix{}$ & LNA and antenna gain\tabularnewline
Analogue (RF) tile beamformer\textsuperscript{a}  & $\Nst\Ntst$ & $\Nbt\Cvar{}$ & beamformer\tabularnewline
Digitiser & RF: $\Nst\Ntst\Nbt$

Dig: $\Nst\Nest$ & $\Cfix{}$+$\Rsample\Cvar{}$ & $\Cfix{}$: analogue signal conditioning, clock distribution

$\Cvar{}$: analogue to digital converter\tabularnewline
Digitiser--bunker link\textsuperscript{b} & Dig: $\Nst\Nest$ & $\Cvar{}\Rdig$ & copper communication from digitiser and fibre transmission electronics\tabularnewline
Station beamformer\textsuperscript{}\textsuperscript{c} & $\Nst$ & RF: $\Nbstavg\Ntst\Cvar{}$

Dig: $\Nbstavg\Nest\Cvar{}$ & digital processing, inter-connections and control\tabularnewline
Station--CPF link transmission & $\Nst$ & $\Rst\Cvar{}$ & fibre transmission\tabularnewline
\bottomrule
\end{tabular}

\centering{}\begin{tablenotes}
\small
\item[a] Optional block. If analogue tile beamforming is included, subsequent quantities and costs are denoted `RF'. If not, the system is all-digital beamforming, denoted by `Dig'.
\item[b] Optional block. Assumes no digital beamforming at the tile. Alternative architectures are discussed in \prettyref{sub:signal-architectures}. 
\item[c] Approximate cost scaling, see \prettyref{app:beamforming-computational-cost}.
\end{tablenotes}
\end{threeparttable}
\end{table}

\begin{table}[!t]
\caption{Unit costs for the power demand of \SKAiLow{} station sub-systems.
\label{tab:blocks-power-cost}}
\begin{threeparttable}%
\begin{tabular}{>{\raggedright}p{0.35\textwidth}>{\raggedright}p{0.4\textwidth}>{\raggedright}p{0.12\textwidth}}
\toprule 
Block name & Power demand estimate (mW) & Block unit\textsuperscript{a}\tabularnewline
\midrule
Active antenna element & $\Cfix{}$: 180 & per element\tabularnewline
RF tile beamformer & $\Cvar{}$: 400 per output beam  & per tile\tabularnewline
Digitiser & $\Cfix{}$: 180

$\Cvar{}$: 200 per GS/s & per signal\tabularnewline
Digitiser--bunker link transmission & $\Cvar{}$: 12.4 per Gbps & per link\tabularnewline
Station beamformer\textsuperscript{b} & $\Cvar{}:$ 18.6 per input per output beam & per station\tabularnewline
Station--CPF link transmission & $\Cvar{}$: Not available & per link\tabularnewline
\bottomrule
\end{tabular}

\centering{}\begin{tablenotes}
\small
\item[a] All elements, beams, inputs and signals are dual polarisation.
\item[b] Dual-band station beamformer power demand is 29\% and 71\% of this value, for the low and high-bands respectively.
\end{tablenotes}
\end{threeparttable}
\end{table}

\section{Station performance considerations\label{app:station-performance}}

\subsection{Sensitivity requirements and inter-element spacing\label{app:sensitivity-requirements}}

Telescope sensitivity is a key requirement on the system. The required
sensitivity is usually derived from the required minimum detectable
flux density and the telescope time available for each pointing (patch
of sky being observed). An exception is some time-domain astronomy
where further integration of a single pointing does not increase sensitivity.
Sensitivity is generally given as the metric $\AonT=\Aearr/\Tsys$,
where $\Aearr$ is the effective area of the telescope array and $\Tsys$
is the system temperature. For the electronically steered aperture
arrays, significant variations in $\AonT$ are caused by:\nomenclature[sA~T]{$\AonT$}{\TAonT{}}\nomenclature[sTsys]{$\Tsys$}{\TTsys{}}
\begin{itemize}
\item inter-element spacing
\item scan angle
\item strong sources in the sidelobes.
\end{itemize}
Additionally, the effect of these parameters on sensitivity is frequency
dependent. The effect of inter-element spacing is the most relevant
to this analysis, and the basic trends are considered here. While
the other parameters are also important, a complete analysis is beyond
the scope of this document. 

The inter-element spacing defines the frequency at which the antenna
elements transition from `dense' to `sparse'. There is no single
definition for when an array is dense or sparse. The broad definition
used in this work is that an array is dense when the aperture is fully
sampled, such that effective area is approximately constant with frequency:
$\Aearr(\nu)={\rm constant}$. When the array is sparse, the effective
area of each isolated element contributes to the array effective area,
hence  $\Aearr\propto\lambda^{2}$. There is also a transition region
between dense and sparse, which occurs at an inter-element spacing
of $0.5-1.5\lambda$ for dipole-like antennas \citep{BraCap06}, or
at a spacing typically greater than $2\lambda$ for more directive
antennas \citep{Rog08-DAM70}. 

\begin{figure}[!t]
\noindent \begin{centering}
\includegraphics[width=0.76\textwidth]{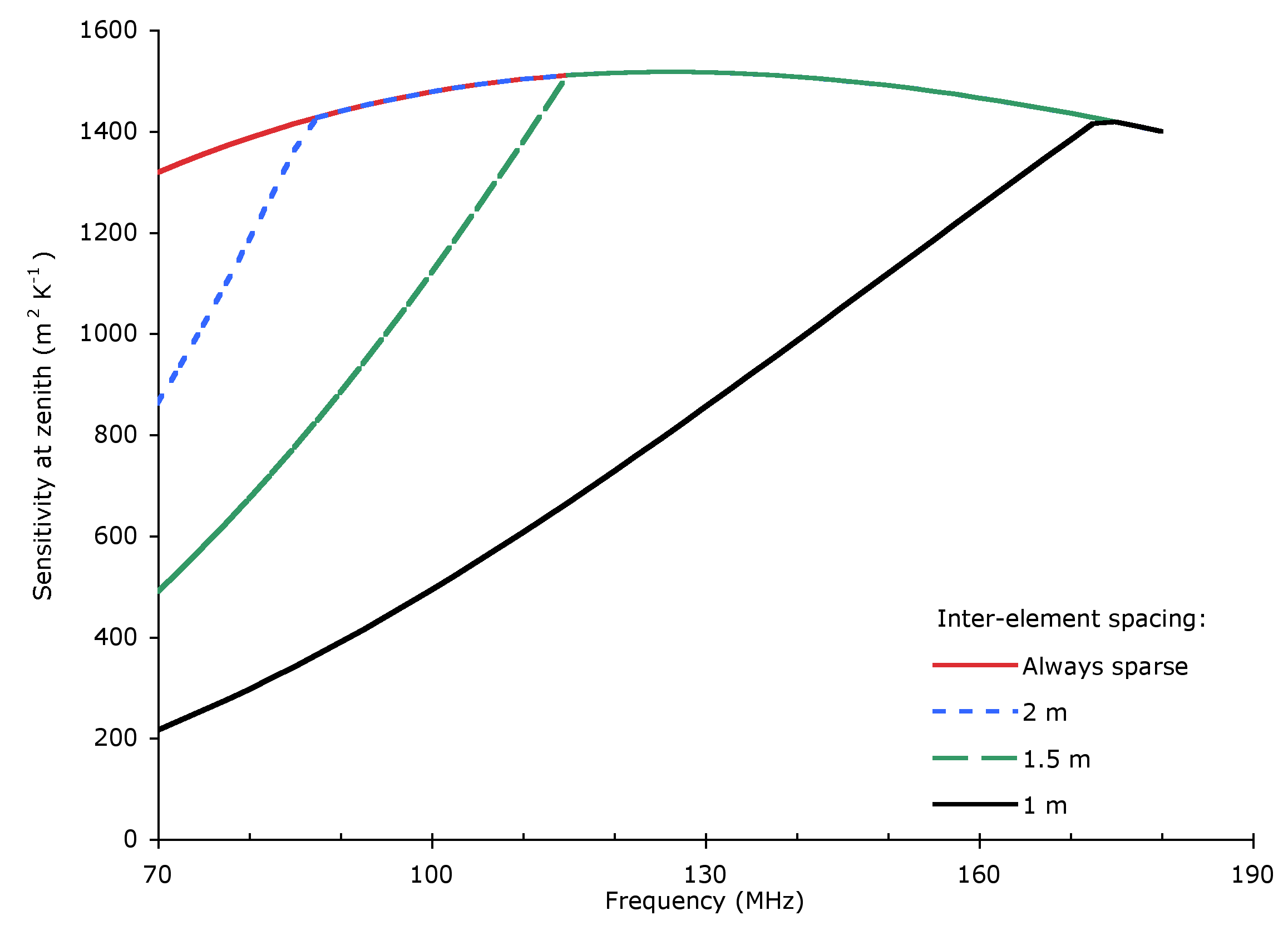}
\par\end{centering}

\caption{Approximate \SKAiLow{} sensitivity ($\AonT$) at zenith as a function
of frequency (70--180\,MHz) and inter-element spacing, using Equations
\ref{eq:Aearr} and \ref{eq:Tsys}.\label{fig:sensitivity-simple}}
\end{figure}

To give an indication of how the inter-element spacing affects the
representative systems, \prettyref{fig:sensitivity-simple} plots
$\AonT$ as a function of frequency, using the first-order analysis
of the problem in \citet{NijPan09} and the HLSD. The array effective
area is given by
\begin{equation}
\Aearr=\begin{cases}
\Nst\frac{\pi}{4}\Dst^{2} & \nu<\nu_{{\rm transition}}\,{\rm (dense)}\\
\Nst\Nest\frac{\lambda^{2}}{3} & \nu>\nu_{{\rm transition}}\,{\rm (sparse)},
\end{cases}\label{eq:Aearr}
\end{equation}
 where $\ftrans$ is the frequency of the dense--sparse transition.
The system temperature is approximated by the sum of the receiver
and sky noise temperatures:\nomenclature[s~nutrans]{$\ftrans$}{\Tftrans{}}
\begin{equation}
\Tsys=150+60\lambda^{2.55}.\label{eq:Tsys}
\end{equation}
The `always sparse' curve in \prettyref{fig:sensitivity-simple}
reflects the isolated antenna element case, where $\nu>\ftrans$ is
always true. The frequency at which this curve peaks depends somewhat
on the receiver and sky noise models. At the higher frequencies, the
arrays are sparse for all four spacing values and $\AonT$ is independent
of the inter-element spacing. At the lower frequencies, the inter-element
spacing defines the discontinuity. This is the transition frequency;
the point at which the aperture becomes fully sampled and $\AonT$
begins to drop below the always sparse curve.

\begin{figure}[!t]
\noindent \begin{centering}
\includegraphics[width=0.76\textwidth]{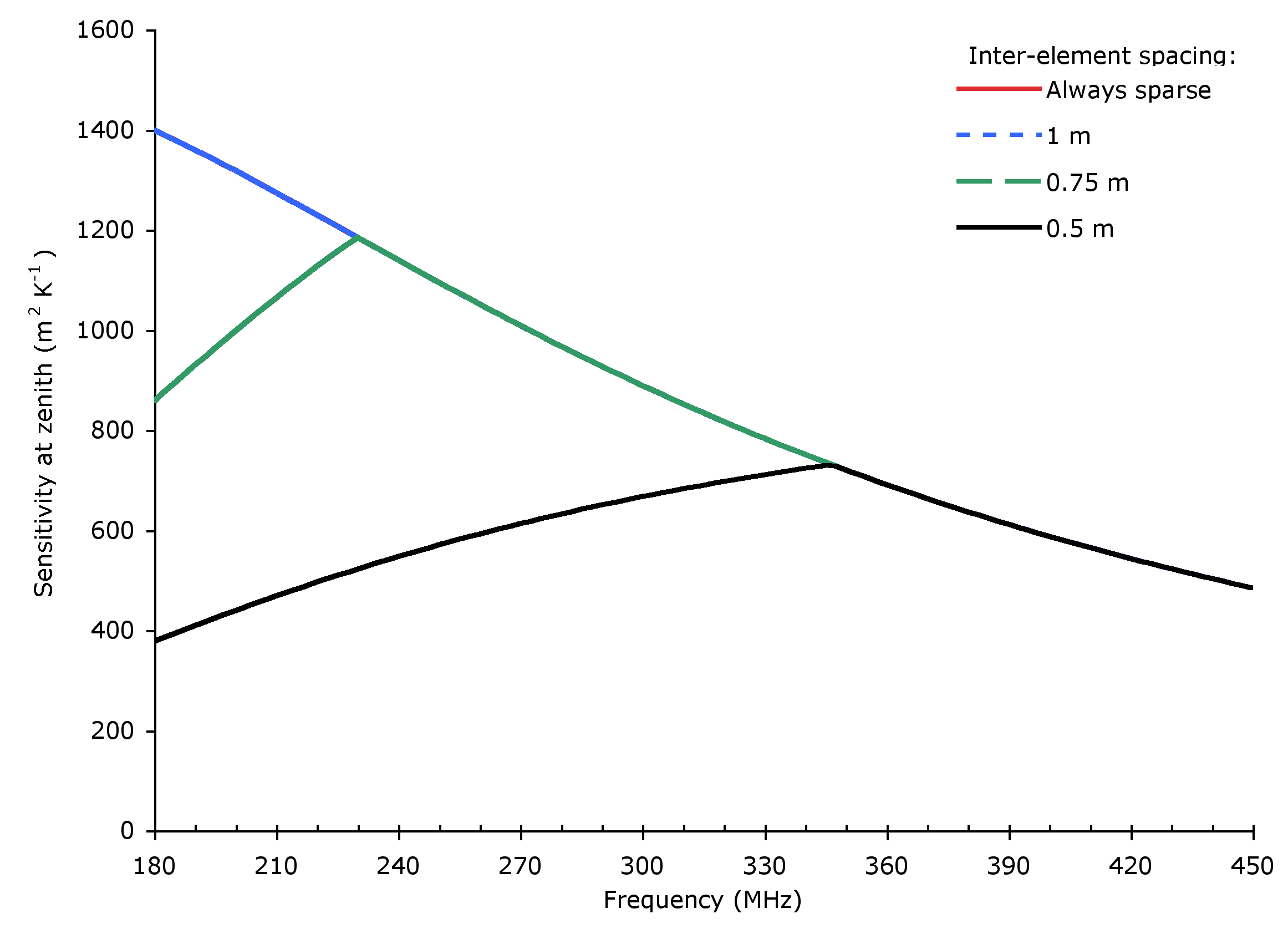}
\par\end{centering}

\caption{Approximate \SKAiLow{} sensitivity ($\AonT$) at zenith as a function
of frequency (180--450\,MHz) and inter-element spacing, using Equations
\ref{eq:Aearr} and \ref{eq:Tsys}.\label{fig:sensitivity-simple-highband}}
\end{figure}

\prettyref{fig:sensitivity-simple} applies to the single-band implementation
and the low (70--180\,MHz) band of the dual-band implementation.
The sensitivity for the high (180--450\,MHz) band is plotted in \prettyref{fig:sensitivity-simple-highband}.
In this figure, the discontinuities exist for the 0.5\,m and 0.75\,m
inter-element spacing. The 1\,m spacing transition frequency is lower
than 180\,MHz, so the $\AonT$ curve is the same as the always sparse
curve. The representative single-band $\AonT$ is also equal to the
always sparse curve, due to its 1.5\,m inter-element spacing. Although
the canonical dual-band array is designed for comparable performance
to the single-band array, the 0.75\,m spacing means the dual-band
array has less sensitivity between 180 and 230\,MHz.

In reality, the layout of the tile elements within a station will
affect the form of the discontinuity and the slope of the curve at
frequencies below this discontinuity. Intra-station layouts such as
golden ratio spiral (GRS)%
\footnote{GRS is a form of spatial taper, with the density of elements reducing
with increased radius from the centre.%
}, fractal patterns and irregular arrays will have a distribution of
inter-element spacing, i.e. a minimum ($\deemin$), maximum ($\deemax$)
and average ($\deeavg$ ) spacing. The station will be dense once
the frequency is low enough that those elements with spacing of $\deemax$
become dense. The station will be sparse once the frequency is high
enough that those elements with spacing of $\deemin$ become sparse.
Between these frequencies, this distribution of inter-element spacing
causes the station to be `semi-sparse', and neither case in \prettyref{eq:Aearr}
is applicable. Such layouts can broaden the discontinuity at the dense-sparse
transition and reduce the slope of the $\AonT$ curve, as can be seen
in Figures 22 to 27 of \citet{AAV11-Concept}.\nomenclature[sd_eemin]{$\deemin$}{\Tdeemin{}}\nomenclature[sd_eemax]{$\deemax$}{\Tdeemax{}}

Although this simple analysis gives some indication of how $\AonT$
changes with inter-element spacing, there is scope for more detailed
investigation. The station effective area changes as a function of
zenith angle $\theta$ and azimuth angle $\phi$, due to a changing
beam pattern. The beam pattern itself, pointed at a particular direction
$\tp$, is dependent on the intra-station layout pattern (this also
defines the station beam FoV). For example, \citet{CapWij06} compares
regular and irregular layouts of uniform aperture distribution for
sparse AA stations. Strong astronomical sources in the sidelobes of
these station beams will also greatly influence $\Tsys$ (hence $\AonT$),
as shown in \citet{WijCap11}. Also, spatial tapering would increase
the $\AonT$ at lower frequencies. But the extent of this increase,
and the related frequency-dependent effect on beam pattern (hence
$\Ost$) requires further investigation. 

The station effective area calculations also depend on the gain or
directivity of the antenna element. For example, log-periodic, conical
spiral and Vivaldi elements are discussed in \citet{AAV11-Concept}.
These have higher directivity (at $\theta=0$) than the proposed element
in the HLSD. However, the directivity as a function of scan angle
depends on the antenna design. Thus the station directivity (hence
$\Aest$) must be considered down to the maximum scan angle (zenith
angle) $\tmax$, which \citet{AAV11-Concept} specifies as $\tmax=45\textdegree$.
\prettyref{app:station-performance-further-work} details further
work to refine station sensitivity estimates. \nomenclature[s~fetamax]{$\tmax$}{\Ttmax{}}


\subsection{Filling factor and station calibration\label{app:filling-factor-calibration}}

The ability to calibrate the \SKAiLow{} telescope also has system-wide
implications. To achieve the desired performance, the instrumental
response of the telescope needs to be accurately characterised via
calibration \citep[e.g.][]{WijTol10} and calibrating the station
beams is one aspect of this. \citet{WijNij11} discusses station calibration
and determines that 3--5 calibration sources need to be detectable
in the station beam, assuming that the station size meets the requirements
outlined in \citet{WijBre11}. A metric which is related to the number
of detectable sources is the station filling factor $\FFst$, given
by
\begin{equation}
\FFst=\frac{\Aest}{\Agst},\label{eq:FFst-AonA}
\end{equation}
where $\Aest$ is the station effective area, and $\Agst$ is the
geometrical (physical) area occupied by the station. For a station
with a uniform taper (aperture distribution), a fractional bandwidth
of 20\,\% and an antenna and receiver noise of 50\,K, \citet{WijNij11}
require a filling factor of 0.2--0.4 at 400\,MHz.\nomenclature[sFFst]{$\FFst$}{\TFFst{}}\nomenclature[sAgst]{$\Agst$}{\TAgst{}}\nomenclature[sAest]{$\Aest$}{\TAest{}}

The competing effects of array sparseness to maintain sensitivity
at low frequencies and array density for station calibration at high
frequencies is problematic. \citet{Wij11-468} propose that a dual-band
implementation should be used, unless a solution can be found by either
optimising the intra-station layout of the single band array to meet
both requirements, or using multiple beams at the higher frequencies
to detect the required calibration sources. The dual-band array has
more flexibility to adjust the high-band design to meet the filling
factor requirement.

This problem is reflected in the single-band implementation considered
in this analysis, where $\FFst=0.08$ at 400\,MHz for a zenith pointing.
For the representative dual-band implementation, $\FFst=0.33$ at
the same frequency and pointing. Although in this case, the filling
factor is clearly too low for the single-band implementation, the
required and calculated filling factors will vary, depending on the
intra-station layout and element gain of the single-band design. Determining
whether single-band solution can meet the filling factor requirements
is a current work in progress within the SKA aperture array community.

\subsection{Trading  \texorpdfstring{$\Nest$}{Ne/st} for \texorpdfstring{$\Nst$}{Nst} and the relationship with station diameter\label{app:trade-Dst-Nst}}

For constant array effective area $\Aearr$, there is a trade-off
between \aa{} station diameter $\Dst$ and the number of stations
$\Nst$, where $\Nst\propto1/\Dst^{2}$. This is analogous to previous
investigations trading dish diameter for the number of dishes \citep[e.g.][]{ChiCol07,SchAle07}.
However, this does not completely describe the independent variables
in the trade-off. A parabolic dish (with \spfeed{}) can be thought
of as a densely sampled aperture, with the mechanical structure performing
the beamforming. If the \aa{} is dense, such that the aperture is
fully sampled (see \prettyref{app:sensitivity-requirements}), then
$\Nst\propto1/\Dst^{2}$. But when the antenna elements are sufficiently
spaced such that the AA is sparse over some or all frequencies, then
$\Dst$ is a function of two independent parameters, $\Nest$ and
the average inter-element spacing $\deeavg$, as shown below. 

Array effective area is given in \prettyref{eq:Aearr}, but it can
be more generally described by 
\begin{equation}
\Aearr(\nu)=\Nst\Aest(\nu),\label{eq:Aearr-general}
\end{equation}
where $\Aest$ is the station effective area at some frequency $\nu$.
Assuming an irregular intra-station element layout with an approximately
uniform element distribution, 
\begin{equation}
\Aest(\nu)\approx\Nest\Aee(\nu),\label{eq:Aest}
\end{equation}
where $\Aee(\nu)$ is the antenna element effective area\nomenclature[sAee]{$\Aee$}{\TAee}.
For an isolated antenna element, 
\begin{equation}
\Aee{\rm (isolated)}=\frac{\GGe\lambda^{2}}{4\pi},
\end{equation}
where $\GGe$ is the antenna gain\nomenclature[sG~e]{$\GGe$}{\TGGe}.
However, in the presence of neighbour elements, $\Aee$ cannot be
greater than the available physical area \citep{NijPan09}, thus 
\begin{equation}
\Aee\approx{\rm min\left\{ \frac{\GGe\lambda^{2}}{4\pi},\, available\, geometric\, area\right\} },\label{eq:Aeff-ele-min}
\end{equation}
where, for a uniform intra-station element distribution, the available
geometric area is approximated by~$\deeavg^{2}$. Restating \prettyref{eq:Aeff-ele-min}
in terms of the frequency-dependent station filling factor (\prettyref{eq:FFst-AonA})
gives
\begin{equation}
\Aee=\deeavg^{2}\FFst,
\end{equation}
where 
\begin{equation}
\FFst\approx{\rm min\left\{ \frac{\GGe\lambda^{2}}{4\pi\deeavg^{2}},\,1\right\} }.\label{eq:FFst-detailed}
\end{equation}
As expected, $\FFst\le1$; a filling factor of 1 signifies that the
array is dense.

In its full form, \prettyref{eq:Aearr-general} is thus
\begin{equation}
\Aearr(\nu)\approx\Nst\Nest\deeavg^{2}\FFst(\nu).\label{eq:Aerr-full-form}
\end{equation}
When the function $\Aearr$ is held constant, 
\begin{equation}
\Nst\propto\frac{1}{\Nest\FFst\deeavg^{2}}.\label{eq:Nst-Nest-FF-trade}
\end{equation}
If $\deeavg$ and the function $\FFst$ also remain constant (i.e.
$\GGe$ does not vary in $\FFst$), then
\begin{equation}
\Nst\propto\frac{1}{\Nest}.\label{eq:Nst-Nest-trade}
\end{equation}
This is not applicable for intra-station layouts with non-uniformly
distributed antenna elements (see \prettyref{app:sensitivity-requirements}),
because the approximation for $\Aest$ (\prettyref{eq:Aest}) does
not hold.

The relationship with station diameter can also be determined, assuming
that the array is dense at the lowest frequency. \prettyref{eq:Aeff-ele-min}
shows that $\Aee\apprle\deeavg^{2}$, therefore the maximum station
effective area is $\Aest{\rm (max)}\approx\Nest\deeavg^{2}.$ From
geometry, $\Aest{\rm (max)}\propto\Dst^{2}$, hence 
\begin{equation}
\Dst\propto\sqrt{\Nest}\deeavg.\label{eq:Dst-prop-Nest-deeavg}
\end{equation}
This relationship is important, because it shows that station diameter
is not an independent parameter, but is influenced by both the number
of elements in the station and the average spacing between the elements. 

The veracity of \prettyref{eq:Nst-Nest-FF-trade} can be confirmed
by considering the extreme cases. If the array is completely dense,
such that $\FFst=1$, then 
\begin{equation}
\Nst\propto1/\Dst^{2}.
\end{equation}
If the array is completely sparse, then 
\begin{equation}
\FFst=\frac{\GGe\lambda^{2}}{4\pi\deeavg^{2}}
\end{equation}
and 
\begin{equation}
\Nst\propto\frac{1}{\Nest\Aee{\rm (isolated)}}.
\end{equation}

\subsection{Further work to refine station performance metrics\label{app:station-performance-further-work}}

Matching the top-level science requirements to the telescope performance
requires an understanding of the telescope's $\AonT$ and processed
FoV performance. The $\AonT$ performance of a higher-frequency dish-based
aperture synthesis telescope is well-understood \citep[e.g.][]{CraNap89}.
The $\AonT$ performance of an aperture synthesis telescope composed
of aperture array stations, in the lower frequency, sky noise dominated
regime, is more complex \citep[e.g ][]{WijCap11}. 

Considering the factors affecting sensitivity discussed in \prettyref{app:sensitivity-requirements},
the requirements and performance metrics can be specified as $\AonT(\theta,\phi,\nu)$
and $\Oreq(\theta,\phi,\nu)$, or at least a minimum $\AonT$ and
$\Oreq$ as a function of frequency over some range of $\tp$. Some
simple rules-of-thumb are used in this analysis, but more accurate
estimates could be obtained through array layout simulation software,
such as Xarray%
\footnote{\url{http://sites.google.com/site/xarraytool/}%
} and OSKAR \citep{DulMor09}. 

For each layout, simulation of a small set of interdependent input
parameters and performance metrics would be useful inputs for the
parametric analysis. They are:
\begin{itemize}
\item input parameters

\begin{itemize}
\item antenna element pattern
\item intra-station layout and inter-element spacing
\item number of elements per station
\item number of elements per tile (if used)
\end{itemize}
\item performance metrics

\begin{itemize}
\item effective area as a function of frequency and direction 
\item station beam FoV as a function of frequency and direction.
\end{itemize}
\end{itemize}
A layout will also have advantages and disadvantages which cannot
be captured in the parametric analysis (e.g. station beam pattern,
sidelobes and calibration); these must be considered separately. 

Improved accuracy will also be obtained if the simulations take into
account effects such as
\begin{itemize}
\item mutual coupling between elements
\item the non-ideal gain of the antenna element, as a function of $\nu$,
$\theta$ and $\phi$ (this parametric analysis assumes ideal gain)
\item LNA response
\item the effect on gain due to the analogue beamformer, as a function of
$\nu$, $\theta$ and $\phi$.
\end{itemize}

\section{Smaller station diameter example\label{app:station-diameter-example}}

We present a simple comparative example showing the effect of smaller
stations, where the diameter of every single-band, low-band or high-band
station is halved. The system details for the single and dual-band
implementations of this half-diameter example are listed in  \prettyref{tab:smaller-station-comparison-dual}. 

\begin{table}[!t]
\caption{\SKAiLow{} station details for the half-diameter station example.\label{tab:smaller-station-comparison-dual}}

\centering{}%
\begin{tabular}{l>{\raggedright}m{6em}cc}
\toprule 
 & \multirow{2}{6em}{\centering{}Single-band} & \multicolumn{2}{c}{Dual-band}\tabularnewline
\cmidrule{3-4} 
 &  & Low band & High band\tabularnewline
\midrule
Diameter & \centering{}90\,m & 90\,m & 45\,m\tabularnewline
Number of elements per station & \centering{}2\,800 & 2\,800 & 2\,800\tabularnewline
Number of stations & \centering{}200 & 200 & 200\tabularnewline
Average spacing between elements & \multicolumn{1}{c}{1.5\,m} & \multicolumn{1}{c}{0.75\,m} & \multicolumn{1}{c}{1.5\,m}\tabularnewline
\bottomrule
\end{tabular}
\end{table}

\subsection{Station hardware costs\label{app:Nst-Nest-trade-station}}

\begin{figure}[!t]
\begin{centering}
\includegraphics[height=0.68\textwidth]{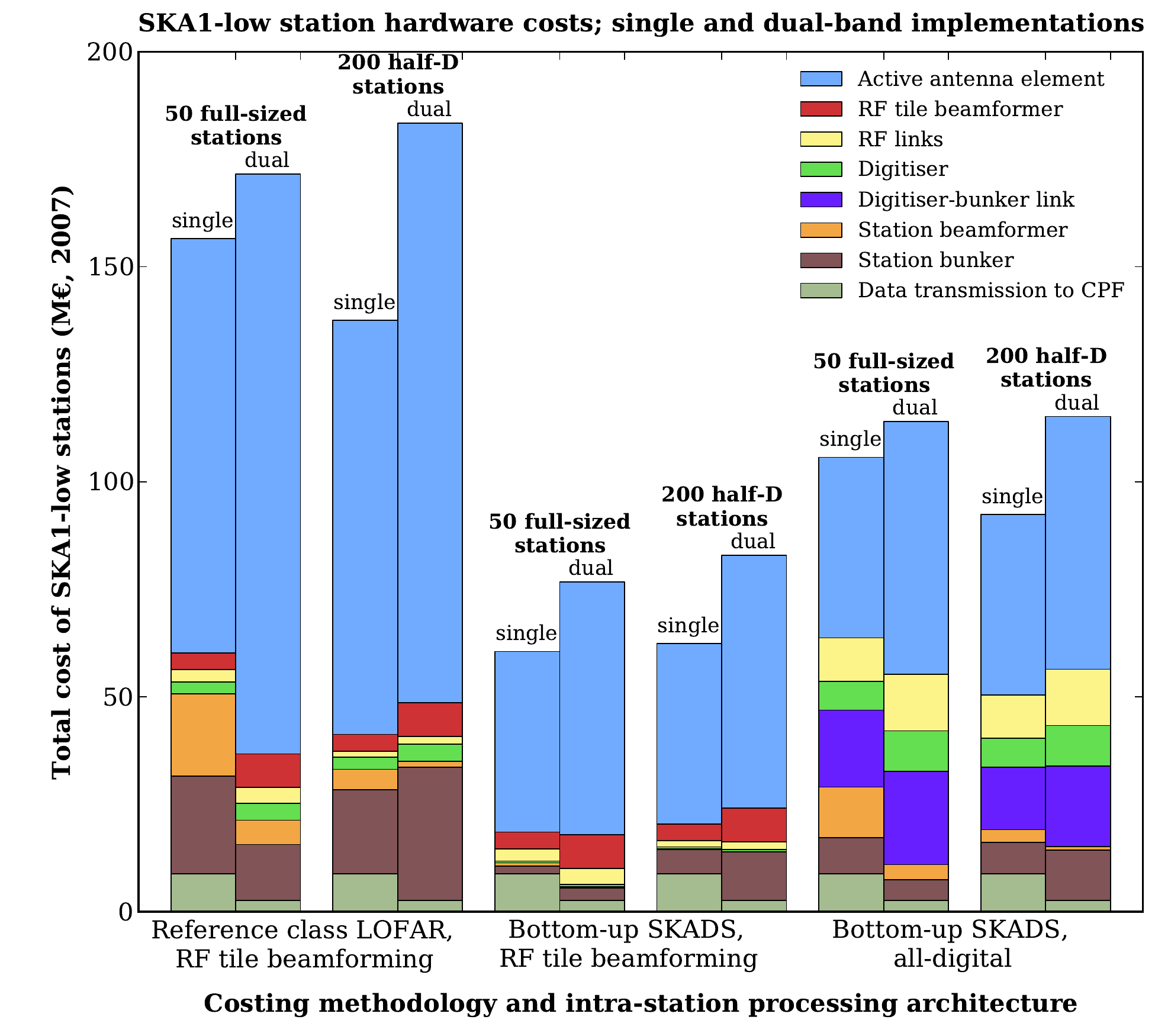}\includegraphics[height=0.68\textwidth]{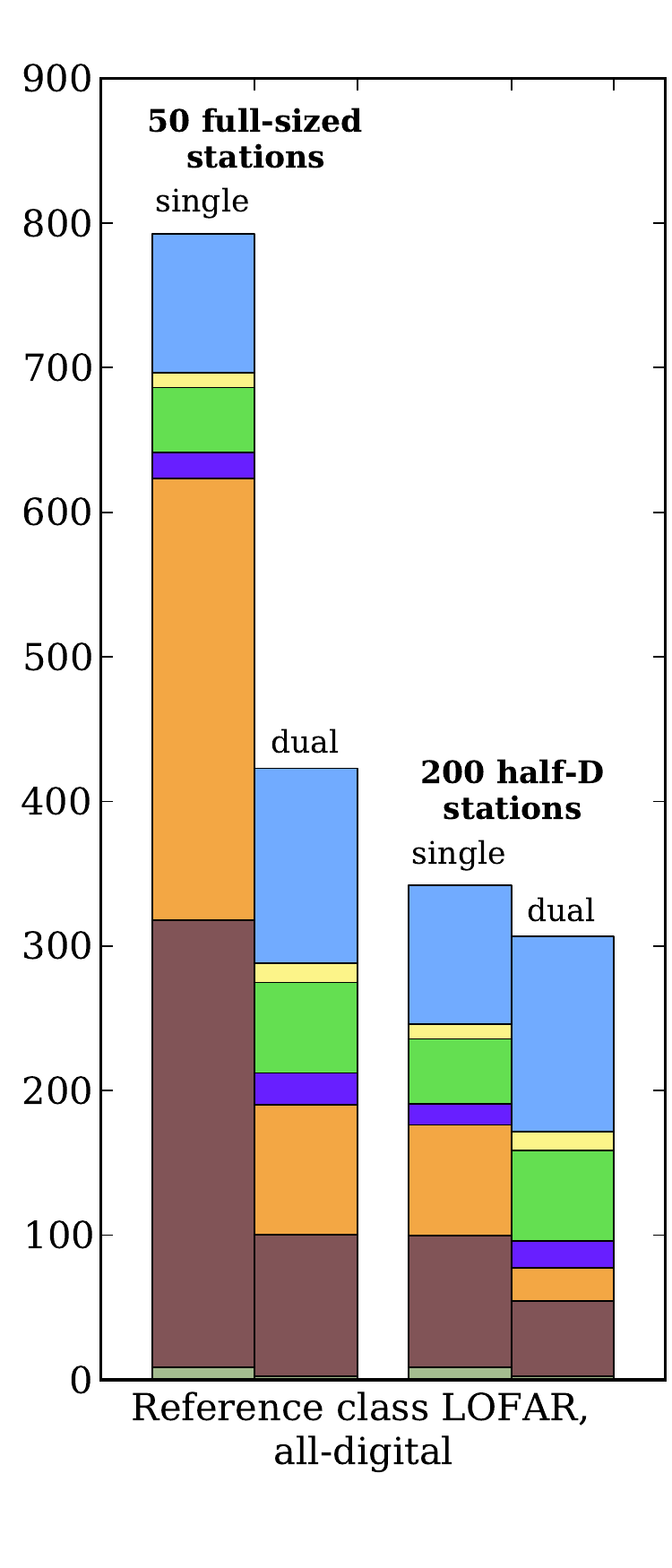}
\par\end{centering}

\caption{\SKAiLow{} station hardware cost, for 50~full-sized stations (per
band) and 200~stations of half the diameter, hence a quarter of the
number of elements per station. Other details as per \prettyref{fig:station-costs}.
\label{fig:cost-SKAlow-stations-smaller}}
\end{figure}

Costing the smaller station diameters requires consideration of the
location of the station processing node. As in the dual-band implementation,
we assume that each smaller station continues to have its own station
beamformer processing node---the bunker. The only change to the cost
model (\prettyref{app:models-SKAlow}) is to halve the unit cost of
the tile--digitiser RF link to \euro40 per link, due to the smaller
station diameter. For the all-digital architecture, the digitiser--bunker
link cost equation already has a distance-dependent term, so remains
the same. Also, the dual-band cost multipliers (\prettyref{app:dual-band-costs})
do not change. 

\prettyref{fig:cost-SKAlow-stations-smaller} plots the hardware cost
of all the \SKAiLow{} stations in the half-diameter example alongside
that of the full-sized stations previously shown in \prettyref{fig:station-costs}.
Because the number of elements per station differ by a factor of 4,
the cost of all stations is a more useful comparison than the cost
per station. The total number of elements in the array remain constant,
hence the total cost of the active antenna elements, RF links (for
the all-digital architecture) and digitisers do not vary. Compared
to the full-sized stations, the smaller average distance to the bunker
reduces the total cost of the RF links in the RF tile beamforming
architecture and the digitiser--bunker links in the all-digital architecture.

The average number of station beams is varied to maintain the required
FoV. For this reason, the total processing cost of $\Nst$ beamformers
(\prettyref{eq:cost-BF-total}, p~\pageref{eq:cost-BF-total}) can
be approximated by
\begin{equation}
P_{{\rm BF-total}}\propto\Nest,
\end{equation}
which is a factor of 4 reduction for the half-diameter stations. But
despite the decrease in intra-station signal transport and beamformer
processing costs, the station bunker puts upwards cost pressure on
the total cost. Although a part of the station bunker cost varies
with station beamformer processing, the fixed-cost portion of each
bunker begins to become significant with the smaller stations. Because
there are more bunkers (one for each station), the total fixed cost
portion of the bunker is higher. This trend is further exacerbated
in the dual-band implementation, where twice as many bunkers are used
(one for each low-band and high-band station). 

In the present analysis, the fixed cost portion of the bunker is set
at 20\,\% of the source data cost estimates (\skadsCost{} and \lofarCost{});
the other 80\,\% is assumed to scale linearly with the amount of
station processing. Although the fixed cost is a zeroth-order estimate,
its purpose is to recognise that there may be inefficiencies in providing
environmental conditioning (cooling), RFI shielding and power to many
smaller controlled environments (the bunkers), as well as higher construction,
testing and deployment costs. Potential solutions such as shared processing
nodes are discussed in \prettyref{app:shared-processing-nodes}.

\subsection{System implications of trading \texorpdfstring{$\Nest$}{Ne/st} for \texorpdfstring{$\Nst$}{Nst}\label{app:Nst-Nest-trade-system}}

As in the single and dual-band comparison (\prettyref{sec:variable-cost-system-implications}),
there are other costs that vary with the $\Nst$ vs. $\Nest$ trade-off.
The only costs in \prettyref{sec:variable-cost-system-implications}
which vary are the correlator and imaging processor costs. The constant
$\Aearr$ means that the total area occupied by the array and the
total number of antenna elements in the array do not change, thus
the site preparation and antenna element deployment costs do not change.
Of course, there will still be twice as many antenna elements in the
dual-band implementation and it will occupy an extra 25\,\% area,
as shown in \prettyref{tab:attribute-comparison}. The \nip{} cost
does not change, because it depends on the diameter of the core, which
remains the same.

The correlator and imaging processor costs do increase as the number
of elements per station are traded for more stations, as derived in
\prettyref{app:central-processing}. These increases apply equally
to both the single and dual-band implementations. The correlator processing
(\prettyref{eq:correlator-processing-Nest}) is 
\begin{equation}
\Pcorr\propto\Nest^{1-\alpha},
\end{equation}
where $1<\alpha<2$ depending on the correlator design and technologies
used. The correlator output data rate, and hence imaging processor
cost (\prettyref{eq:correlator-data-out-Nest}) is 
\begin{equation}
\Rcorrout\propto\frac{1}{\Nest^{3/2}}.
\end{equation}
For the half-diameter station example, where there are a quarter of
the number of elements per station, the correlator cost increases
by up to 400\,\%. Although the number of correlations increases by
$\Nst^{2}$ (i.e a factor of 16), the number of beams required to
produce the $\unit[20]{deg^{2}}$ processed FoV is reduced by a factor
of 4 for the half-diameter stations, resulting in only a factor of
4 increase in correlation cost.  For the post-correlation imaging
processor, the cost increases by 800\,\%. 

\begin{figure}[!tp]
\begin{centering}
\includegraphics[width=0.95\textwidth]{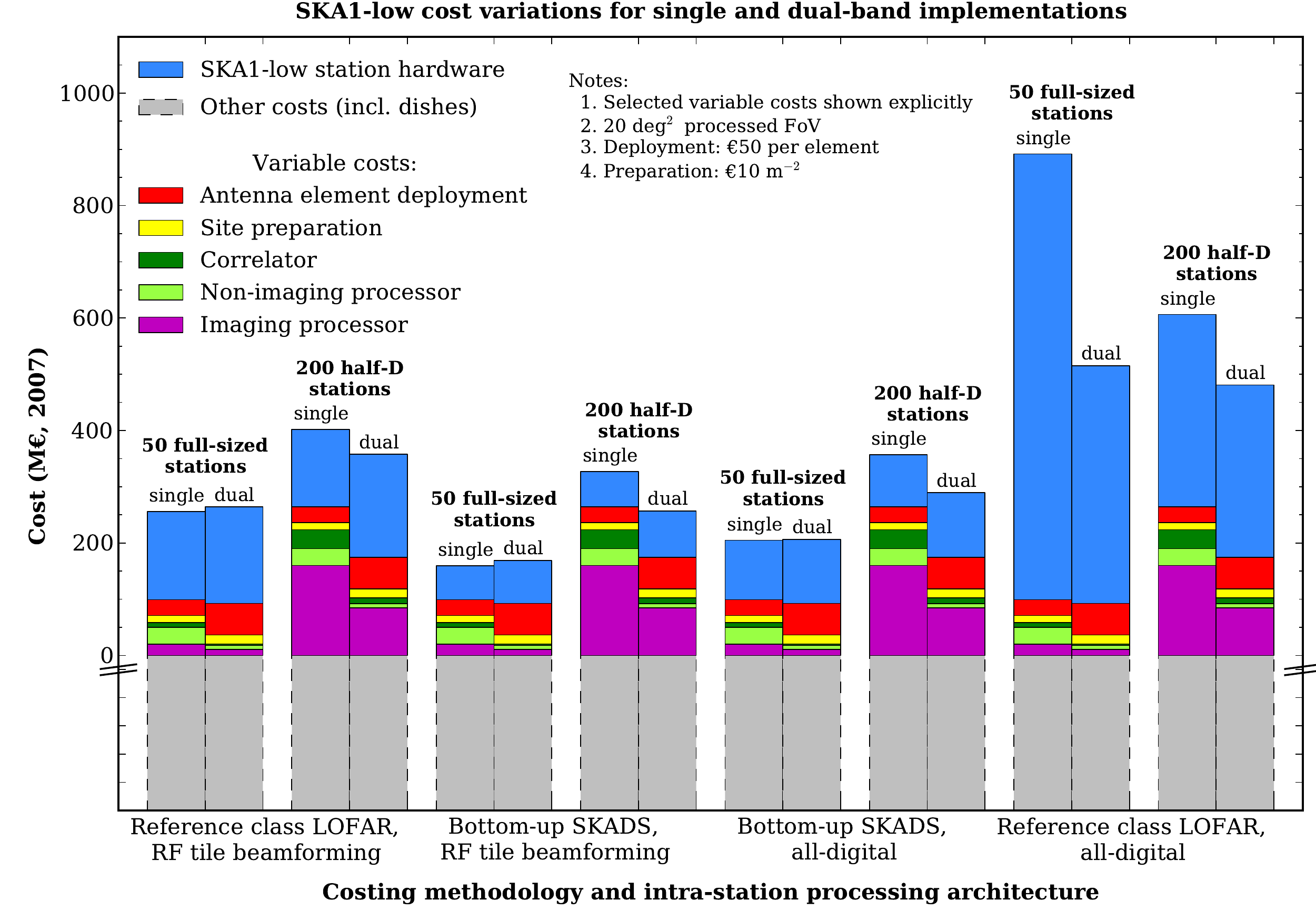}
\par\end{centering}

\centering{}\caption{Comparison of significant variable costs (excluding power) for 50~full-sized
stations (per band) and 200~stations of half the diameter. The correlator
processing cost is assumed to be scale as $1/\Nest$. Other details
as per \prettyref{fig:total-cost-50dep-10site}.\label{fig:total-cost-50dep-10site-smaller-stations}}
\end{figure}

\prettyref{fig:total-cost-50dep-10site-smaller-stations} plots the
variable costs, for the zeroth-order cost estimates used in \prettyref{sec:variable-cost-system-implications}.
A value of $\alpha=2$ is used for the correlator processing cost.
The correlator and imaging processor costs are significant for the
half-diameter example, more so for the single-band implementation.
Although the cost for the \lofarDig{} scenario remains significantly
larger than the others, the cost of single-band implementation of
that scenario has reduced considerably.

The principal cost differences arise from whether the processing is
distributed amongst the station beamformers or centralised at the
correlator and image processor, however the magnitude of the cost
differences depend on the implementation. The power demand, which
scales with the amount of processing, also follows the same trend.
The power demand of the station is not plotted, because it should
be considered in context with the power demand of the correlator and
imaging processor, which is beyond the scope of the present analysis.

\subsection{Station diameter and shared processing nodes\label{app:shared-processing-nodes}}

To control the bunker cost, which is higher due to the increased
number of stations, shared station nodes may be the preferable option
for some architectures. Conceptually, this leads to two different
sized stations: a `logical' and `physical' station, to borrow
from software engineering terminology \citep[e.g.][]{Kru95}. The
logical (functional) station is the beamformed station, where the
beams are input to the correlator. The physical (infrastructure) station
has one node (bunker) shared between a number of the beamformed stations.
In the centrally condensed part of the array, where there is a `sea'
of antenna elements, the logical stations could be re-configurable
in diameter, to suit the scientific application~\citep{Ale11}.

\begin{figure}[!t]
\centering{}\includegraphics[scale=1.3]{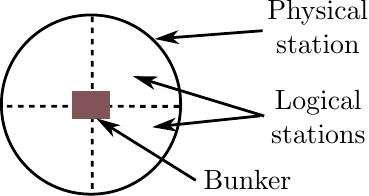}\caption{Schematic of logical stations within a physical station, and the shared
processing node (bunker).\label{fig:shared-processing-node}}
\end{figure}

As shown in \prettyref{fig:shared-processing-node}, an obvious alternative
in the half-diameter station example is to have 4 logical stations
per physical station, which would mean the physical station size is
similar to the full-diameter stations. The would be viable if the
cost reduction of a single larger bunker outweighs the higher intra-station
link costs. The link costs are higher due to the larger average distance
to the bunker; inefficiencies and losses could be introduced in the
longer RF and power cables, or alternatively, more expensive cables
or digital transmission equipment may be required. A higher ratio
of logical to physical stations may be cost-effective, especially
for the high band of the dual-band array, which is already smaller
in diameter.

There are some real-world examples of \aa{} telescopes which have
differing logical and physical stations. The \MWA{} (MWA) is a telescope
which will have 128 \aa{} tiles, each with an RF beamformer, distributed
out to a distance of 1.5\,km from the array centre%
\footnote{\url{http://www.mwatelescope.org/instrument/specs.html}%
}. A receiver node digitises and applies a coarse filterbank to the
single tile beam from each RF beamformer and transmits the digital
data to the correlator \citep{LonCap09}. The receiver node serves
8 tiles, rather than each beamformer having its own node. In this
context, the tiles, as inputs to the correlator, are the logical stations
and the 8 tiles connected to the receiver node forms the physical
station. The LOFAR stations also have different logical and physical
stations. The `core' stations at the inner region of the telescope
are each composed of a set of low band antennas (LBA) and two sets
high band antennas (HBA). These antennas are served by a single processing
node \citep{GunBen10}.

The required geographical layout of stations, which determines \uvc{},
is another trade-off consideration. To maintain comparable infrastructure
costs, the half-diameter station example assumes that the 4 smaller
stations are located adjacent to each other, as will already be the
case in the centrally condensed part of the array. But for stations
placed outside of the centrally condensed part of the array, `clustering'
or grouping of antennas is proposed as a method to reduce the infrastructure
costs \citep{BolMil11-Config}. Beyond the 5\,km `inner' region
of \SKAi{}, the HLSD describes clusters of receptors, containing
5 dishes and a single \SKAiLow{} station. Each physical station could
be divided into 4 adjacently located smaller logical stations. However,
if the purpose of reducing station diameter (thereby increasing the
number of stations) is to improve \uvc{} through more diverse placement
of stations, then such a strategy will not be particularly useful.
Estimating the extra infrastructure cost to separate the stations
with larger distances is beyond the scope of this analysis.

\section{Reduced fixed beam--bandwidth product\label{app:beam-bandwidth-trade}}

If the \aa{} system has a fixed processing capacity, then the processed
FoV and processed bandwidth are tradable quantities. There are a few
contemporary examples of this approach to \aa{} system design:
\begin{itemize}
\item The trade-offs in the SKADS \aa{} designs are based on the capacity
to transmit data from the station to the \cpf{} being a primary limitation
\citep{BolAle09}.
\item The LOFAR station processing is reconfigurable such that processed
bandwidth can be traded for station beams, thus maintaining a set
data rate from the station \citep{VosGun09}.
\item The MWA design has 220\,MHz of sampled bandwidth available, but at
any one time it only transports 30.72\,MHz of this bandwidth to the
central processing hardware \citep{LonCap09}.
\item The \LWA{} (LWA) design constrains the bandwidth for multiple station
beams, due to limitations in transporting data from the stations to
the correlator \citep{EllCla09}.
\end{itemize}
The amount of FoV that can be processed over some bandwidth can be
constrained by fixing the beam--bandwidth product, $\beamBW$\nomenclature[sNbstavg~BW]{$\beamBW$}{\TbeamBW{}},
where $\Nbstavg$ is the average number of beams formed over the bandwidth
$\BW$. Assuming that the data consists of many channelised beams
(\prettyref{app:FX-architecture}), the processing capacity of the
station beamformer, station--CPF data transmission and the central
processing sub-systems can then be determined by $\beamBW$.

\subsection{Strawman details\label{app:strawman-beam-bandwidth-product}}

An alternative analysis by Faulkner%
\footnote{AA-low system \& station architecture for SKA\textsubscript{1}, 18
January 2012%
} of the \SKAi{} \DRM{} (\DRMi{}) version 2.0 \citep{SKA11-DRM20}
considers the beam--bandwidth product required for individual \DRMi{}
chapters. We do not examine the details of such an analysis, but simply
consider a strawman example where the largest beam--bandwidth product
is defined by requirement to observe $\unit[20]{deg^{2}}$ processed
FoV over the 70--180\,MHz band. However, we note that a frequency
split at 200\,MHz (hence a 70--200\,MHz low band) for the dual-band
implementation would provide a better fit to the science relevant
to \SKAiLow{}  in the current \DRMi{} (Chapters 2--5); three of
those science chapters specify either a maximum or minimum frequency
of 200\,MHz.

For this strawman, $\unit[20]{deg^{2}}$ FoV (70--180\,MHz) equates
to $\Nbstavg=44$ over the band; this applies to both the single and
dual-band implementations, because the station diameters are equal
between 70 and 180\,MHz. The beam--bandwidth product is thus 4.8\,GHz.
In contrast, to achieve $\unit[20]{deg^{2}}$ FoV (70--450\,MHz)
for the single and dual-band representative implementations, the beam-bandwidth
capacity is 80\,GHz and 24\,GHz respectively.

The same 4.8\,GHz of beam--bandwidth product can be applied to the
dual-band array, although the strawman requires further design decisions
to be assumed. We maintain the original assumption of separate low
and high-band cores \prettyref{ass:low-high-cores}, hence the same
4.8\,GHz beam--bandwidth requirement applies to each station in each
core. The stations beyond the core are co-located \prettyref{ass:trenching-cabling-co-location},
but we now assume that the station node hardware (beamformer, bunker
and station--CPF data transmission) is shared between each pair of
low and high-band stations; this is conceptually similar to the LOFAR
station design \citep{GunBen10}. A different design of the dual-band
strawman could have a single core populated with both low and high-band
stations, with shared processing. However, this would result in a
larger core, which will have consequences on the science applications,
such as low surface brightness density and non-imaging processing
observations.

Because the station beamforming computational capacity and the maximum
rate of station--CPF data transmission is defined by the beam--bandwidth
product, observations with other processed FoV and bandwidth combinations
are possible, although the observations cannot be concurrent. For
example, an observation over the 180--450\,MHz frequency range with
the single-band array results in a processed FoV of $\Ostproc=\unit[1.3]{deg^{2}}$.
The same observation with the dual-band results in a processed FoV
four times larger, due to the smaller diameter of the high-band station.
 \textbf{}

\subsection{Station hardware costs\label{app:reduced-FoV-station-hardware-costs}}

\begin{figure}[!t]
\includegraphics[height=0.68\textwidth]{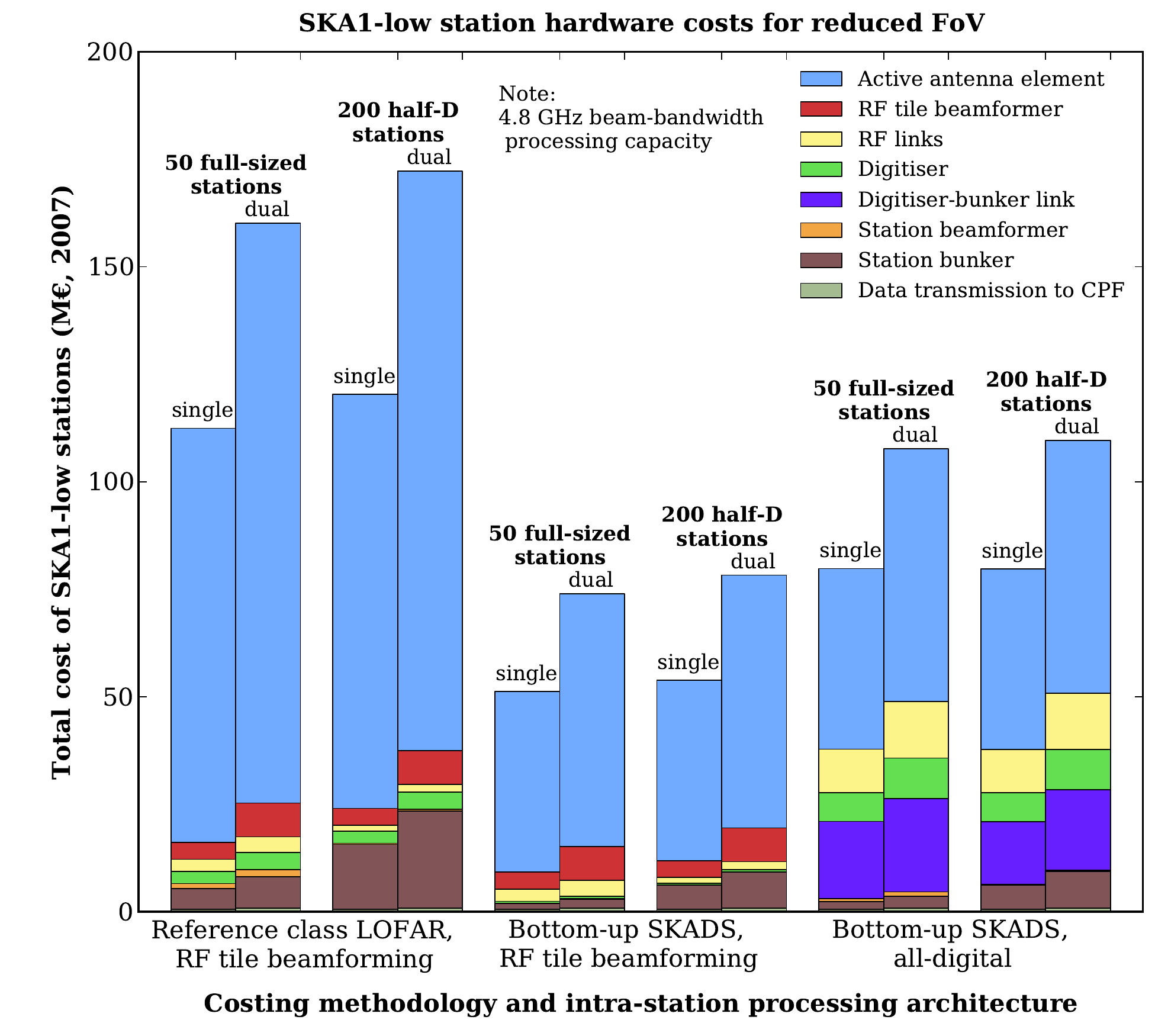}\includegraphics[height=0.68\textwidth]{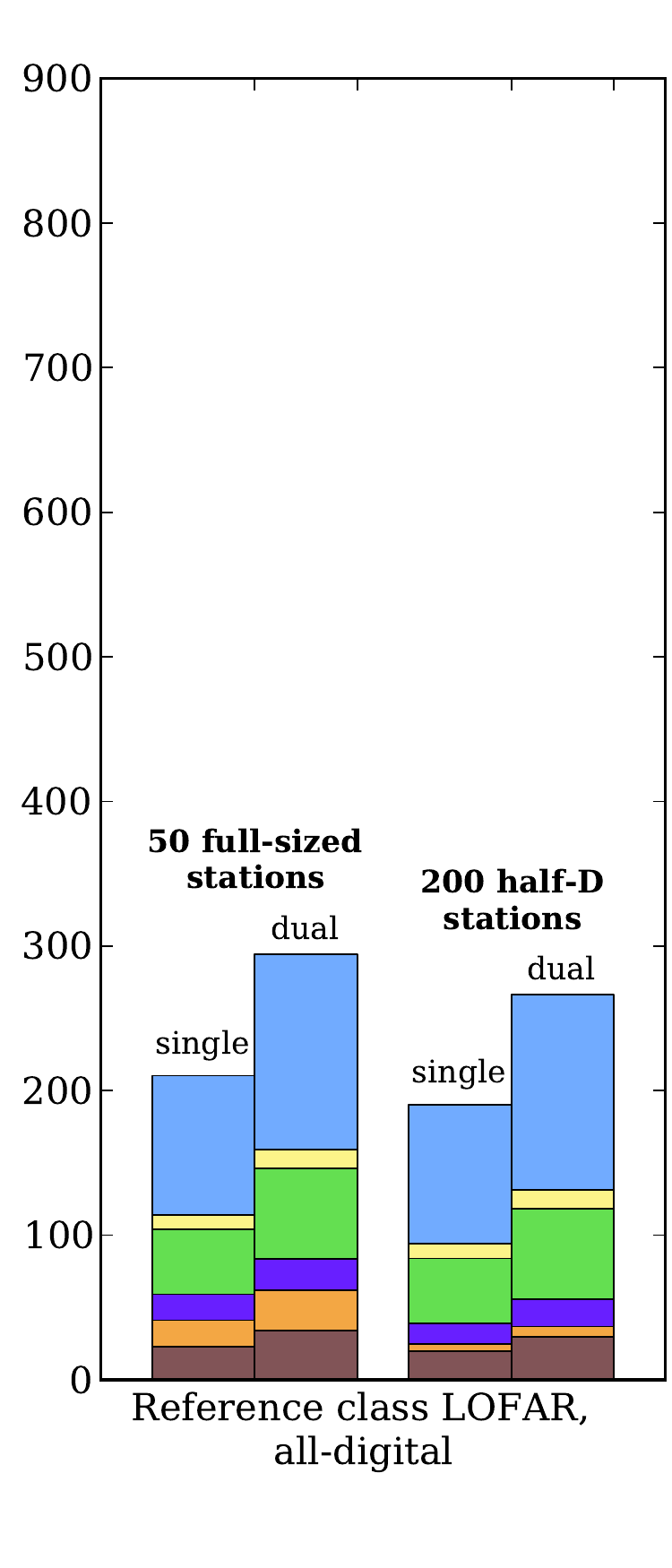}\caption{\SKAiLow{} station hardware cost for a beam--bandwidth product limited
to 4.8\,GHz, which is equivalent to a processed FoV of $\unit[20]{deg^{2}}$
over a 70--180\,MHz band. Other details as per \prettyref{fig:cost-SKAlow-stations-smaller}.
\label{fig:cost-SKAlow-stations-smaller-low-FoV}}
\end{figure}

\prettyref{fig:cost-SKAlow-stations-smaller-low-FoV} shows the station
hardware costs for the fixed 4.8\,GHz beam--bandwidth product. For
all scenarios, the single-band implementation is cheaper than the
dual-band, at approximately 70\,\% of the dual-band cost. The station
beamformer and station--CPF data transmission costs are now insignificant.
This is because the beam--bandwidth capacity of the station node hardware
in \prettyref{fig:cost-SKAlow-stations-smaller-low-FoV} is reduced
by a factor of approximately 17 and 5, compared to the representative
single and dual-band implementations respectively. The station bunker
cost does not reduce to the same extent as the station beamformer,
due to the fixed cost portion of the bunker (see \prettyref{app:Nst-Nest-trade-station}).
Meanwhile, the cost of the station hardware sub-systems located in
the signal path prior to the station beamformer remain the same; they
are independent of changes to the beam--bandwidth product. The net
result is a reduction in station hardware cost for each scenario,
compared to when processed FoV of $\unit[20]{deg^{2}}$ over 70--450\,MHz
is required. 

The previous results, calculated for the representative implementations,
show that the dual-band implementation puts downward pressure on the
station beamformer and station-CPF transmission costs, which counteracts,
to varying extent, the increase in the cost of other station hardware
sub-systems in the dual-band implementation. But for this strawman,
which is limited by the beam--bandwidth product, the downward pressure
is insignificant. Thus in all the scenarios, the cost of the dual-band
station hardware is higher. The increase in cost not only depends
on the cost data source and intra-station architecture, but given
the dominance of the active antenna element costs, the cost multipliers
(\prettyref{app:dual-band-costs}) used to determine the active antenna
element costs for the dual-band implementation will have a strong
influence.

\subsection{System implications\label{app:reduced-FoV-system-implications}}

\begin{figure}[!tp]
\begin{centering}
\includegraphics[width=0.95\textwidth]{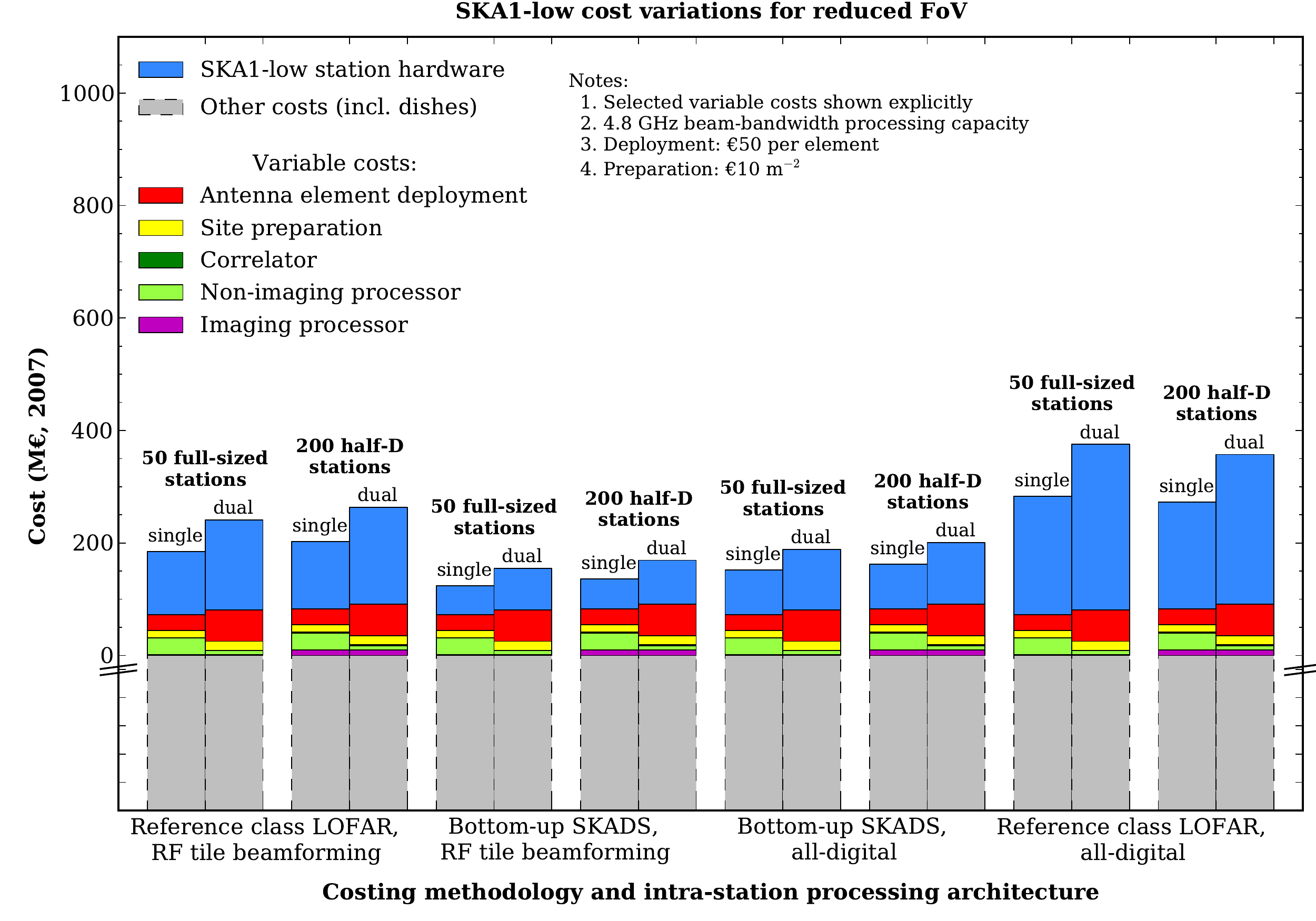}
\par\end{centering}

\centering{}\caption{Comparison of significant variable costs (excluding power) for a beam--bandwidth
product limited to 4.8\,GHz. Other details as per \prettyref{fig:total-cost-50dep-10site-smaller-stations}.\label{fig:total-cost-50dep-10site-smaller-stations-low-FoV}}
\end{figure}

A fixed beam--bandwidth product also has implications for the downstream
processing. The processing capacity\textbf{ }of the correlator and
imaging processor sub-systems scales linearly with the beam--bandwidth
product (see \prettyref{app:central-processing}). The cost of these
sub-systems is thus $4.8/80=6.0\%$ of the single-band costs listed
in \prettyref{tab:other-variable-costs} (p \pageref{tab:other-variable-costs}).
Because the processing capacity of these sub-systems is now defined
by the same beam--bandwidth product for each implementation, the cost
does not change between the single and dual-band implementation. However,
the cost scaling effects due $\Nst$ vs. $\Nest$ trade (\prettyref{app:Nst-Nest-trade-system})
still apply: more stations of smaller diameter increase the correlator
and imaging processor costs.

\prettyref{fig:total-cost-50dep-10site-smaller-stations-low-FoV}
shows the significant variable costs for \SKAiLow{}. For all scenarios
in \prettyref{fig:total-cost-50dep-10site-smaller-stations-low-FoV},
the single-band implementation is cheaper than the dual-band, at approximately
80\,\% of the dual-band cost. The station hardware, correlator and
imaging processor costs are smaller than when a processed FoV of $\unit[20]{deg^{2}}$
over 70--450\,MHz is required (\prettyref{fig:total-cost-50dep-10site-smaller-stations-low-FoV});
the latter two are no longer significant costs. The antenna element
deployment and site preparation costs do not change, because the number
and location of the antenna elements is independent of changes to
the beam--bandwidth product. The \nip{} also remains unchanged,
because the strawman presented in this section still has a separate
high-band core, with half the diameter of the single-band core.

\pagebreak{}

\addtocontents{toc}{\protect\setcounter{tocdepth}{2}}
\end{document}